\documentclass[lettersize,journal]{IEEEtran}
\usepackage{amsmath,amsfonts}
\usepackage{algorithmic}
\usepackage{algorithm}
\usepackage{array}
\usepackage[caption=false,font=normalsize,labelfont=sf,textfont=sf]{subfig}
\usepackage{textcomp}
\usepackage{stfloats}
\usepackage{url}
\usepackage{verbatim}
\usepackage{graphicx}
\usepackage{cite}
\usepackage{subfig}
 \usepackage{ booktabs }
 
 \usepackage{ multirow}
\usepackage{color}

\begin{document}

\title{VEDA: Uneven light image enhancement via a vision-based exploratory data analysis model}

\author{Tian Pu,
        Shuhang Wang,~\IEEEmembership{}
        Zhenming Peng,~\IEEEmembership{}
        and~Qingsong Zhu *~\IEEEmembership{}
        \IEEEcompsocitemizethanks{\IEEEcompsocthanksitem Tian Pu, and Zhenming Peng are with the School of Information and Communication Engineering, University of Electronic Science and Technology of China,
e-mail: putian@uestc.edu.cn; zmpeng@uestc.edu.cn
\IEEEcompsocthanksitem Shuhang Wang is with the 
Center for Ultrasound Research\&Translation, Massachusetts General Hospital, Harvard Medical School, email: swang38@mgh.harvard.edu
\IEEEcompsocthanksitem Qingsong Zhu is with the 
Shenzhen Institutes of Advanced Technology, Chinese Academy of Sciences, Shenzhen, email: qs.zhu@foxmail.com (corresponding author)
}

}

\markboth{IEEE Journal Template}
{Shell \MakeLowercase{\textit{et al.}}: A Sample Article Using IEEEtran.cls for IEEE Journals}


\maketitle

\begin{abstract}
Uneven light image enhancement is a highly demanded task in many industrial image processing applications. Many existing enhancement methods using physical lighting models or deep-learning techniques often lead to unnatural results. This is mainly because: 1) the assumptions and priors made by the physical lighting model (PLM) based approaches are often violated in most natural scenes, and 2) the training datasets or loss functions used by deep-learning technique based methods cannot handle the various lighting scenarios in the real world well. In this paper, we propose a novel vision-based exploratory data analysis model (VEDA) for uneven light image enhancement. Our method is conceptually simple yet effective. A given image is first decomposed into a contrast image that preserves most of the perceptually important scene details, and a residual image that preserves the lighting variations. After achieving this decomposition at multiple scales using a retinal model that simulates the neuron response to light, the enhanced result at each scale can be obtained by manipulating the two images and recombining them. Then, a weighted averaging strategy based on the residual image is designed to obtain the output image by combining enhanced results at multiple scales. A similar weighting strategy can also be leveraged to reconcile noise suppression and detail preservation. Extensive experiments on different image datasets demonstrate that the proposed method can achieve competitive results in its simplicity and effectiveness compared with state-of-the-art methods. It does not require any explicit assumptions and priors about the scene imaging process, nor iteratively solving any optimization functions or any learning procedures.
\end{abstract}


\section{Introduction}
\IEEEPARstart{M}{any} image/video processing applications, such as surveillance, daily photography and aerial imaging, require high-quality images that preserve scene contrasts faithfully \cite{chen2010natural}. However, images taken in unevenly-lit scenes often fail to meet this requirement due to limited dynamic ranges. 

To address this problem, uneven light image enhancement methods are highly demanded and extensive researches have been made over the past decades. In the following, we classify and briefly review the related studies from the viewpoint of uneven light image enhancement.
\subsection{Histogram specification based enhancement}

Histogram specification (HS) is one of the most widely used techniques. Early HS based methods attempt to achieve desired output histogram shapes under the assumption that visually-pleasing images have ideal histogram characteristics. However, due to the lack of universal criteria for determining the ideal histogram for natural images, these methods often lead to detail loss and over-stretching of contrasts. Later studies improve the performance by applying restrictions on contrast stretch, such as hue and range preservation \cite{NikolovaFast}, gamma correction \cite{Shih2013Efficient}, saliency preservation \cite{Gu2015saliency}, and contrast limitation \cite{2014XuGeneralized}. Recently, swarm optimization technique is introduced to the HS, focusing on brightness and feature preservation\cite{ashkumar2020salp}. However, the enhanced results are prone to unnatural artifacts.
\vspace{-2mm}
\subsection{Physical lighting model based enhancement}

Taking the imaging process of natural images into account is a viable approach for image enhancement, and many existing methods vary in their construction of physical lighting models. An especially worth citing category of techniques is enhancement methods based on Retinex theory, which assumes that an image is the element-wise product of the illumination image and the reflectance image. Early Retinex-based methods take the reflectance image as the enhanced result \cite{JobsonA}, but the results often suffer from unnatural appearance. To overcome this issue, subsequent advances aim to modify the estimated illumination image instead of removing it. Wang et al. \cite{Shuhang2013Naturalness} propose a naturalness preserved enhancement method by designing a bright-pass filter to recover the illumination image. Liang et al. recover illumination by iteratively solving a diffusion filtering equation. This approach is effective in preserving texture details\cite{LiangContrast}. Wang and Luo present a multi-layer model to decompose an image into a reflectance layer and a cascaded sequence of illumination \cite{WangNaturalness}. Contrast can be enhanced by adjusting the dynamic range of each illumination layer. The variational approach for Retinex, originally proposed by Kimmel et al. \cite{RonA}, provides a unified framework to formulate the illumination estimation as an optimization problem. Since decomposing an image into the illumination and the reflectance is a mathematically ill-posed problem, and there are no ground truth reflectance and illumination images for natural scenes, various assumptions on the illumination\cite{RonA}, or both the illumination and the reflectance \cite{XA}, \cite{wang2014variational},\cite{XuSTAR} are imposed on the Retinex framework. Within this framework, the subsequent studies differ mainly in modeling such assumptions and priors through different norms and regularization terms, focusing on different aspects of illumination recovery, such as edge-aware smoothing, illumination structure preservation, and local texture extraction \cite{XA, XuSTAR,Guo2017LIME,lin2022low,zhou2023TMMsaretinextmm}. Variation-based methods require solving complex objective functions.

Unlike Retinex-based techniques, some studies assume that the observed image is the result of an ideal image being degraded by the air light scattering model, which is typically used in image dehazing methods. Following this idea, several methods employ dehazing techniques and related priors to enhance images by treating the inverted unevenly-lit images as hazy images \cite{dong2011fast}, \cite{zhang2012enhancement}. Yu and Zhu recently propose a method to enhance images by iteratively estimating the ambient light and the light-scattering attenuation rate \cite{yu2019low}. In another recent work, Wang et al. propose an enhancement method by estimating the absorption light and the transmittance \cite{wang2019low}. Both methods produce impressive results.

The performance of the PLM-based methods depends on the assumptions or priors about real scenes. For example, a common assumption is that scene lighting varies smoothly in space. However, this assumption is easily violated in unevenly-lit scenes \cite{li2018structure}. In addition, since image enhancement is closely related to subjective preferences, assumptions or priors based on physical lighting may lead to discrepancies between the enhanced results and the human perception of scenes. 

\begin{figure*}[h]
\vspace{-8mm}

	\centering
	\includegraphics[scale=1]{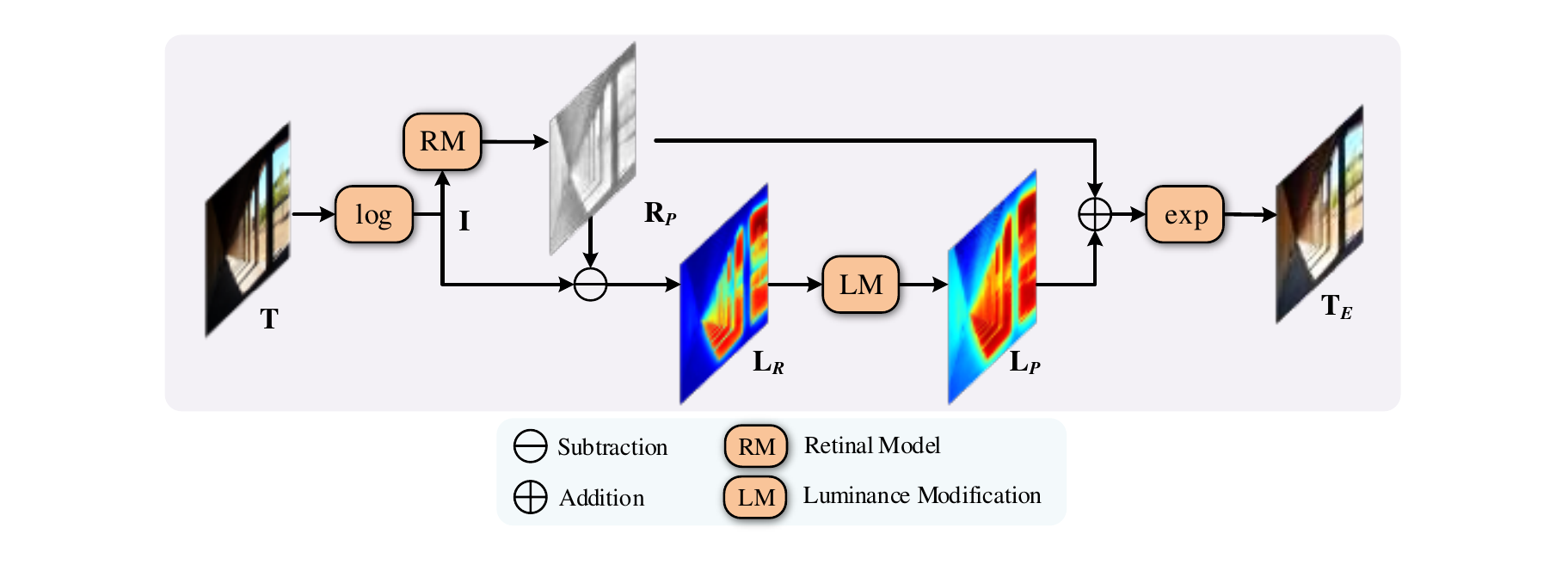}
	\vspace{-6mm}
	\caption{The flowchart of the proposed method. All computations are element-wise.}

 	\label{ref_fig_1}
 	\vspace{-2mm}
\end{figure*}
 
\vspace{0mm}
\subsection{Deep-learning based enhancement}

Recent advances in convolutional neural networks(CNNs) have shown that deep-learning techniques can help with image enhancement. Based on paired training images, some highly-cited studies include: Wei et al. propose RetinexNet to enhance low-light images \cite{wei2018deep}. Wang et al. propose a convolution neural network to enhance underexposed images by learning the illumination map from paired low-light/normal images \cite{wang2019underexposed}. Xu et al. construct a hierarchical feature mining network to enhance low-light images\cite{Xu2022hfmnet}. Zhang et al. propose KIND and its upgraded network KIND++\cite{zhang2019kindling,zhang2021beyond}. KIND and KIND++ decompose the input image into the reflectance and illumination images and adjust both images to achieve the enhanced result. In contrast to the paired image based methods, EnlightenGAN proposed by Jiang et al. is an unsupervised deep-learning based image enhancement method trained on unpaired datasets\cite{jiang2021enlightengan}. Guo et al. propose a deep curve estimation network named ZeroDCE and its light-weighted version ZeroDCE++ \cite{guo2020zero, guo2021zeropp}. ZeroDCE and ZeroDCE++ achieve image enhancement by learning elegantly designed mapping curves instead of learning image-to-image mapping, so the training procedure does not require paired data. More recently, Zhao et al. propose a zero-reference RetinexDIP network to achieve image enhancement by learning the reflectance and illumination images \cite{zhao2022retinexdip}. Based on a semantically contrastive learning network, Liang et al. propose an impressive SCL-LLE method for image enhancement \cite{liang2022semantically}. 

The performance of learning based methods heavily hinges on the training datasets, the carefully fine-tuned loss functions, the image-specific mapping curves, and the expensive hardware resources.

Is it possible to enhance unevenly-lit images neither requiring learning procedures nor imposing sophisticated priors or assumptions on scenes? The ease with which the HVS perceives real scenes across a wide range of lighting variations suggests that visual models may help to solve this problem of naturally enhancing the images taken in unevenly-lit conditions. In this paper, we propose an enhancement method using a vision-based exploratory data analysis model (VEDA). The major contributions of our work are as follows:
\begin {itemize}
\item 
We propose a new perspective on image decomposition for image enhancement. The key idea is to decompose the image into two images: the output image of a retinal model (RM), which contains the scene contrasts, and the residual image, which contains most of the scene lighting variations. The enhanced image is obtained by manipulating the two images. To the best of our knowledge, little work has been reported on decomposing an image into the contrast and the residual for image enhancement. Our method is strikingly simple. The processing flow consists of merely a few simple computations, without any explicit assumptions or priors about physical lighting or surface reflectance, without iteratively solving any complex optimization functions, and without any learning procedures.
\item
We propose a residual image based weighted averaging strategy to combine enhanced results at multiple scales. A similar residual image based weighting scheme can also be leveraged to reconcile noise suppression and detail preservation.
\item
Experiments on a variety of image datasets demonstrate the performance of our method is competitive with several state-of-the-art methods.
\end {itemize}

The remainder of this paper is organized as follows: Section \ref{method} illustrates the proposed method.  Section \ref{expe} presents the experimental results. Section \ref{discussion} conducts a discussion. The conclusion is drawn in Section \ref{conclusion}.

\section{Image enhancement by the VEDA}\label{method}
\subsection{Motivation and Overview}\label{overview}

The motivation for our method is twofold: firstly, it is widely accepted that the receptive field (RF) in the retina primarily responds to local contrast changes rather than absolute light levels. This mechanism allows the retina to function analogously to an excellent image processing model capable of perceiving details in natural scenes across the wide range of light levels. Secondly, a well-known concept in exploratory data analysis is to view a signal as the sum of the output of a particular model and the residual between the model output and the input \cite{Deng2011A}. Accordingly, we regard an input image as the superimposition of two images: one image is the output of a retina model, and the other one is the residual image, simply given by image = RM + residual. The RM preferentially extracts the scene contrasts, while the residual image contains the lighting variations that cannot be fully processed by the RM. Thus, the enhanced image can be achieved by adjusting these two images and recombining them.

Fig. \ref{ref_fig_1} shows the flowchart of the proposed method. The first step is to convert input image $\bf{T}$ to the logarithmic domain by ${\bf{I}} = \log \left( {{\bf{T}}} \right)$, as the logarithm of the incident light is a rough approximation to the perceived brightness \cite{RonA}. Second, the contrast image ${{\bf{R}}_P}$ is extracted from $\bf{I}$ by the RM, and then the residual image ${{\bf{L}}_R}$ is obtained by subtracting ${{\bf{R}}_P}$ from $\bf{I}$. Third, since the residual image contains most of lighting variations, ${{\bf{L}}_R}$ is manipulated through a luminance modification (LM) unit to achieve ${{\bf{L}}_P}$, referred to as the perceived residual image in this paper. Finally, the enhanced intensity image ${{\bf{T}}_E}$ is generated by converting back the sum of ${{\bf{L}}_P}$ and ${{\bf{R}}_P}$ to the intensity domain. Note that all computations are element-wise and are applied to the V channel in the HSV color space to avoid color shifts.

\subsection{Extracting the contrast image by the retinal model}\label{cm}
A variety of RMs have been proposed to explain various characteristics of the HVS \cite{eshraghian2017biological, hansen2004simple, lindeberg2013computational, wienbar2018the,huang2010cortical, cho2014model,silver2012neural,zhang2016retinal}. After testing many of them, we find that the center-surround shunting equation, which simulates the rate of change in neuron activities \cite{grossberg2020toward,BROWNING2009320}, provides the best performance within the scope of this paper. A simplified version of this time-varying equation takes the form:
\begin{equation}
\label{eq_model}
\frac{d}{{dt}}{\bf{r}} =  - m{\bf{r}} + \left( {g - {\bf{r}}} \right){\bf{C}} - \left( {g + {\bf{r}}} \right){{\bf{S}}_\sigma }
\end{equation}
where $\bf{r}$ denotes the neuron activity of the center-surround RF, $m$ is the decay rate, $g$ is the positive gain, $\bf{C}$ is the center input stimulus which generally has one pixel width for practical image processing, namely ${\bf{C}} = {\bf{I}}$, and ${{\bf{S}}_\sigma }$  is the surround input stimulus given by
\begin{equation}
\label{eq_simg}
{{\bf{S}}_\sigma } = {\bf{I}}*{w_\sigma }\nonumber
\end{equation}
where $*$ is the convolution operator, ${w_\sigma }$  is a Gaussian kernel with a standard deviation $\sigma$. We set the Gaussian kernel width ${W_S}$ as ${W_S} = 6ceil\left( \sigma  \right) + 1$ according to the Pauta criterion, where ${ceil\left( \bullet \right)}$ is the function that rounds the argument value to the nearest integer greater than or equal to the value.

Eq. (\ref{eq_model}) is an initial value problem of an ordinary differential equation. Given $t = 0$ , ${\bf{r}} = 0$, its solution is
\begin{equation}
\label{eq_solution}
{\bf{r}} = g\frac{{{\bf{I}} - {{\bf{S}}_\sigma }}}{{m + {\bf{I}} + {{\bf{S}}_\sigma }}}\left\{ {1 - \exp \left[ { - \left( {m + {\bf{I}} + {{\bf{S}}_\sigma }} \right)t} \right]} \right\}
\end{equation}

Eq. (\ref{eq_solution}) simulates the adaptation mechanism of the RF neurons in response to light stimuli. When the neuron activity reaches to the steady state, i.e., $t \to \infty $ , the time-decaying term vanishes, and the output of the RM, denoted as ${{\bf{R}}_P}$, is achieved by
\begin{equation}\label{eq_cimg}
{{\bf{R}}_P} = {\bf{r}} = g\frac{{{\bf{I}} - {{\bf{S}}_\sigma }}}{{m + {\bf{I}} + {{\bf{S}}_\sigma }}}
\end{equation}

Thus, the RM extracts the scene contrasts by the ratio of the difference-of-Gaussian and the biased sum-of-Gaussian.

\subsection{Yielding the residual image}\label{ri}

\begin{figure*}[t]
\vspace{-6mm}
	\centering
		
	\subfloat[]{\includegraphics[width=35mm]{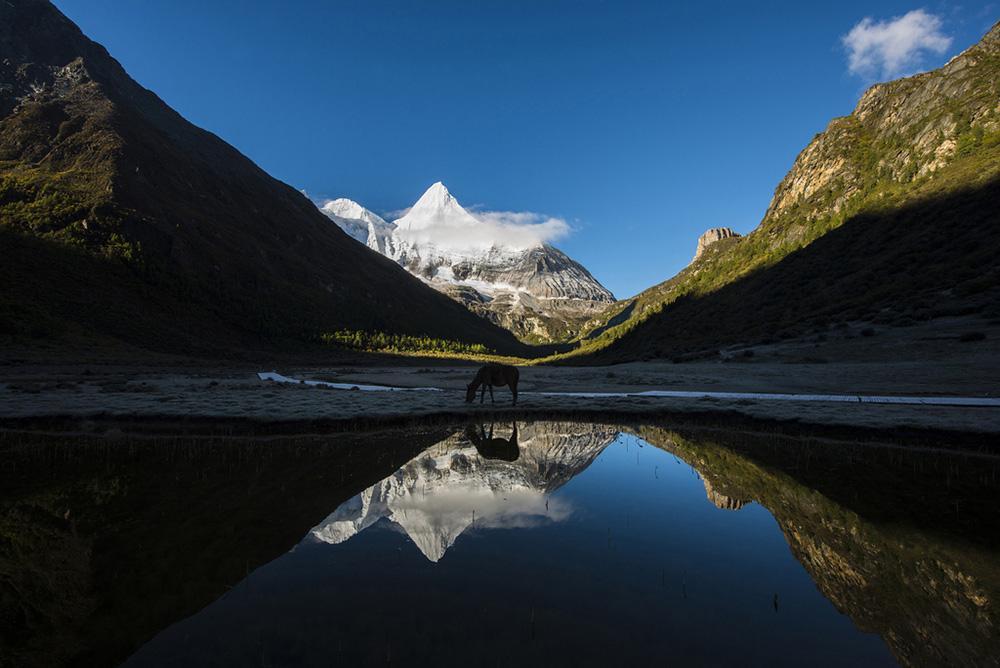}}		
	\hfil
	\subfloat[]{\includegraphics[width=35mm]{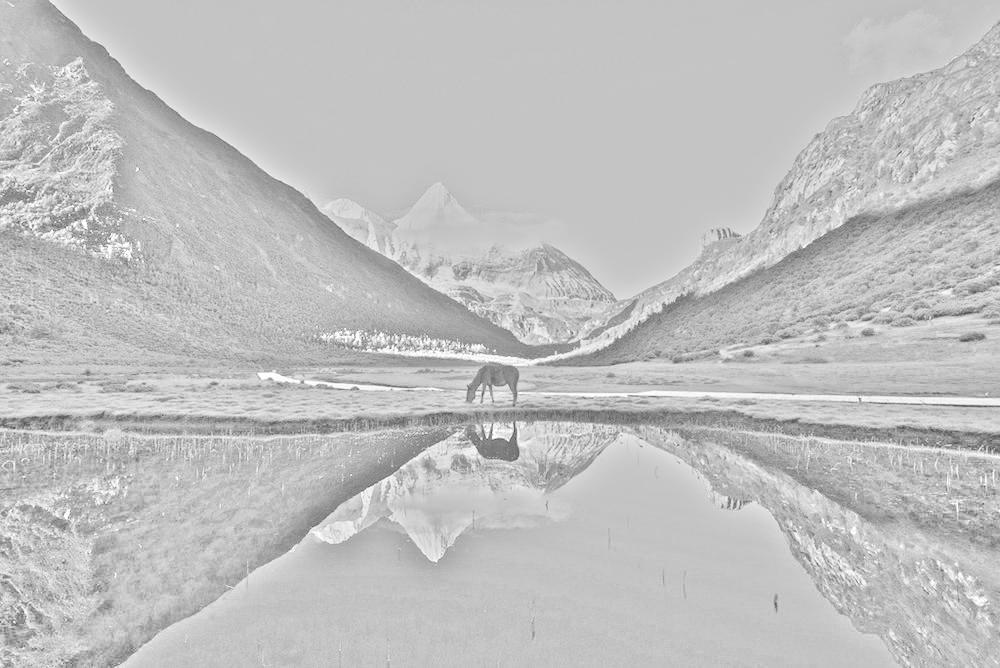}}		
	\hfil
	\subfloat[]{\includegraphics[width=35mm]{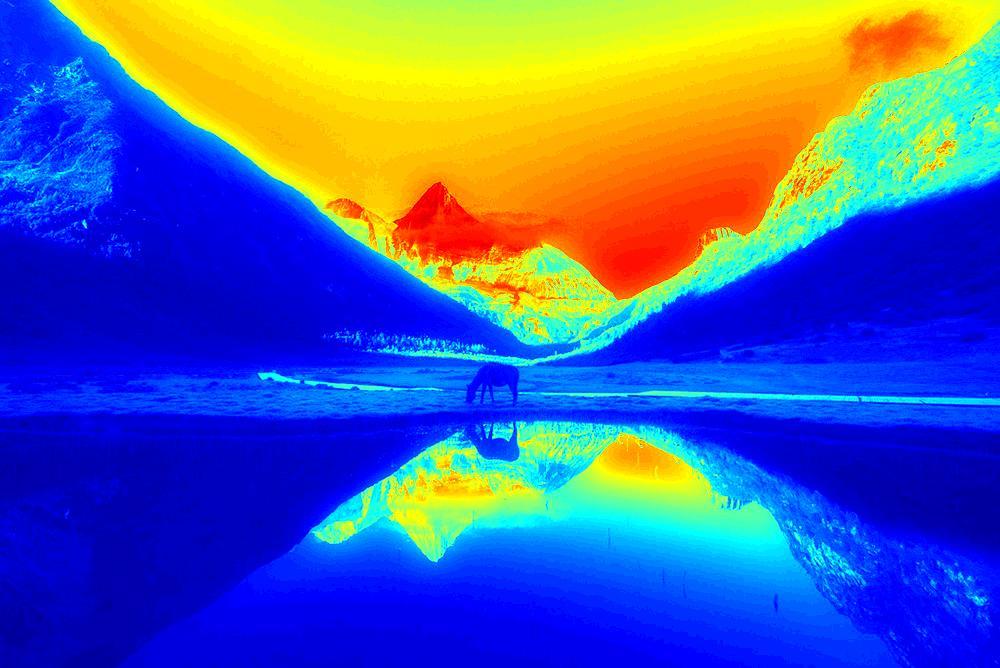}}		
	\hfil
	\subfloat[]{\includegraphics[width=35mm]{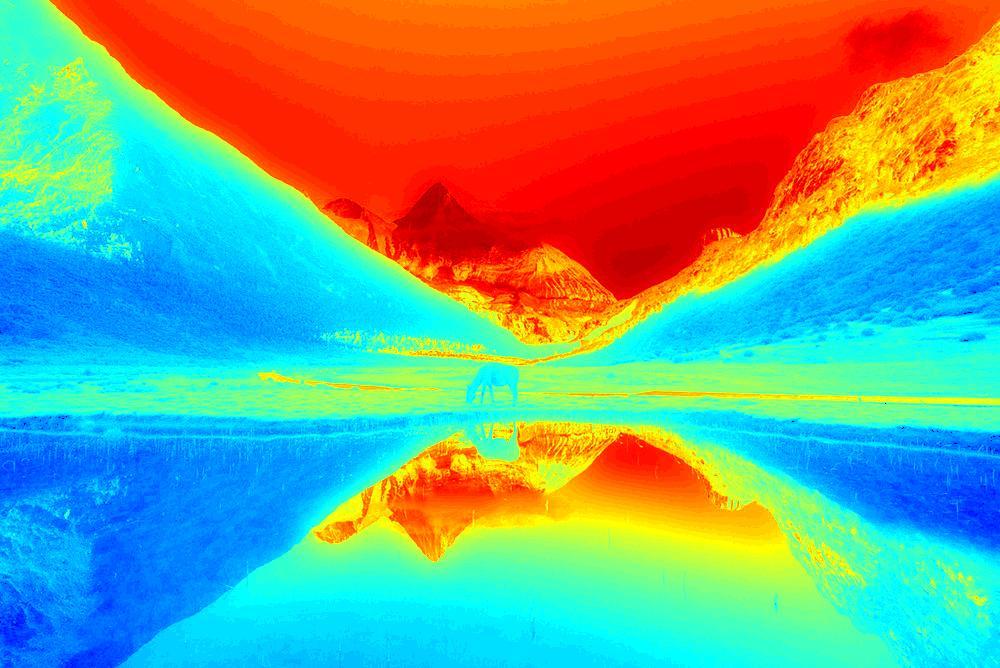}}		
	\hfil
	\subfloat[]{\includegraphics[width=35mm]{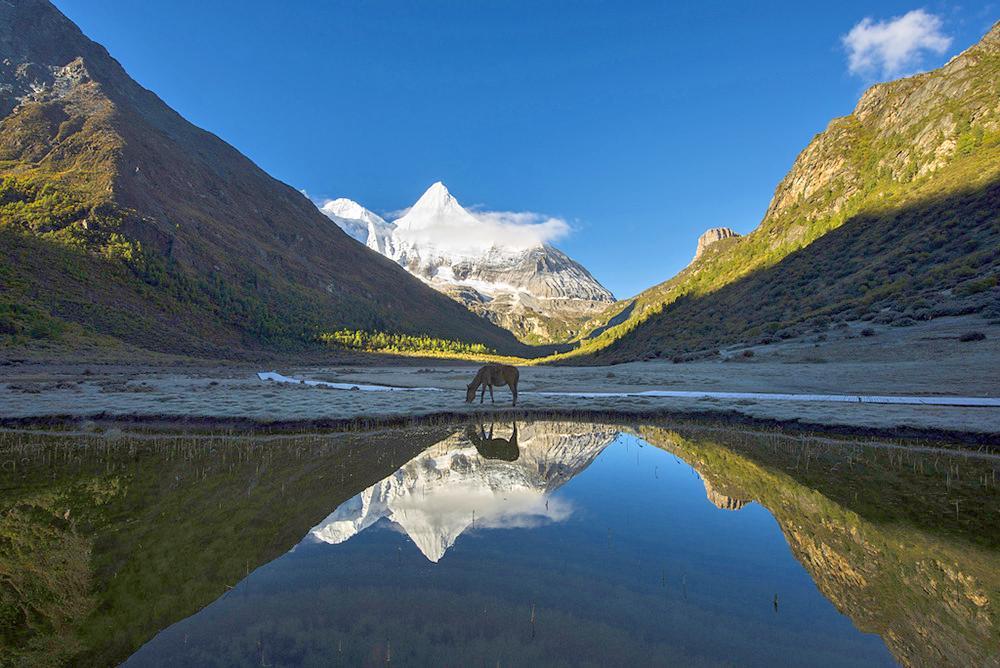}}
		
	\centering
	\caption{Decomposition example. (a) Input. (b) Contrast image. (c) Residual image. (d) Perceived residual image. (e) Enhanced result. }
	\label{fig_decompose}
\vspace{-2mm}
\end{figure*}

\begin{figure*}[t]
\vspace{0mm}
	\centering
	\subfloat[]{\includegraphics[width=29mm]{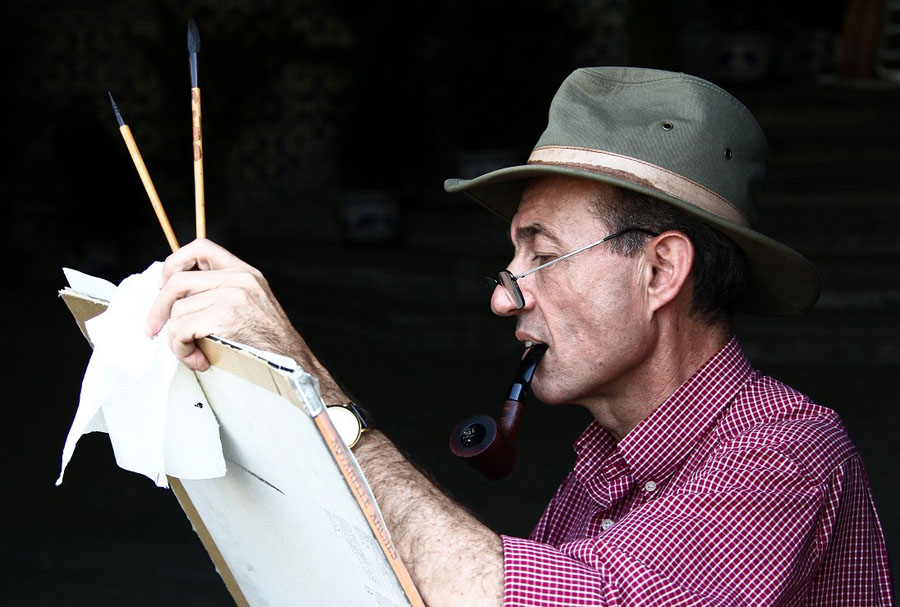}}		
	\hfil
	\subfloat[]{\includegraphics[width=29mm]{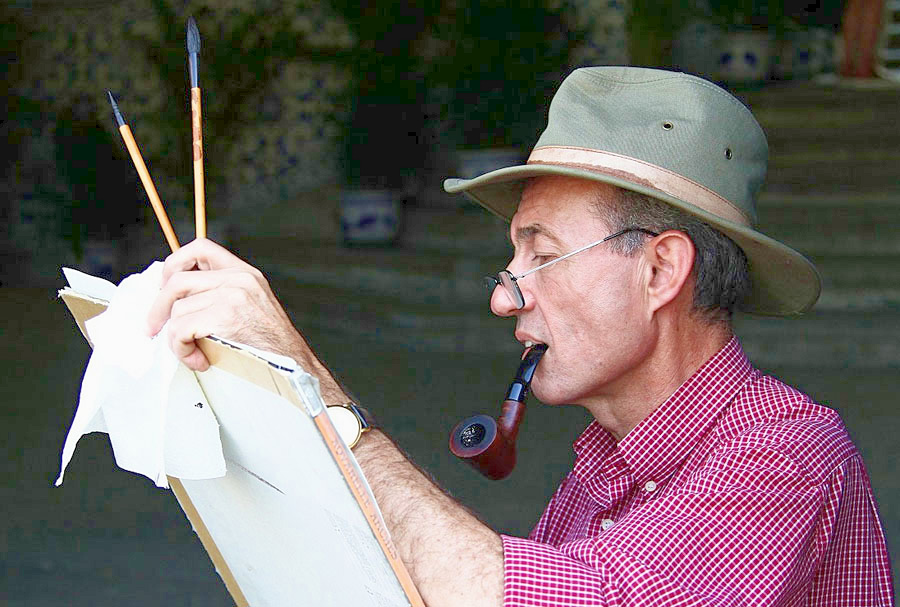}}
	\hfil
	\subfloat[]{\includegraphics[width=29mm]{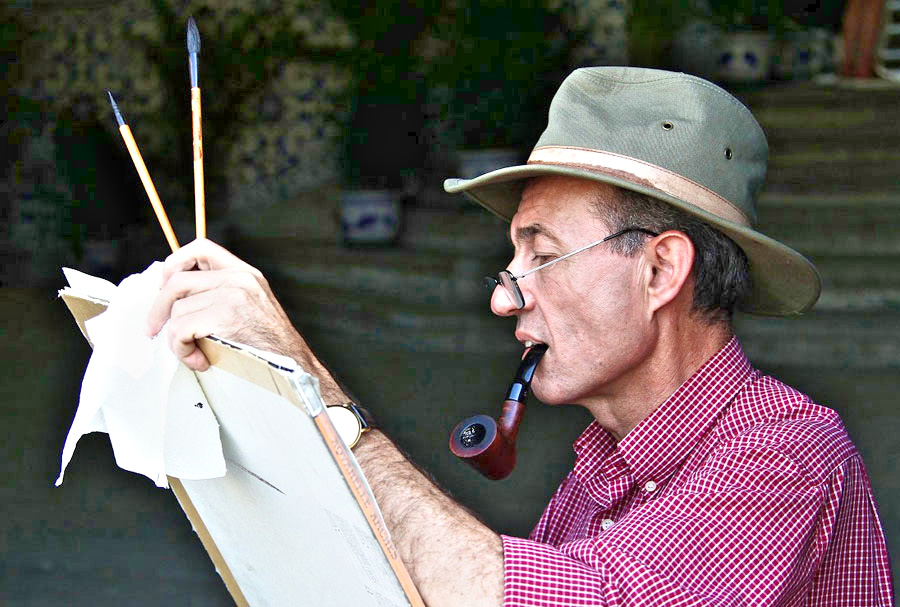}}
	\hfil
	\subfloat[]{\includegraphics[width=29mm]{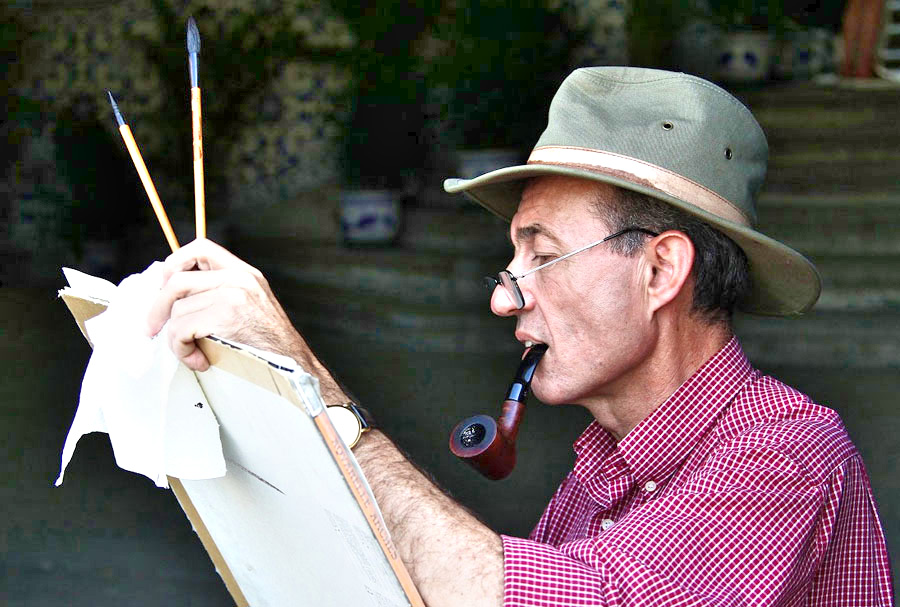}}
	\hfil
	\subfloat[]{\includegraphics[width=29mm]{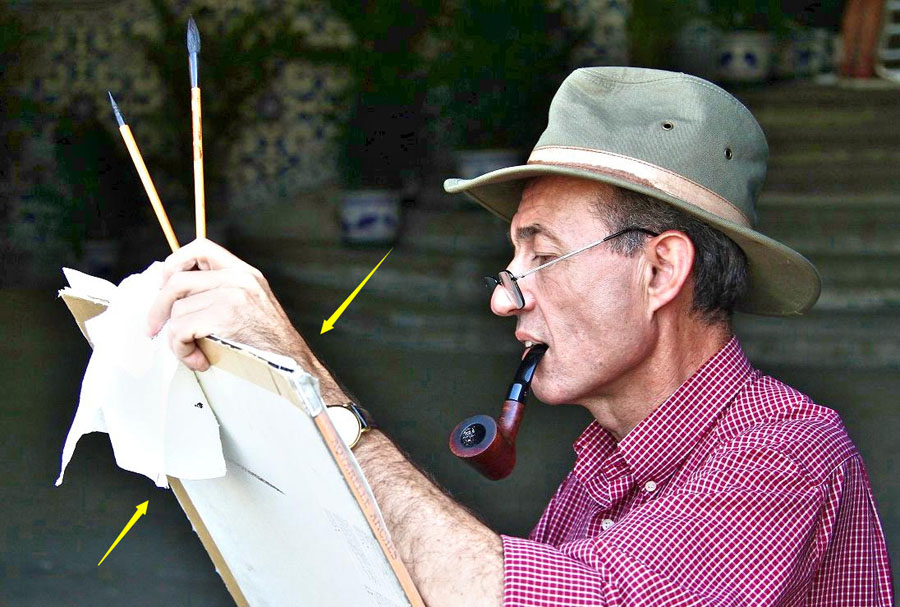}}
	\hfil
	\subfloat[]{\includegraphics[width=29mm]{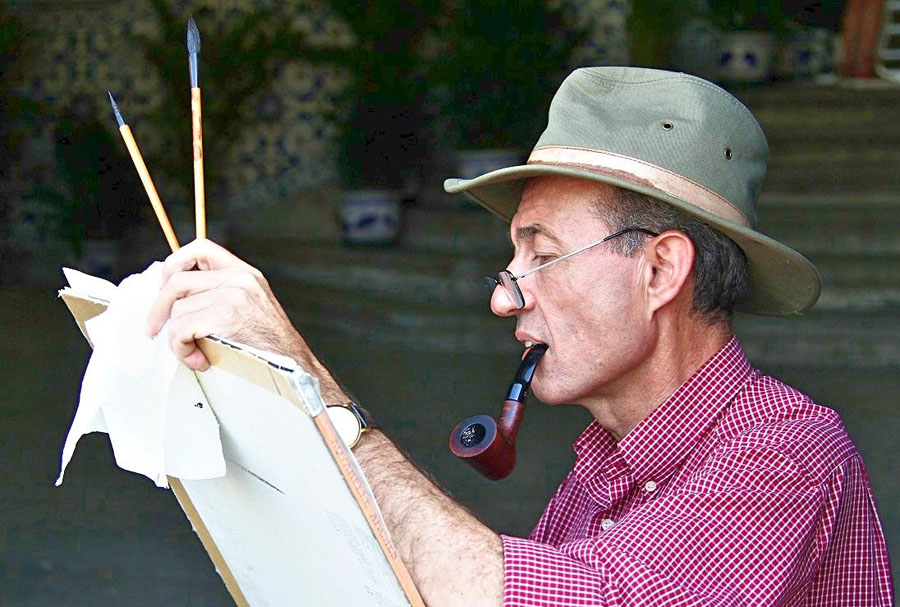}}
	\caption{ Enhanced results at multiple scales and result with WGIF. (a) Input. (b) $\sigma=1$. (c) $\sigma=4$. (d) $\sigma=16$. (e) Multi-scale. (f) Result using WGIF. Larger $\sigma$ values provide better overall contrast than smaller ones, but tend to produce halos. The multi-scale form produces better visual quality than any single scale, but halos are still visible, as indicated by the yellow arrows. WGIF effectively removes halos. Please zoom in to see details.  }
	\label{ref_multiscale}
	\vspace{-2mm}
\end{figure*}

The residual image  ${{\bf{L}}_R}$ is obtained by subtracting the contrast image ${{\bf{R}}_P}$ from ${\bf{I}}$ 
\begin{equation}
\label{eq_rimg}
{{\bf{L}}_R}={\bf{I}} - {{\bf{R}}_P}
\end{equation}
Figs. \ref{fig_decompose}(a), (b), and (c) show an example of decomposing an image into the contrast image and the residual image, respectively. It can be observed that the contrast image retains the visual details, while the residual image mainly retains the scene lighting variations. 

The dynamic range of scene lighting is usually vast. Hence, we develop a luminance modification (LM) function to compress the dynamic range of the residual image. The LM function should be progressive, suppressing larger luminance values more heavily than smaller ones\cite{stroebel2000basic}. In addition, the LM function is desired to be analogous to the visual response to light intensity. After examining a number of functions that satisfy these goals, we find that power law \cite{bolton2008modeling} performs the best, which is a linear function in the logarithmic domain:
\begin{equation}\label{eq_primg}
{{\bf{L}}_P} = LM\left( {{{\bf{L}}_R}} \right) = \gamma {{\bf{L}}_R} + k
\end{equation}
where  $\gamma $ and $k$  are both constants, and ${{\bf{L}}_P}$ is the perceived residual image.

Fig. \ref{fig_decompose}(d) shows the perceived residual image. Compared with (c), the brightness in shaded areas is improved.


\subsection{Achieving the enhanced image}\label{ei}
The single-scale enhanced image ${{\bf{T}}_E}$ is obtained by recombining the contrast image and the perceived residual image, and converting the result back to the intensity domain, given by
\begin{equation}\label{eq_add}
{{\bf{T}}_E} = \exp \left( {{{\bf{R}}_P}}+ {{{\bf{L}}_P}}\right)
\end{equation} 

The enhanced result of Fig. \ref{fig_decompose}(a) is shown in (e). Our method succeeds in bringing out details from the shadowed areas while maintaining good contrasts elsewhere.

Complex images contain contrasts at multiple scales \cite{peli1990contrast}. Therefore, we develop a multi-scale strategy to produce the final result ${{\bf{T}}_{MSE}}$ by taking the weighted average of enhanced images at all scales:
\begin{equation}\label{eq_multi}
{{\bf{T}}_{MSE}} = \sum\nolimits_{n = 1}^N {{\varphi _n}{{\bf{T}}_{E,n}}}
\end{equation}
where $N$ is the number of scales, ${{\bf{T}}_{E,n}}$ is the enhanced image at the $n$th scale, and ${\varphi _n}$ is the weight associated with the $n$th scale. Since the HVS is more likely to see details in brightly-lit regions than in weakly-lit regions, larger weights are assigned to bright regions:
\begin{equation} \label{eq_weight}
{\varphi _n} = \frac{{{{\bf{L}}_{R,\;n}}}}{{\sum\nolimits_{n = 1}^N {{{\bf{L}}_{R,\;n}}} }}
\end{equation}
where ${{\bf{L}}_{R,\;n}}$  is the residual image at the $n$th scale.  

\begin{algorithm}[t] 
	\caption{ Image enhancement based on the VEDA} 
	\label{alg1} 
		
	{\bf {Input:} \\}
	\hspace*{0.5cm} Image ${\bf{T}}$, parameters ${m}$, $g$, $\gamma$, and $k$.\\		
	{\bf Output:} \\
	\hspace*{0.5cm} Enhanced image ${{\bf{T}}_{MSE}}$. \\
	{\bf {Begin}}	
	\begin{algorithmic}[1]
		\STATE Convert ${\bf{T}}$ to the logarithmic domain by ${\bf{I}} = \log \left( {{\bf{T}}} \right)$;
		\STATE Construct the surround images ${{\bf{S}}_\sigma } = {{\bf{S}}_{\sigma 1}},...,{{\bf{S}}_{\sigma n}}$ using the weighted guided image filter at different scales;
						
		\FOR{each ${{\bf{S}}_\sigma }$}
			\STATE Extract the contrast image ${{\bf{R}}_{P}}$ via Eq. (\ref{eq_cimg});
			\STATE Calculate the residual image ${{\bf{L}}_{R}}$ via Eq. (\ref{eq_rimg}); 
			\STATE Achieve the perceived residual image ${{\bf{L}}_P}$ via Eq. (\ref{eq_primg});
			\STATE Obtain the enhanced image ${{\bf{T}}_E}$ via Eq. (\ref{eq_add});			
		\ENDFOR
		
		\STATE Calculate the weight ${\varphi _n}$ via Eq. (\ref{eq_weight});
		\STATE Obtain ${{\bf{T}}_{MSE}}$ via Eq. (\ref{eq_multi});			
	\end{algorithmic} 

	{\bf {End}}		
\end{algorithm}

The difference between ${{\bf{T}}_{E,n}}$ and ${{\bf{T}}_{E}}$ is essentially governed by the surround images associated with different values of $\sigma$. We construct the multi-scale surround images by the sequence $\sigma  \in \left\{ {{\sigma _1},{\sigma _2}, \cdots ,{\sigma _n}, \cdots {\sigma _N}} \right\}$, where ${\sigma _n} = {4^{n - 1}}{\sigma _1}$. In this paper, we set ${\sigma _1} = 1$ to detect small details and the maximum number of scales $N=3$ due to the limited size of the RF. In addition, the values of $g$, $m$, $\gamma$, and $k$ are invariant across scales, respectively. The influence of the four free parameters is discussed in Section \ref{para_set}.

Figs. \ref{ref_multiscale}(b) to (d) show the enhanced results for (a) at different scales. Larger $\sigma$ values achieve better overall contrast than smaller $\sigma$ values, while smaller $\sigma$ values produce better local details. It is worth noting that the difference-of-Gaussian in Eq. (\ref{eq_cimg}) tends to yield halo artifacts around sharp edges, as shown in (c) and (d). The weighted multi-scale averaging is capable of a high degree of halo removal, but cannot eliminate them entirely. To better remove halos, we produce the surround image ${{\bf{S}}_\sigma }$ by the weighted guided image filter(WGIF) \cite{Li2015wgif} instead of the Gaussian filter. The result produced by the WGIF is shown in (f), where the halos in (e) are effectively eliminated.

In summary, the entire procedure of the proposed method is outlined in Algorithm \ref{alg1}.

\section{Experimental results}\label{expe}
In this section, we first present the experimental settings. Then, we study the effect of involved parameters on results. Finally, we make qualitative and quantitative comparisons with several state-of-the-art methods to demonstrate the performance of our method.

\subsection{Experimental settings}
\textbf{Computational environment:} All non-deep-learning methods are run in the MATLAB 2019a environment on a PC with 16G RAM and 2.9GHz Intel i7-10700k CPU. All compared deep-learning based methods are deployed on dual NVIDIA TITAN GTX GPUs. 

\textbf{Compared methods:} The compared methods include: DFE\cite{LiangContrast}, LIME\cite{Guo2017LIME}, MLLP\cite{WangNaturalness}, PLME\cite{yu2019low}, ALSM\cite{wang2019low}, PnpRetinex(PnpRtx)\cite{lin2022low}, NRMOE\cite{kumar2022noref}, EnlightenGAN(EGAN)\cite{jiang2021enlightengan}, SCL-LLE\cite{liang2022semantically}, RetinexDIP(RtxDIP)\cite{zhao2022retinexdip}, KIND++\cite{zhang2021beyond}, and ZeroDCE\cite{guo2020zero}. For a fair comparison, all results are produced by publicly available codes with parameters set as exactly as given in their papers.

\textbf{Datasets:} We compare all methods on two test sets. The first test set, denoted as Testset-1, contains 266 images collected from a variety of publicly available datasets without reference images, including NPE (85 images)\cite{Shuhang2013Naturalness}, MLLP (76 images)\cite{WangNaturalness}, VV (24 images)\cite{web37}, DICM (64 images) \cite{Chul2012Contrast} and MEF (17 images)\cite{MEF2015Ma}. The second test set (Testset-2) is built on the Part2 subset of SICE dataset\cite{Cai2018SICE}, which consists of 229 multi-exposure sequences and their corresponding reference image (Ref.Image). We select the first three under-exposed images in each multi-exposure sequence of Part2 subset for testing since the compared methods are not specially designed for over-exposed images. The test images are resized to 25${\%}$ of their original size due to the memory limitation of a previous study\cite{lin2022low}.

{\bf{Objective Metrics:}} To the best of our knowledge, there is no widely accepted measure to quantitatively assess the quality of enhanced images since image quality assessment (IQA) is highly related to subjective preferences. We evaluate all compared methods on Testset-1 by three representative image quality assessment (IQA) metrics: NIQE\cite{mittal2012NIQE}, BIQI\cite{BIQI2010}, and NFERM\cite{gu2014nferm}. NIQE estimates the deviations between the target image and a statistical model of natural scenes. BIQI measures the human perception of the naturalness of an image. NFERM scores the image quality based on the free-energy based brain theory and HVS features. For these three metrics, smaller values represent better image quality. For Testset-2, three fully referenced IQA metrics: peak signal-to-noise ratio(PSNR), structure similarity(SSIM)\cite{wang2004imageSSIM} and lightness order error(LOE)\cite{Shuhang2013Naturalness} are also adopted. Higher PSNR and SSIM values indicate better image quality and lower LOE values indicate better degree of naturalness preservation\cite{Shuhang2013Naturalness}.

\subsection{Parameter study}\label{para_set}

There are four free parameters in the proposed method: two power law related parameters $\gamma$ and $k$ in Eq. (\ref{eq_primg}), the decay rate $m$, and the gain ${g}$ in Eq. (\ref{eq_cimg}).

\begin{figure*}[h]
\vspace{-8mm}
	\captionsetup[subfloat]{labelsep=none,format=plain,labelformat=empty}
	\flushleft
	\hspace{6mm}	
	\begin{minipage}{126mm}	
	\subfloat[(a)]{\includegraphics[width=125mm]{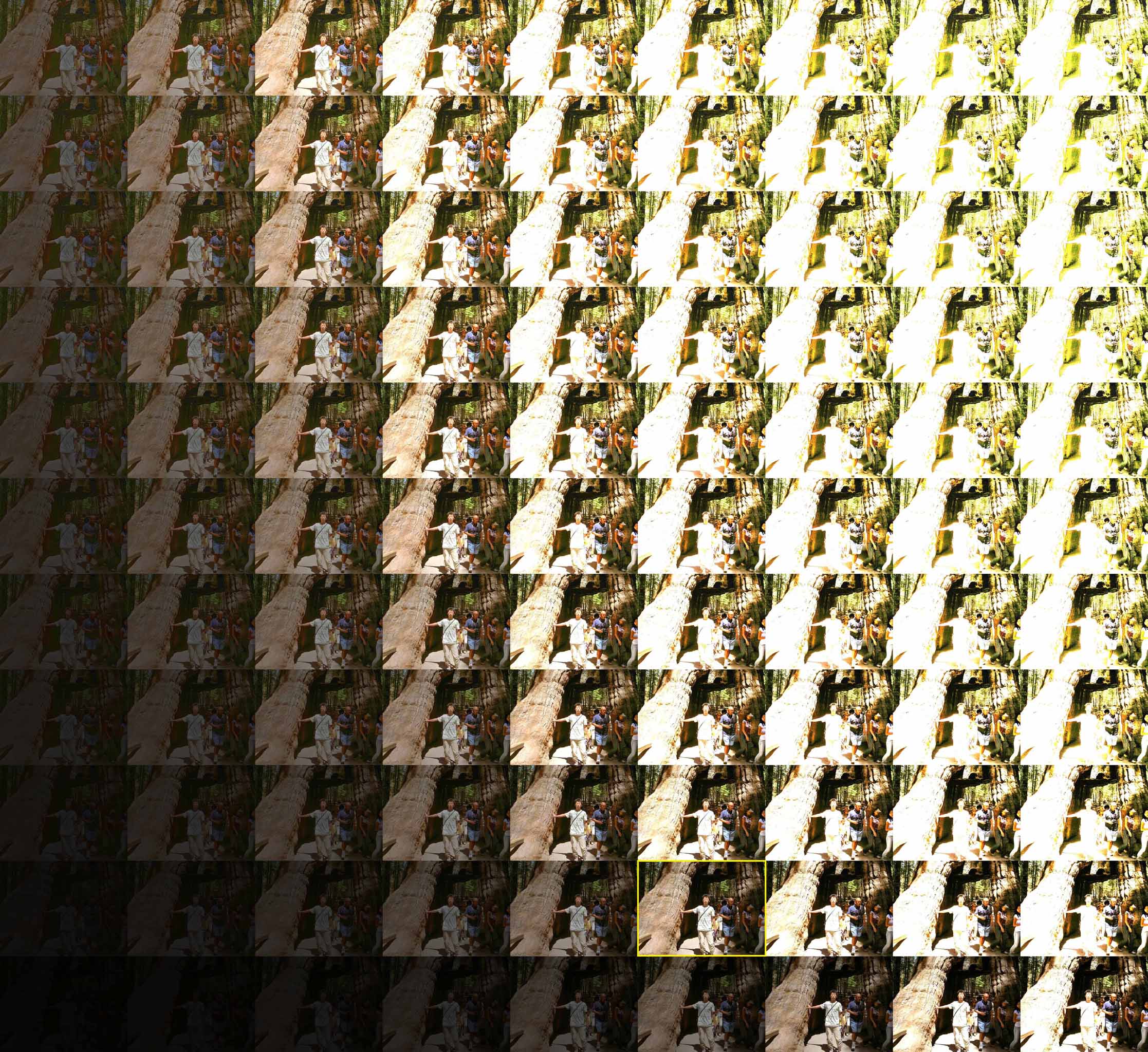}}
	\end{minipage}	
	\begin{minipage}{45mm}	
	\subfloat{\includegraphics[height=39mm]{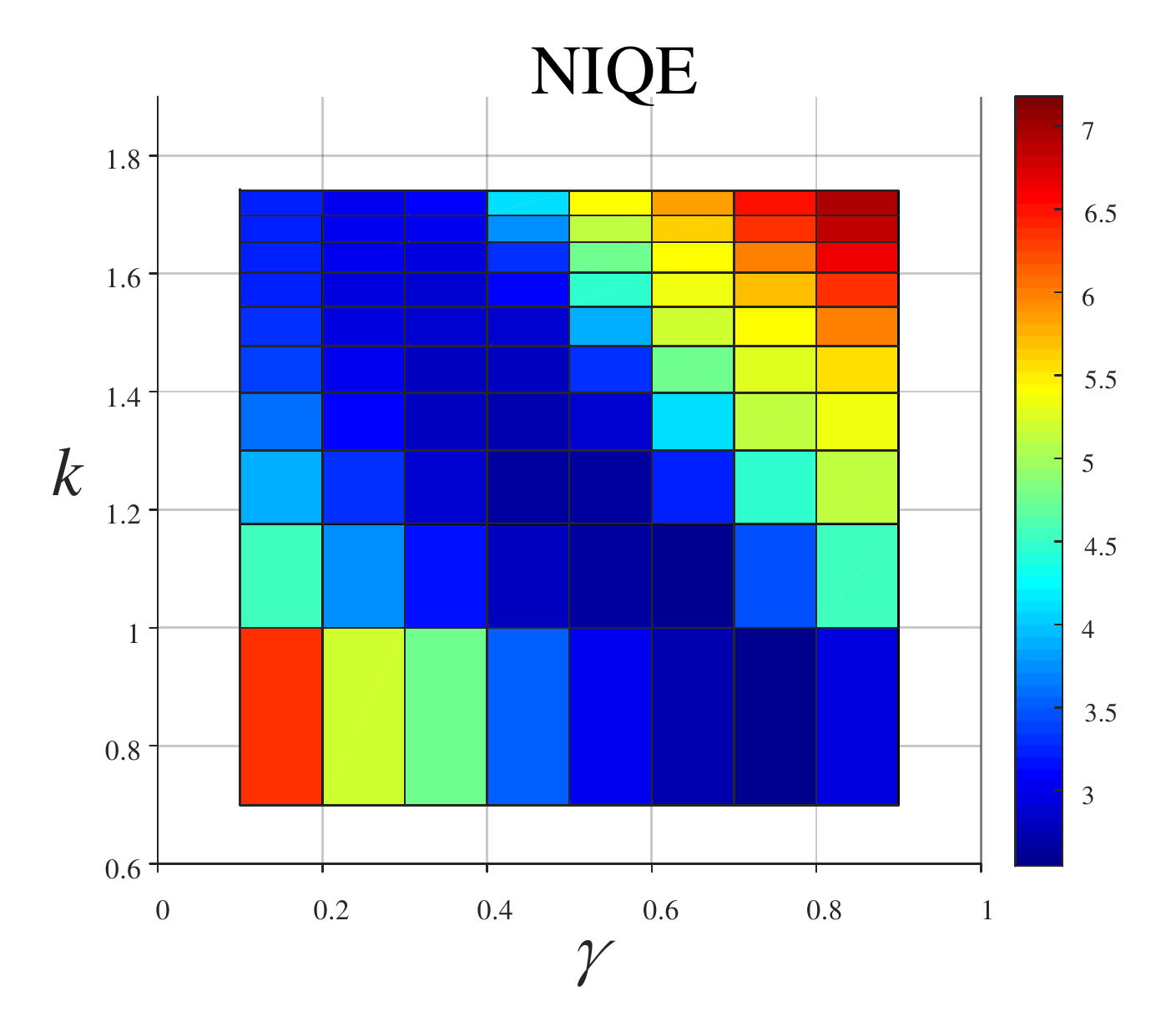}}
	\vspace{-0.2mm}
	\subfloat{\includegraphics[height=39mm]{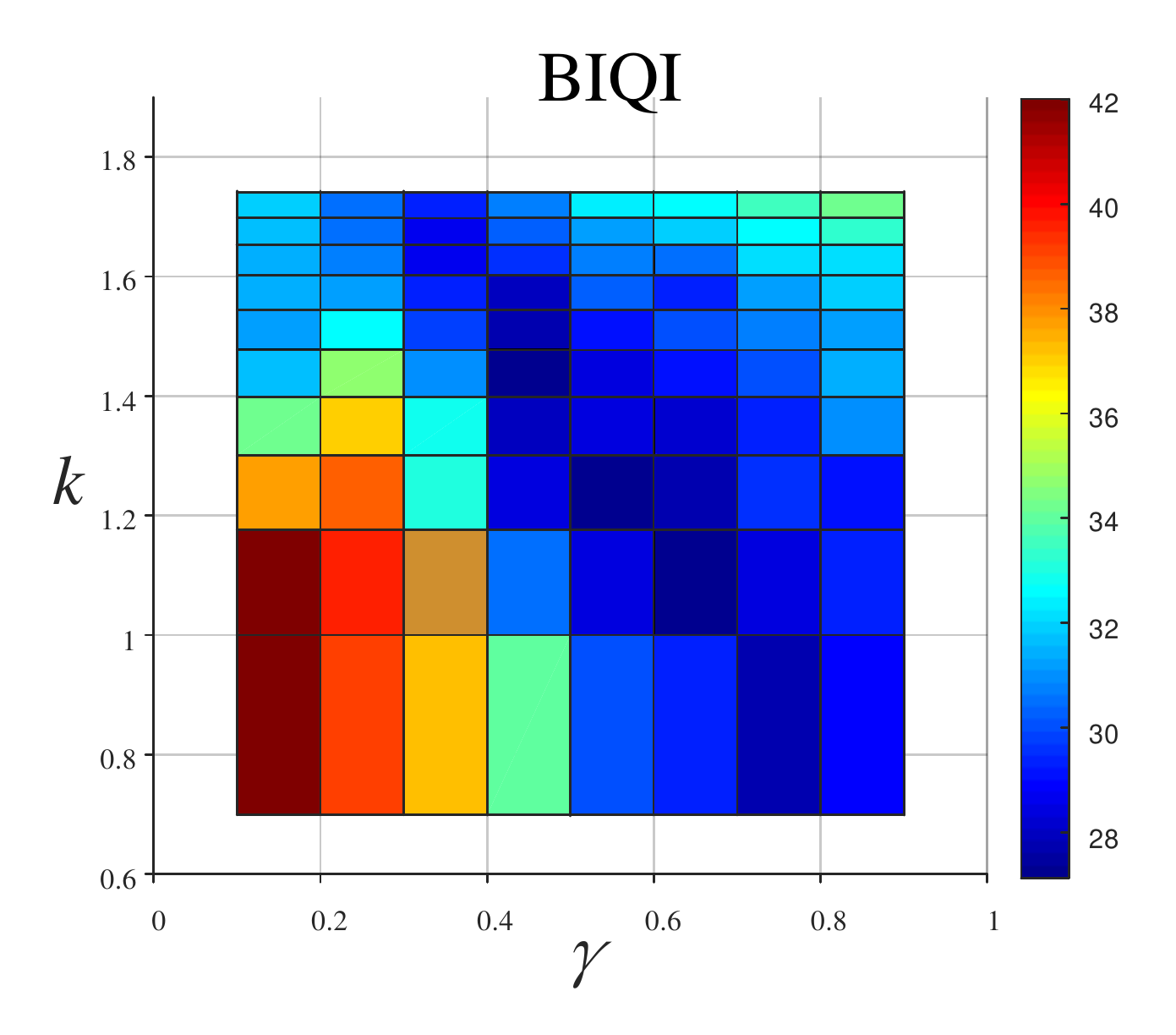}}
	\vspace{-0.2mm}
	\subfloat[(b) NIQE-BIQI-NFERM]{\includegraphics[height=39mm]{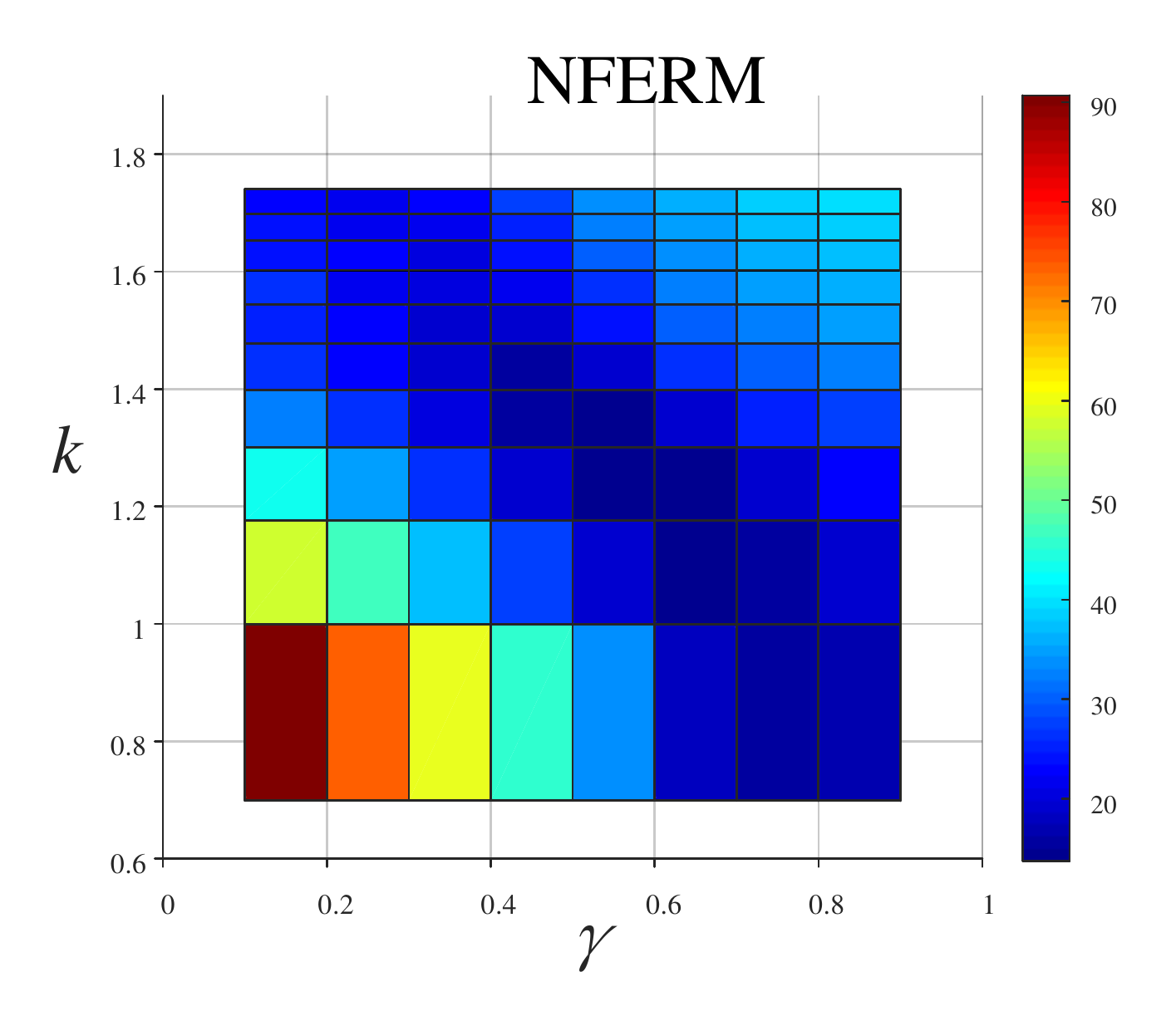}}
	\end{minipage}	
	\vspace{-0mm}			
		
\vspace{-1mm}	
	\caption{The influence of $\left( {\gamma ,k} \right)$ pairs. (a) Results of different $\left( {\gamma ,k} \right)$ pairs. The $\gamma$ values increase from 0.1 to 0.9 in steps of 0.1 from left to right, and the $k$ values increase from $log(5)$ to $log(55)$ in steps of 5 from bottom to top. The image in the yellow box is of the best objective assessment score. }
	\label{ref_params_gamma_k}
\vspace{-3mm}
\end{figure*}

Firstly, both $\gamma$ and $k$ modify the shape of the LM function, which should be a progressively attenuating curve, thus, $\gamma<1$. Smaller $\gamma$ values tend to bring out contrasts in shadow regions, while larger $\gamma$ values result in over-exposure of bright regions. To determine their reasonable values, we run experiments on DCIM dataset and compute the three averaged IQA metrics of the results for different $\left( {\gamma ,k} \right)$ pairs, where ${\gamma}$ ranges from 0.1 to 0.9 in steps of 0.1, and ${k}$ ranges from ${log(5)}$ to ${log(55)}$ in steps of 5. We set ${m=g=1}$ to avoid extensive parameter tuning. Fig. \ref{ref_params_gamma_k}(a) shows an example with different $\left( {\gamma ,k} \right)$ pairs.  Results demonstrates that the pair $\left( {\gamma ,k} \right)=\left( {0.6,log(10)} \right)$ achieves the best average objective assessment score, as evidenced by the lowest NIQE, BIQI and NFERM values in Fig. \ref{ref_params_gamma_k}(b). 

\begin{figure}[h]
\vspace{-2mm}
	\captionsetup[subfloat]{labelsep=none,format=plain,labelformat=empty}
	\centering
	\subfloat[(a) ${m=0.1}$]{\includegraphics[width=20mm]{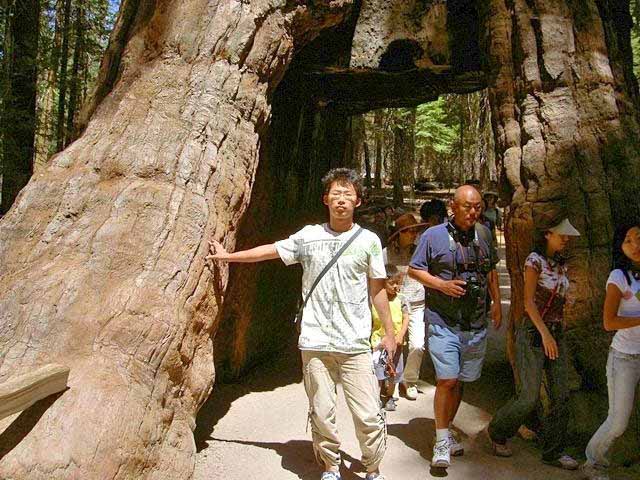}}	
	\hspace{0.2mm}
	\subfloat[(b) ${m=1}$]{\includegraphics[width=20mm]{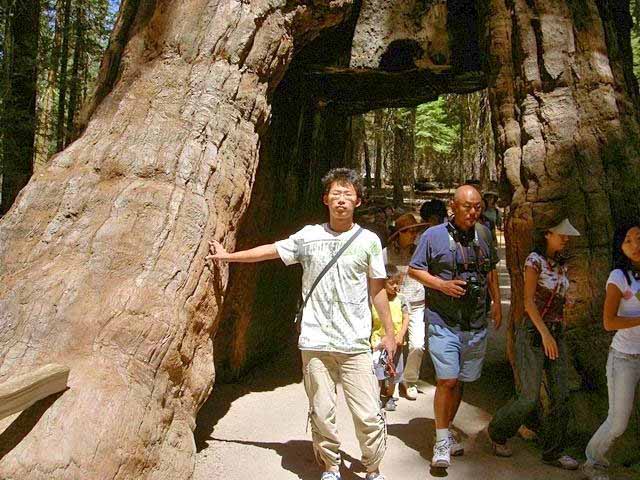}}
	\hspace{0.2mm}
	\subfloat[(c) ${m=5}$]{\includegraphics[width=20mm]{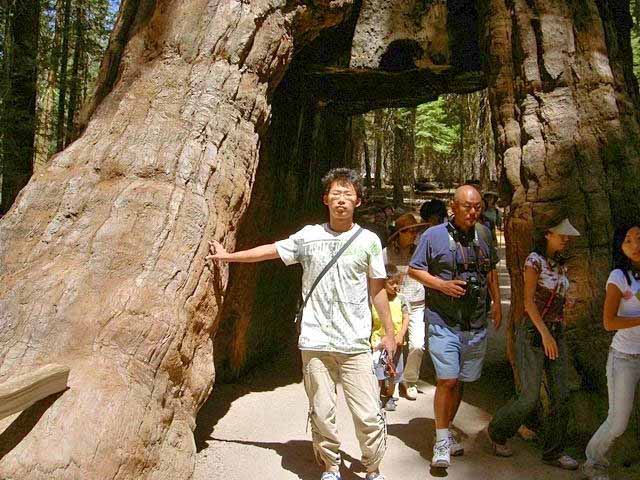}}
		
	\centering
	\vspace{-2mm}
	\subfloat[(d) ${m=10}$]{\includegraphics[width=20mm]{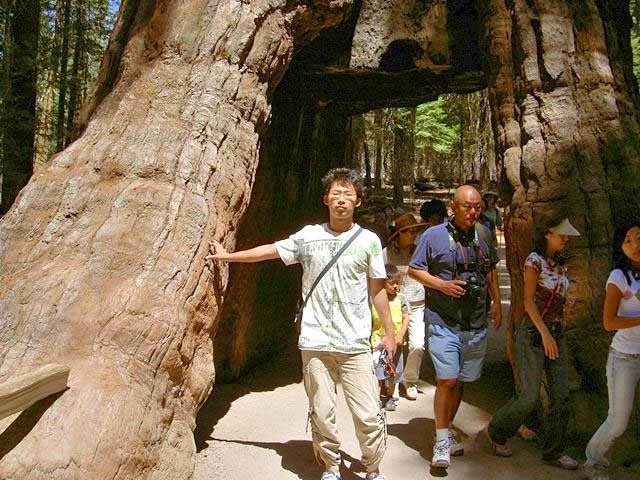}}	
	\hspace{0.2mm}
	\subfloat[(e) ${m=15}$]{\includegraphics[width=20mm]{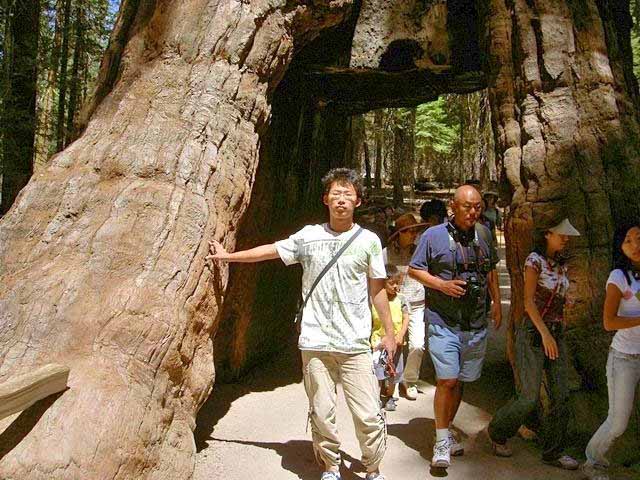}}
	\hspace{0.2mm}
	\subfloat[(f) ${m=20}$]{\includegraphics[width=20mm]{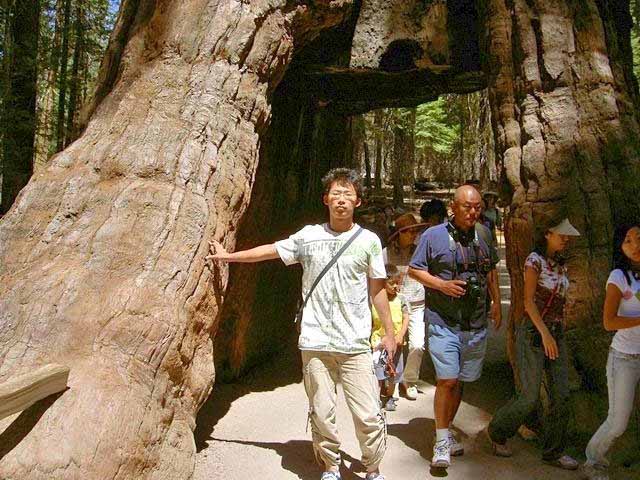}}
			
	\centering	
	\vspace{-2mm}
	\subfloat[(g) NIQE-BIQI-NFERM]{\includegraphics[height=45mm]{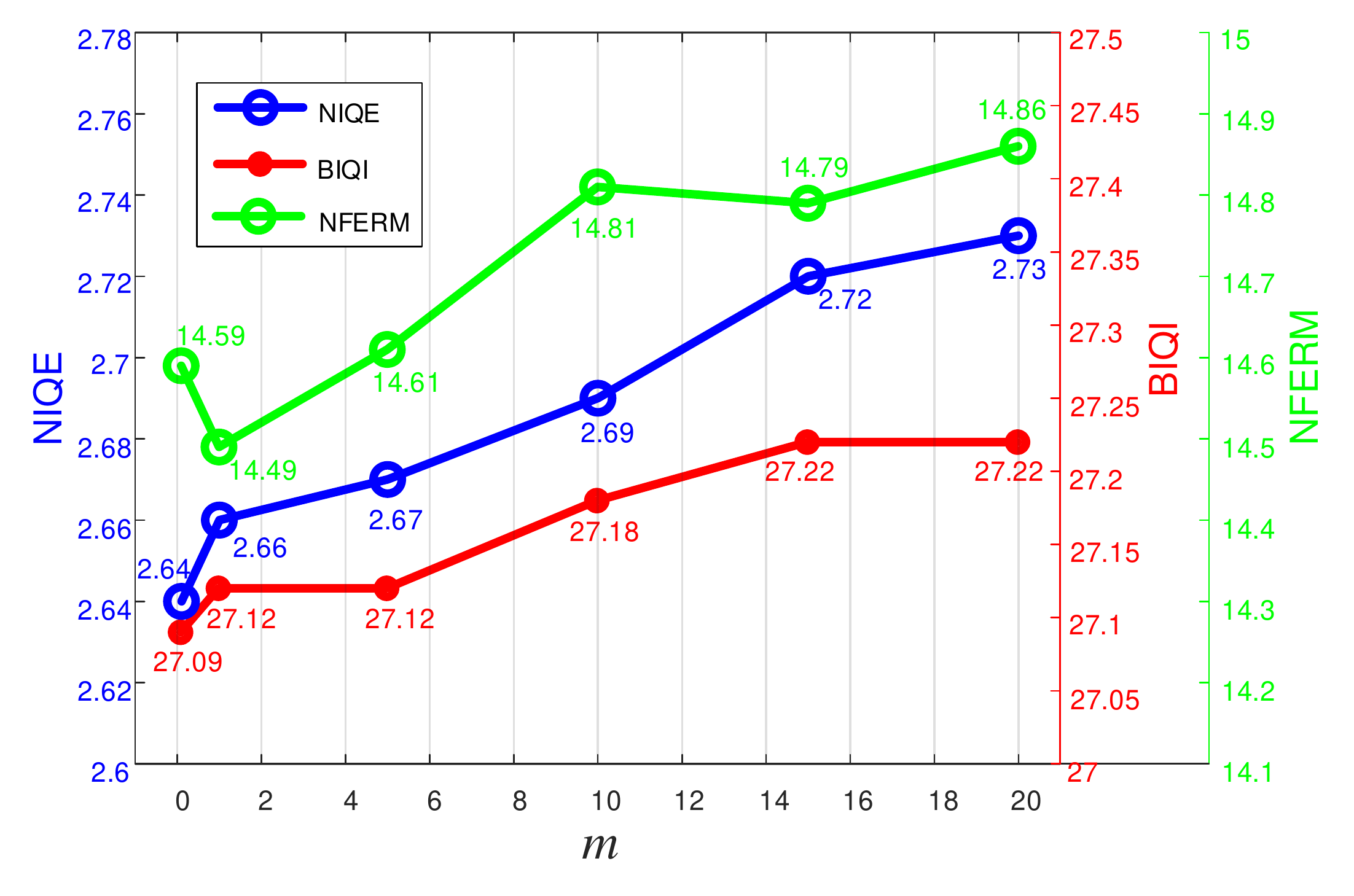}}
\vspace{-1mm}	
	\caption{The influence of ${m}$. The ${m}$ values change from 0.1 to 20 with minor effects on the output quality.	 }
	\label{ref_params_m}
\vspace{-4mm}
\end{figure}

Secondly, since the RM produces ratio-type contrast images rather than high-pass filtered images, the value of ${m}$ should satisfy ${m \ll\left( {{\bf{I}}+{{\bf{S}}_\sigma }}\right)}$. An example is presented in the first row of Fig.\ref{ref_params_m}, where $\left( {\gamma ,k} \right)=\left( {0.6,log(10)} \right)$ and ${g=1}$. It can be seen that the ${m}$ values change from 0.1 to 20 with minor effects on the output quality. The IQA curves in Fig.\ref{ref_params_m} show that larger ${m}$ values produce slightly higher NIQE, BIQI, and NFERM values, indicating slightly lower image quality. We achieve an appropriate balance between the quantitative scores and visual quality by setting ${m=1}$. 

\begin{figure}[t]
\vspace{-3mm}
	\captionsetup[subfloat]{labelsep=none,format=plain,labelformat=empty}
	\centering
	\subfloat[(a) ${g=0.5}$]{\includegraphics[width=20mm]{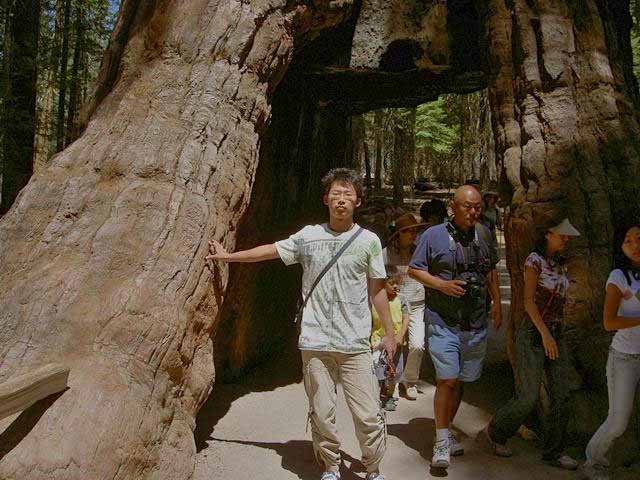}}	
	\hspace{0.2mm}
	\subfloat[(b) ${g=0.75}$]{\includegraphics[width=20mm]{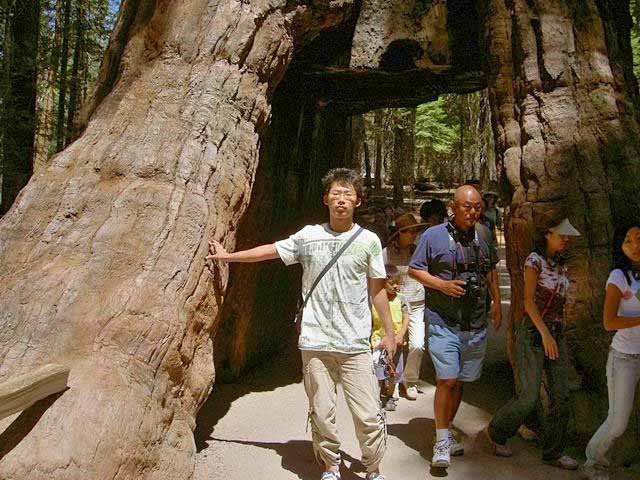}}
	\hspace{0.2mm}
	\subfloat[(c) ${g=1}$]{\includegraphics[width=20mm]{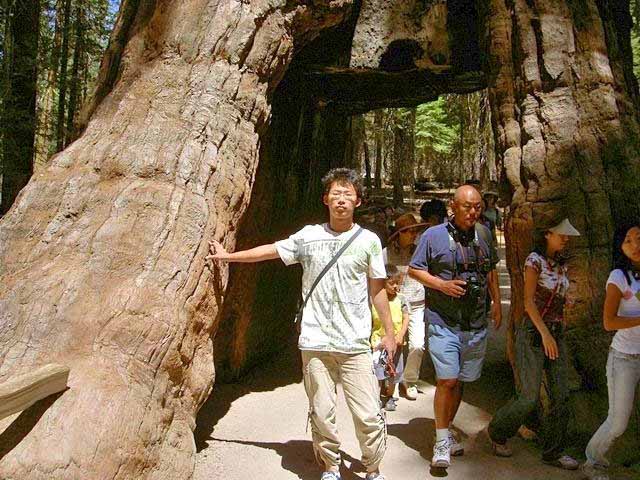}}
	\hspace{0.2mm}	
	\subfloat[(d) ${g=1.25}$]{\includegraphics[width=20mm]{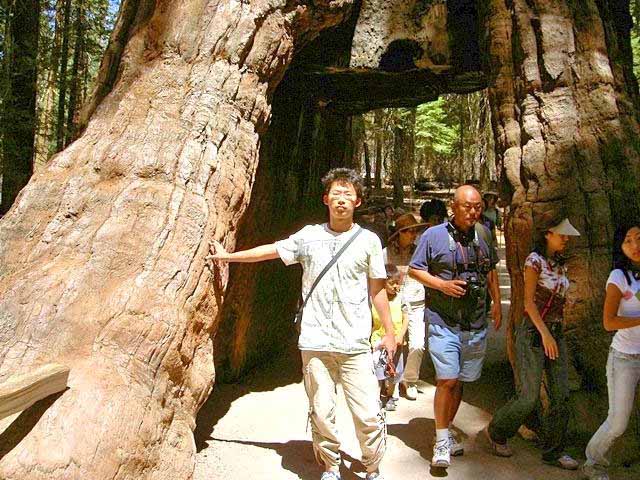}}
	
	\centering
	\vspace{-2mm}	
	\subfloat[(e) ${g=1.5}$]{\includegraphics[width=20mm]{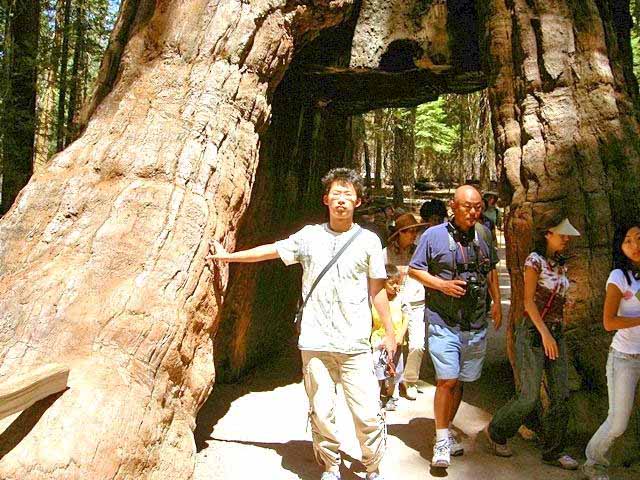}}
	\hspace{0.2mm}
	\subfloat[(f) ${g=1.75}$]{\includegraphics[width=20mm]{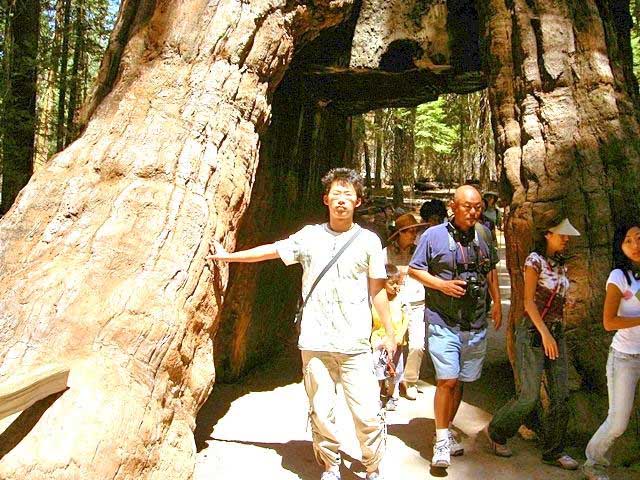}}
	\hspace{0.2mm}
	\subfloat[(g) ${g=2.0}$]{\includegraphics[width=20mm]{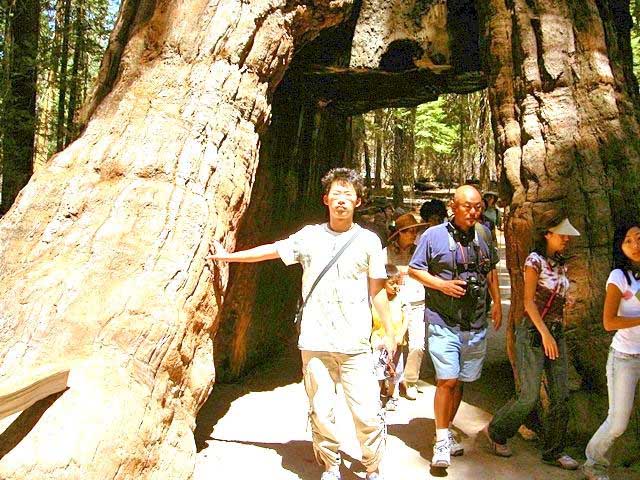}}
				
	\centering	
	\vspace{-2mm}
	\subfloat[(h) NIQE-BIQI-NFERM]{\includegraphics[height=45mm]{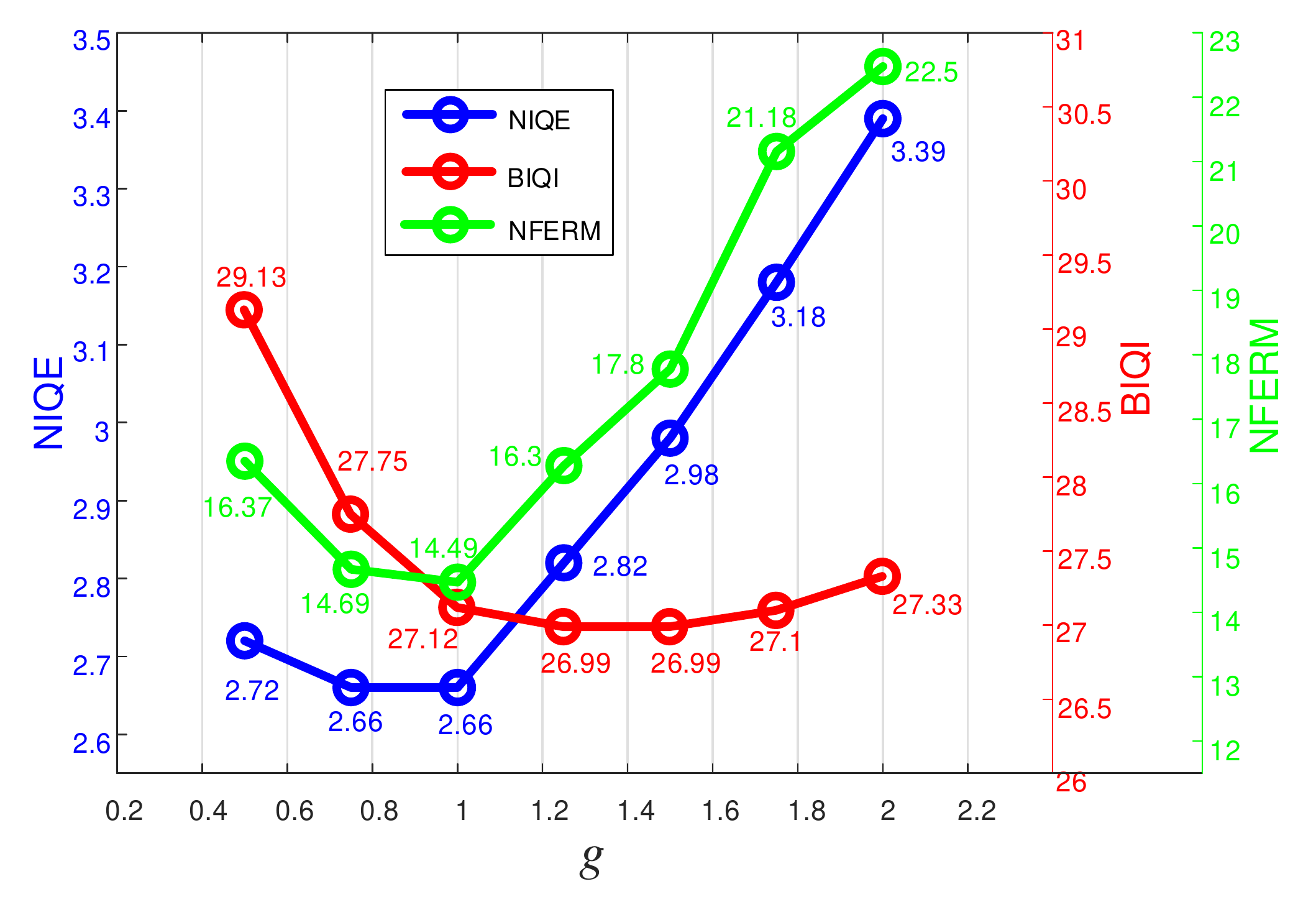}}
\vspace{-1mm}	
	\caption{The influence of ${g}$. Larger ${g}$ values lead to stronger image enhancement, but are more likely to result in over-enhancement. }
	\label{ref_params_g}
\vspace{-2mm}
\end{figure}

Finally, the gain parameter ${g}$ controls the amplitude of the contrast image. Fig.\ref{ref_params_g} presents the visual results and IQA curves when $g$ changes from 0.5 to 2.0 in steps of 0.25 with constant ${m=1}$ and $\left( {\gamma ,k} \right)=\left( {0.6,log(10)} \right)$. Larger ${g}$ values lead to stronger image enhancement, but are more likely to result in over-enhancement. As can be seen in Fig.\ref{ref_params_g}, results with ${g=1}$ have the lowest average NIQE and NFERM values, and almost the lowest BIQI value. 

Based on the visual quality and the quantitative evaluation, we set $\left( {\gamma ,k} \right)=\left( {0.6,log(10)} \right)$, ${g=1}$, and ${m=1}$ for our method.

\subsection{Comparisons}

\begin{figure*}[h]
	\vspace{-6mm}
	\centering                                
	\subfloat[Input]{\includegraphics[width=25.5mm]{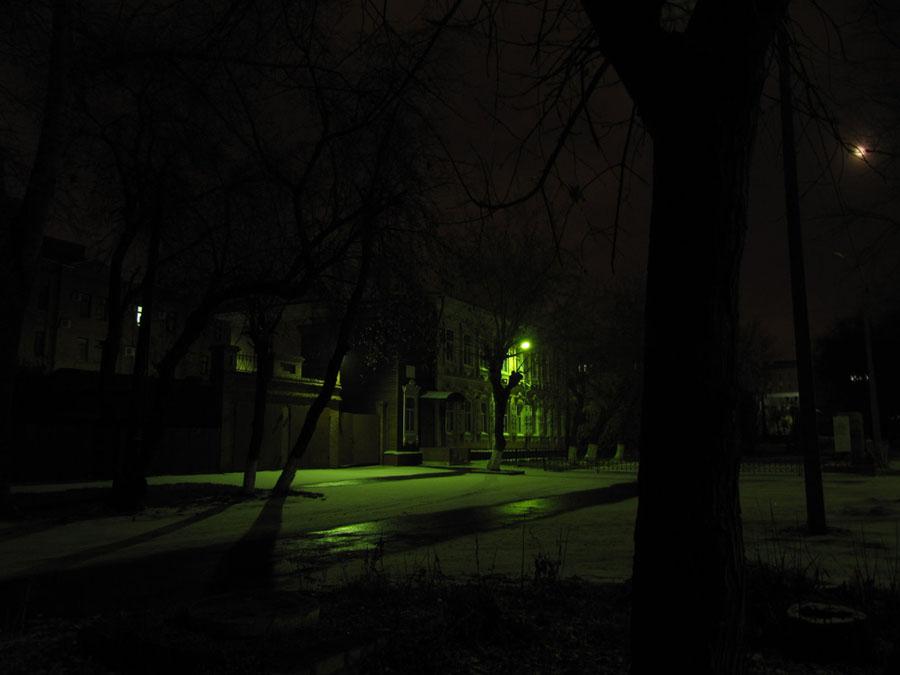}}		
	\hfil                                     
	\subfloat[DFE]{\includegraphics[width=25.5mm]{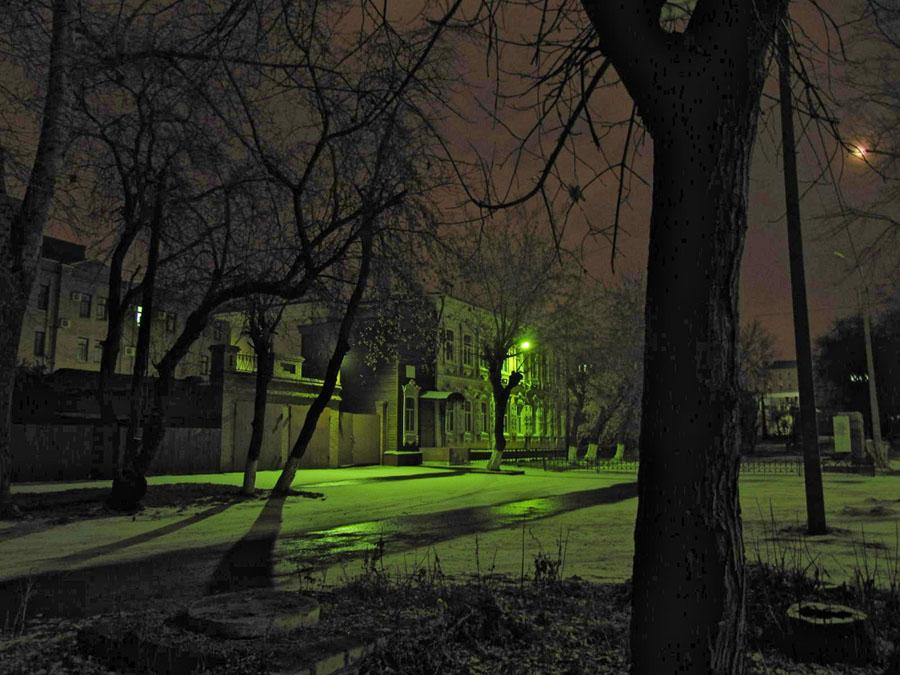}}		
	\hfil                                     
	\subfloat[ALSM]{\includegraphics[width=25.5mm]{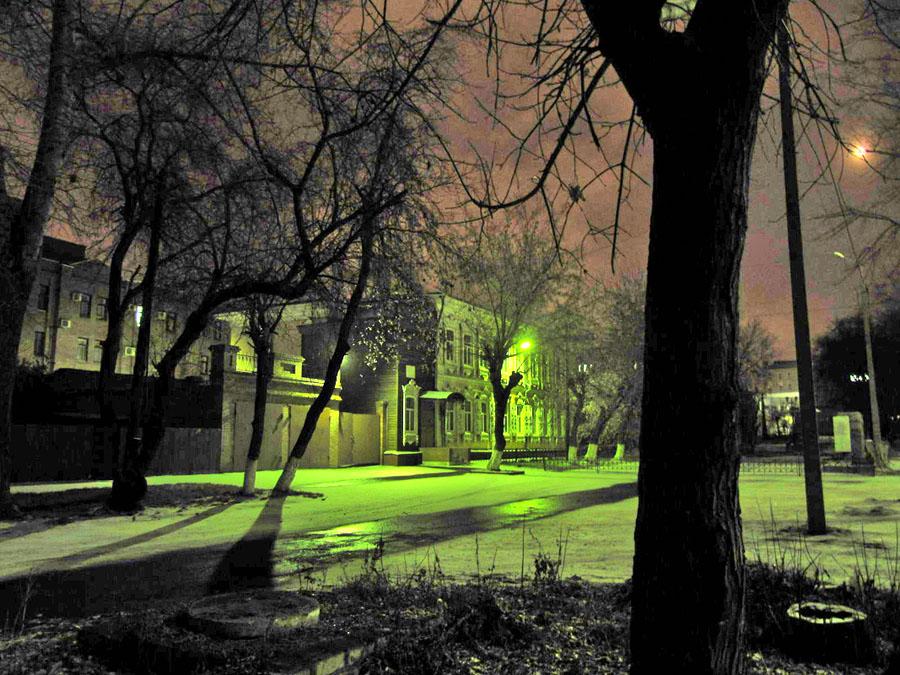}}	
	\hfil	                                  
	\subfloat[LIME]{\includegraphics[width=25.5mm]{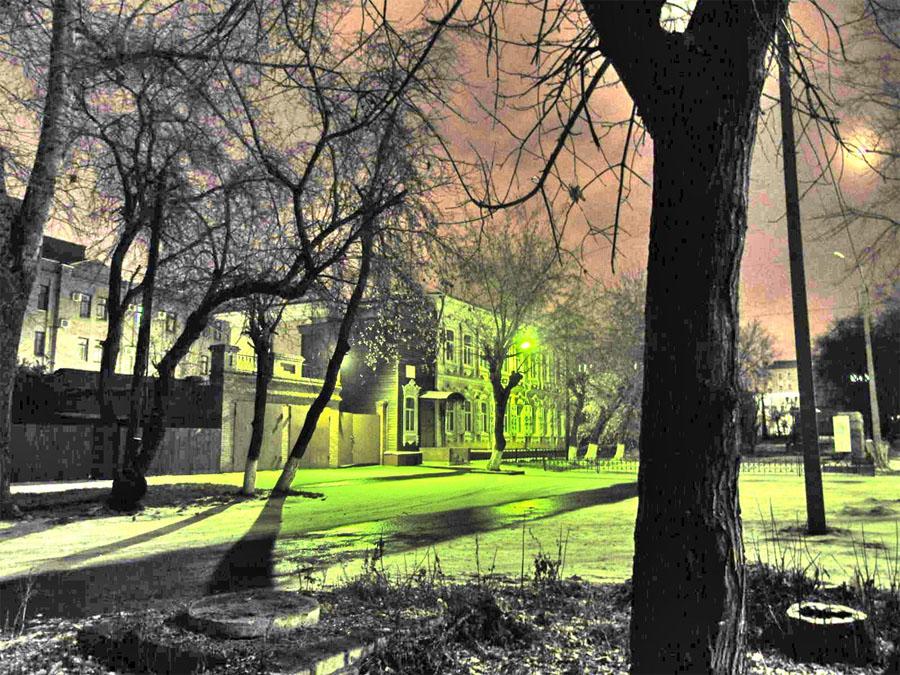}}		
	\hfil                                     
	\subfloat[MLLP]{\includegraphics[width=25.5mm]{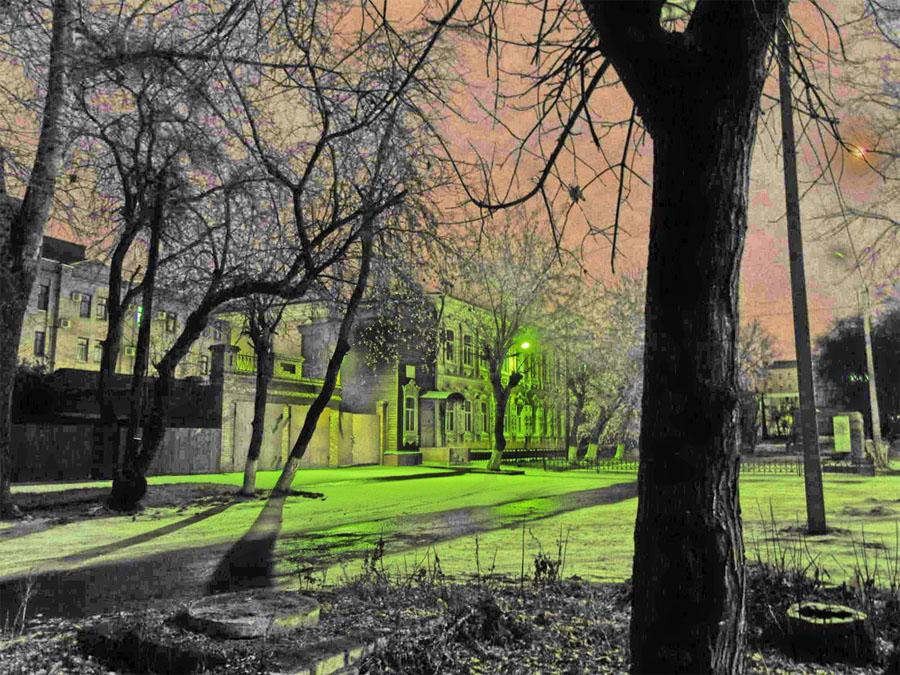}}
	\hfil	                                  
    \subfloat[PnpRtx]{\includegraphics[width=25.5mm]{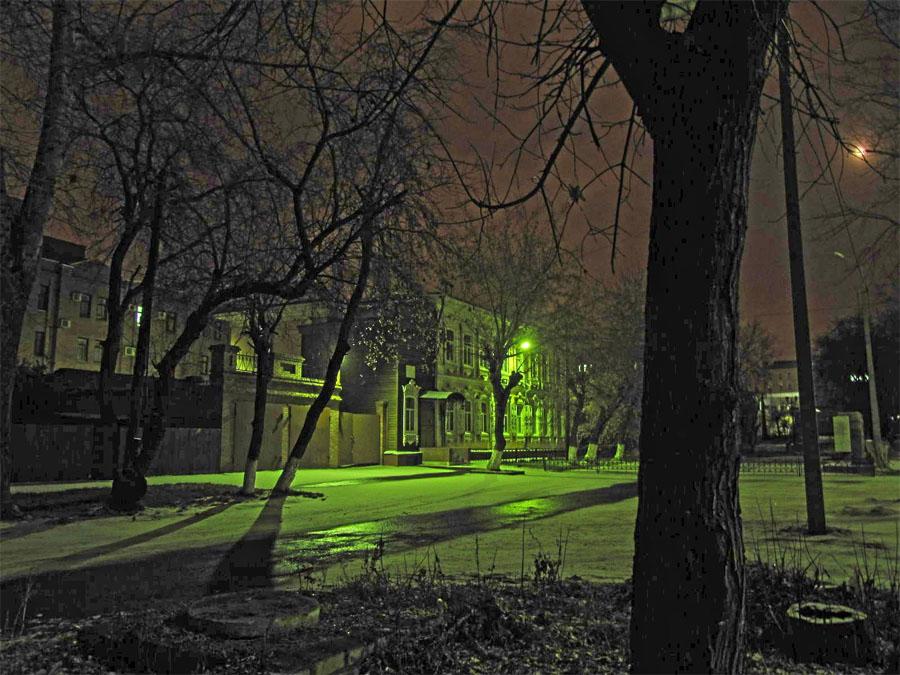}}
	\hfil	                                  
	\subfloat[NRMOE]{\includegraphics[width=25.5mm]{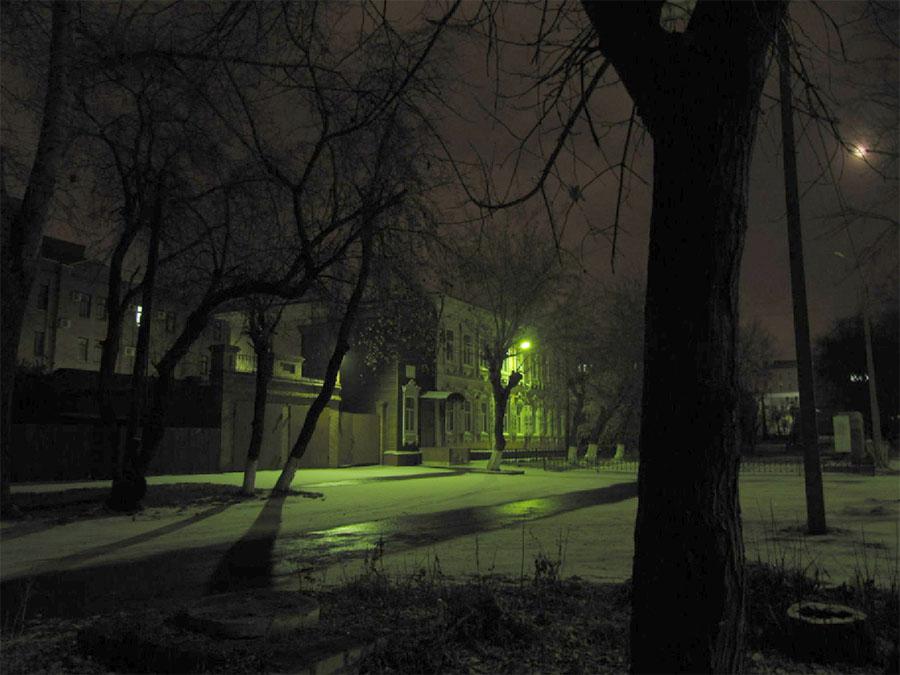}}	
	
	\vspace{-2.5mm}
	
	\centering                                
	\subfloat[PLME]{\includegraphics[width=25.5mm]{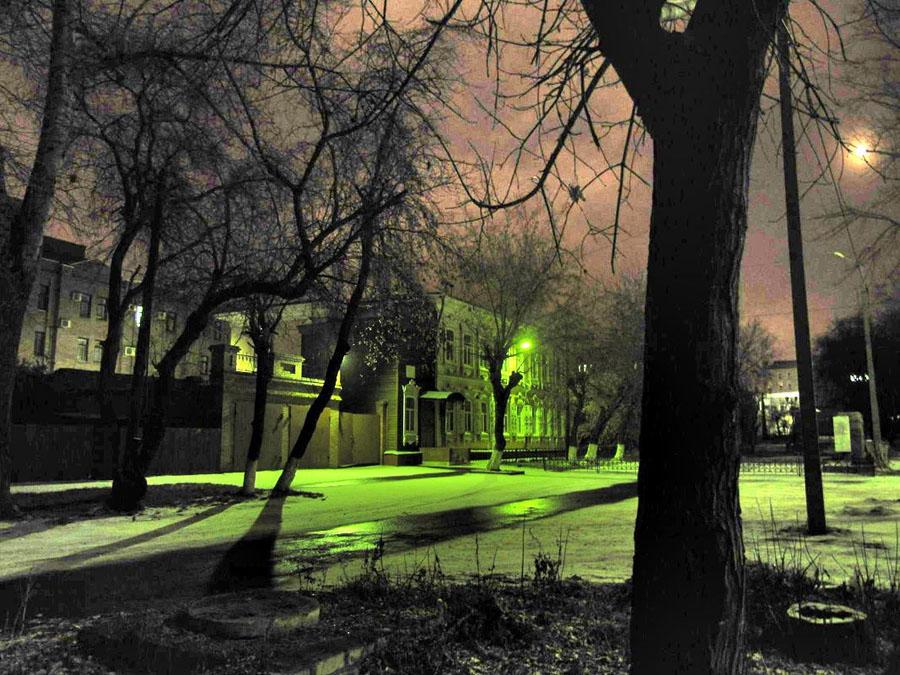}}
	\hfil                                     
	\subfloat[EGAN]{\includegraphics[width=25.5mm]{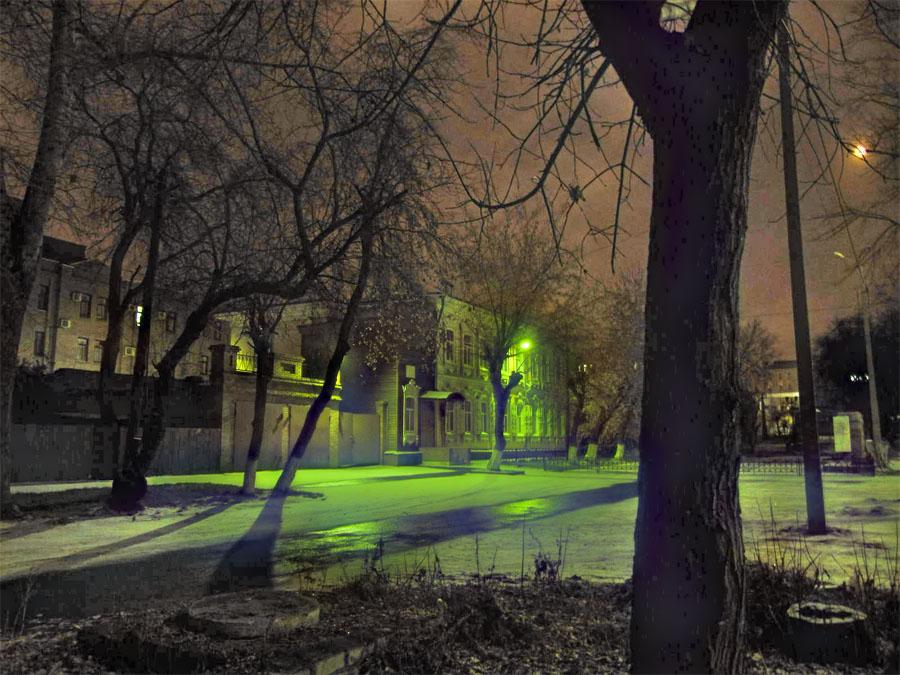}}
	\hfil                                     
	\subfloat[KIND++]{\includegraphics[width=25.5mm]{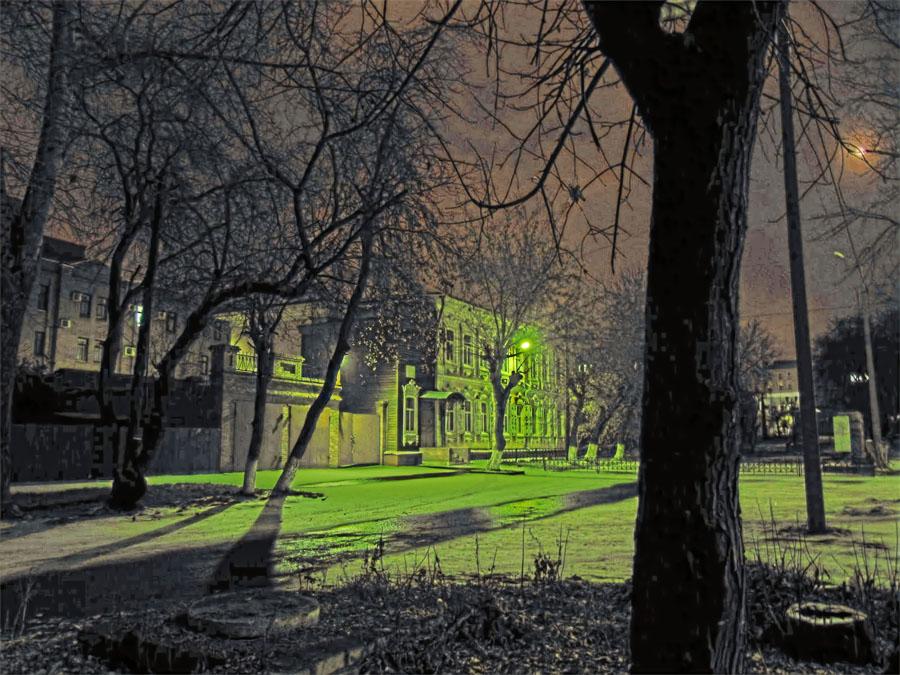}}
	\hfil	                                  
	\subfloat[ZeroDCE]{\includegraphics[width=25.5mm]{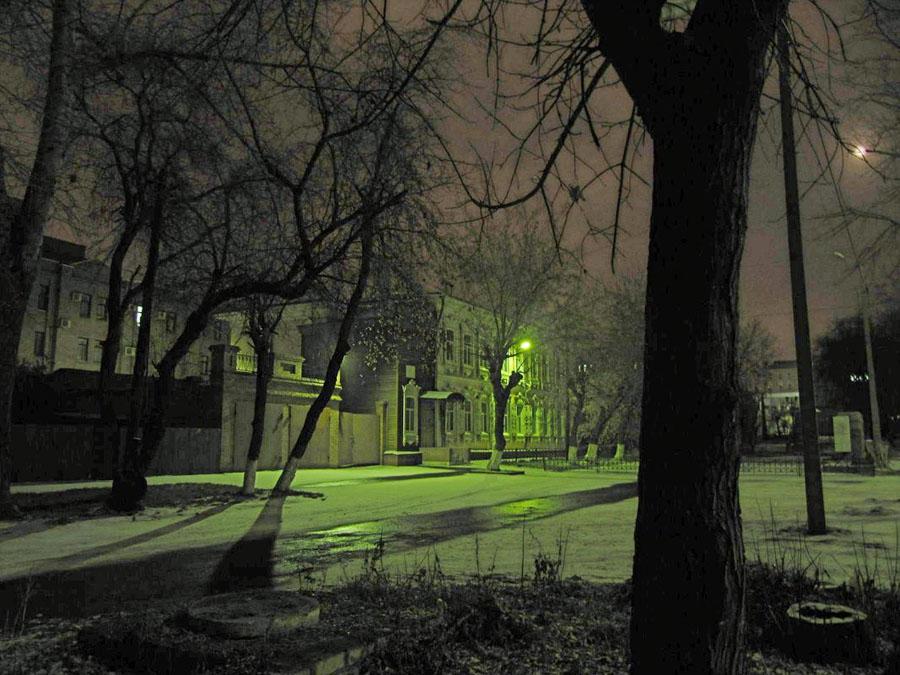}}		
	\hfil                                     
	\subfloat[SCL-LLE]{\includegraphics[width=25.5mm]{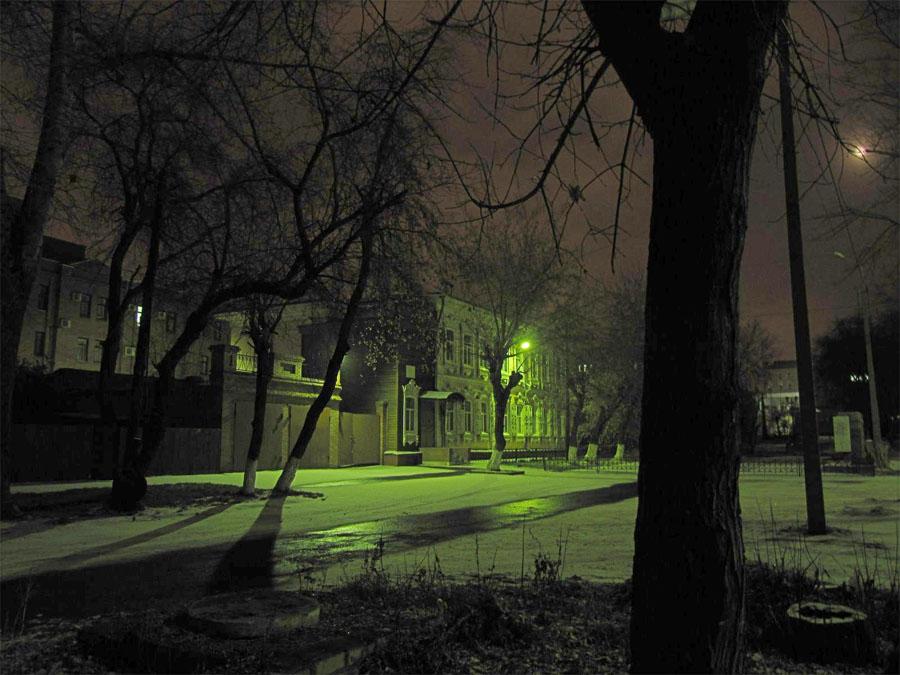}}
	\hfil	                                  
	\subfloat[RtxDIP]{\includegraphics[width=25.5mm]{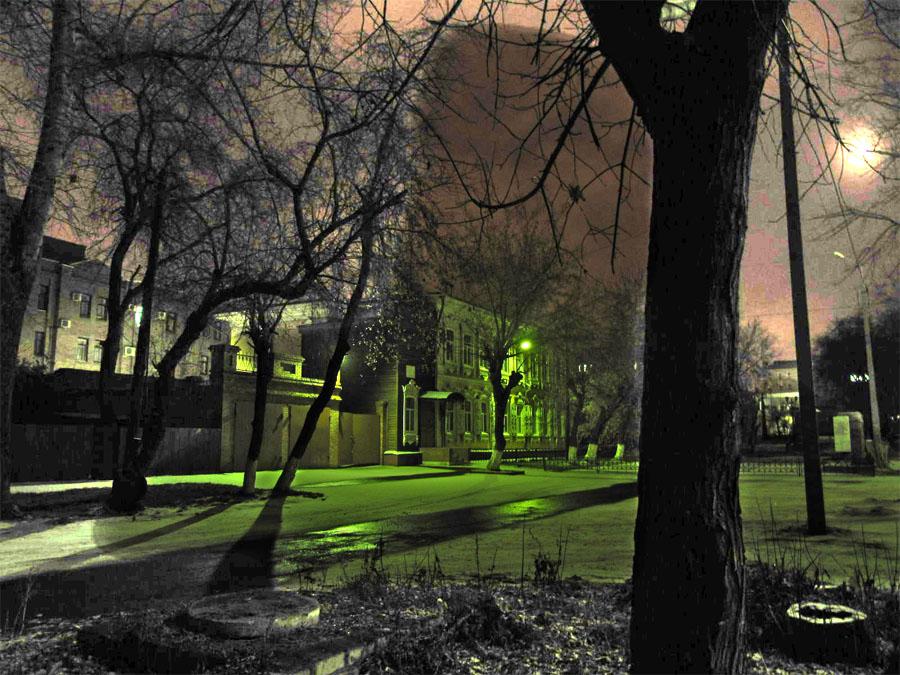}}
	\hfil	                                  
	\subfloat[Proposed]{\includegraphics[width=25.5mm]{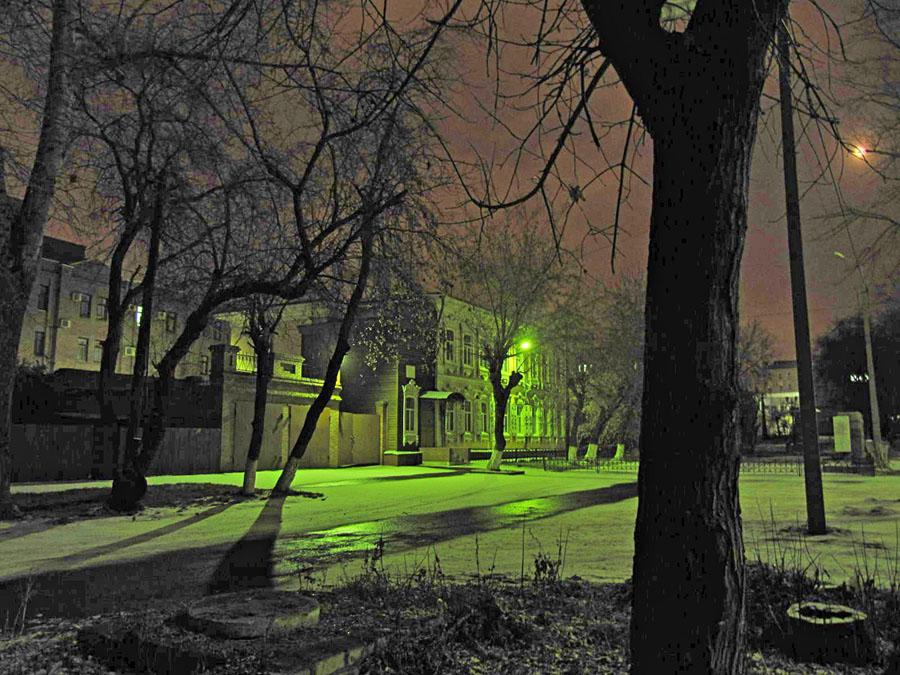}}
	
	\caption{Comparison 1. The images obtained by LIME and MLLP are significantly over-enhanced. The result of RetinexDIP contains noticeable artifacts. Other methods yield graceful results. Please zoom in to see details. }
	\label{Comparison_night}
	\vspace{-2mm}
\end{figure*}

\begin{figure*}[h]
\vspace{-4mm}
	\centering                                
	\subfloat[Input]{\includegraphics[width=25.5mm]{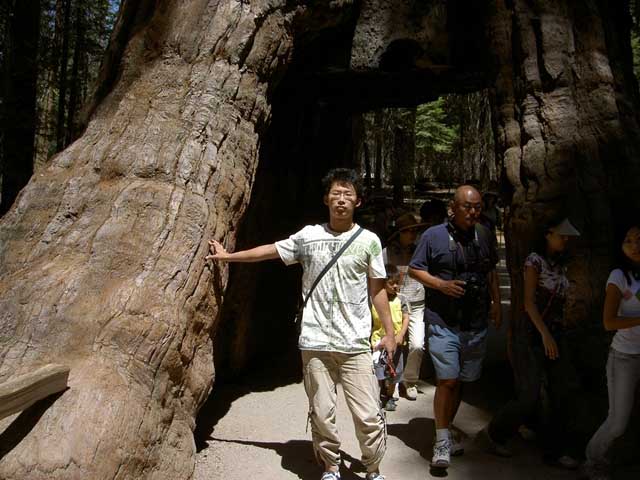}}		
	\hfil                                     
	\subfloat[DFE]{\includegraphics[width=25.5mm]{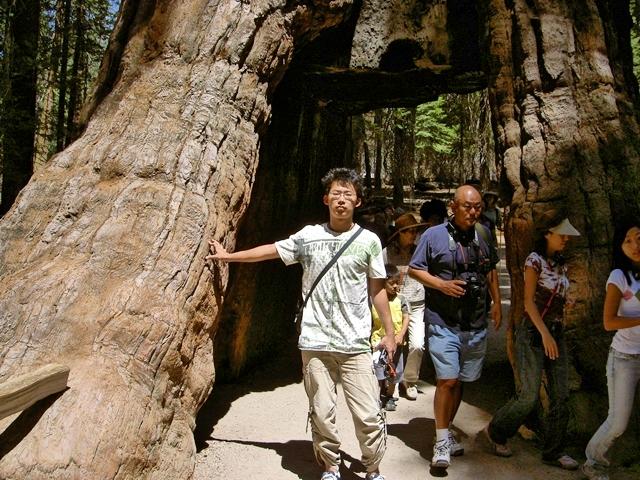}}		
	\hfil                                     
	\subfloat[ALSM]{\includegraphics[width=25.5mm]{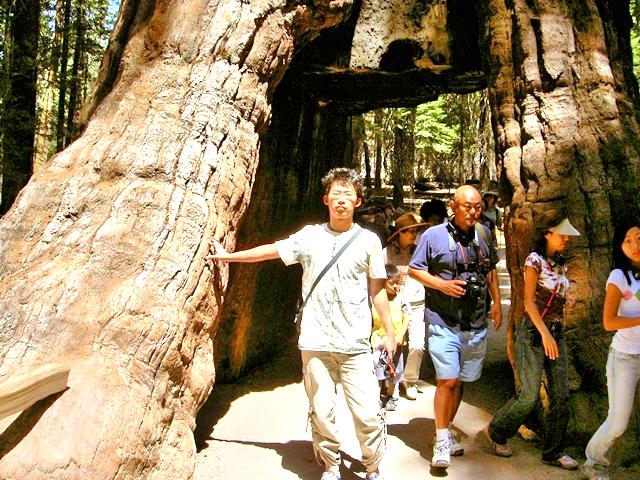}}
	\hfil	                                  
	\subfloat[LIME]{\includegraphics[width=25.5mm]{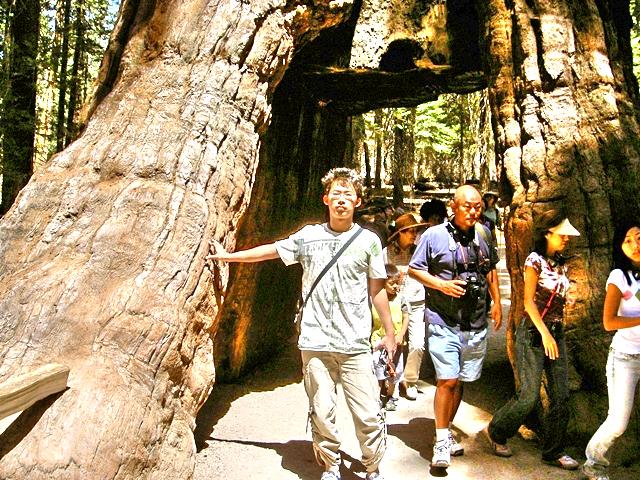}}
	\hfil                                     
	\subfloat[MLLP]{\includegraphics[width=25.5mm]{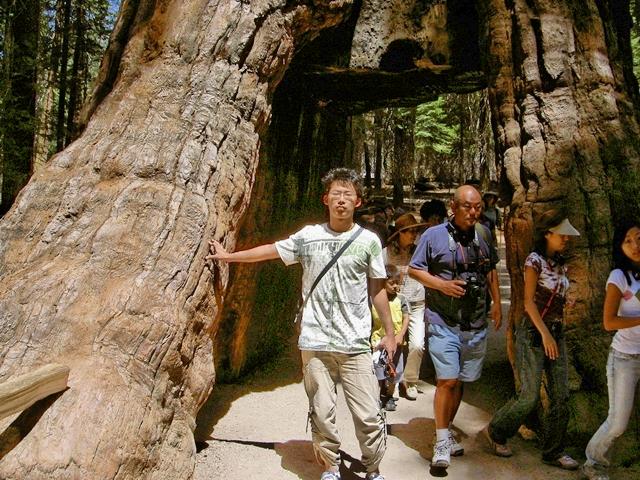}}
	\hfil	                                  
    \subfloat[PnpRtx]{\includegraphics[width=25.5mm]{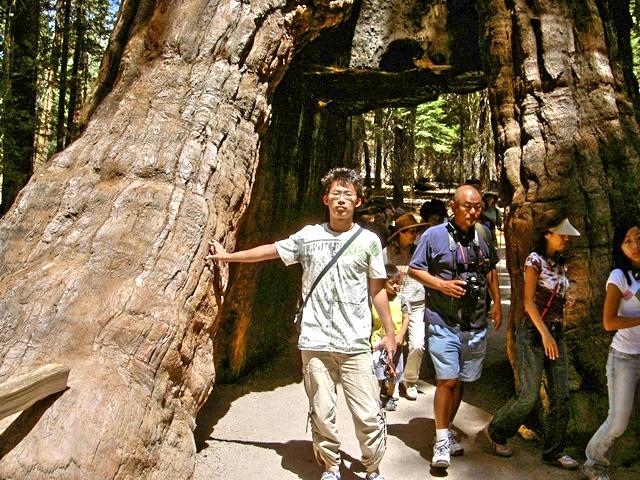}}
	\hfil	                                  
	\subfloat[NRMOE]{\includegraphics[width=25.5mm]{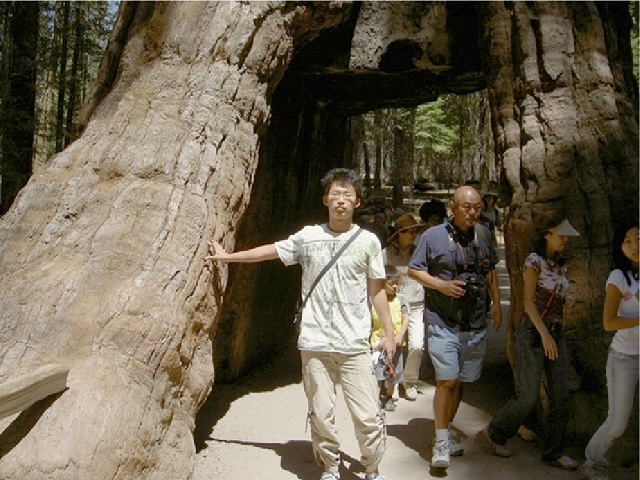}}		
	
	\vspace{-2.5mm}
	
	\centering                                
	\subfloat[PLME]{\includegraphics[width=25.5mm]{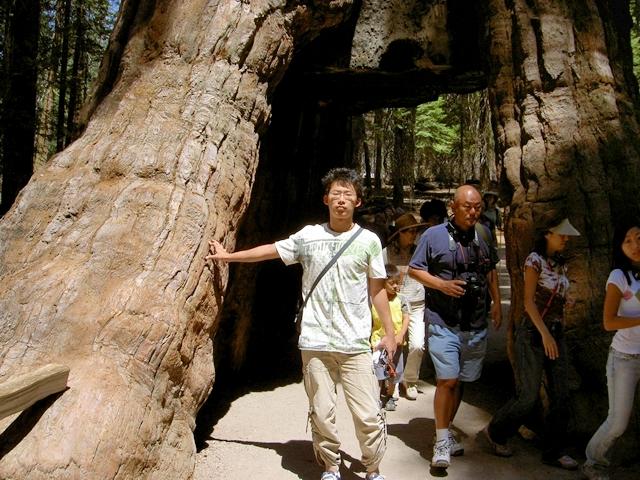}}
	\hfil                                     
	\subfloat[EGAN]{\includegraphics[width=25.5mm]{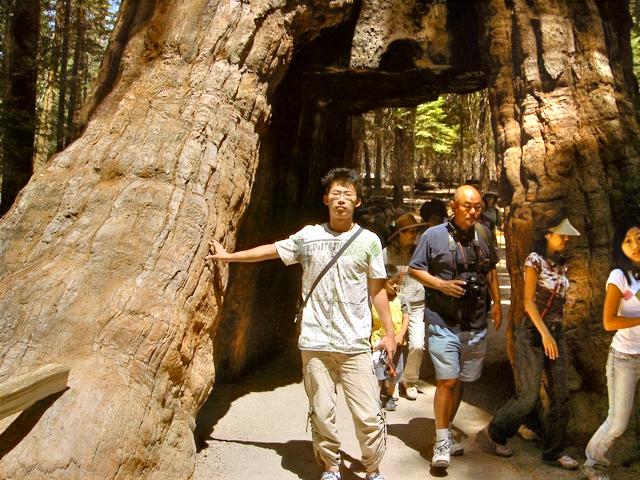}}
	\hfil                                     
	\subfloat[KIND++]{\includegraphics[width=25.5mm]{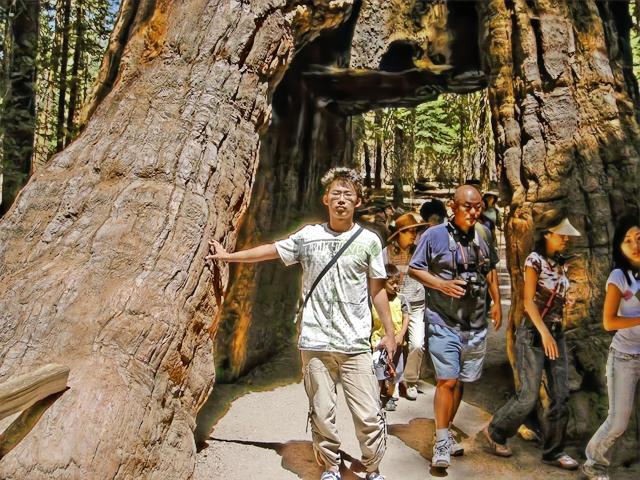}}
	\hfil	                                  
	\subfloat[ZeroDCE]{\includegraphics[width=25.5mm]{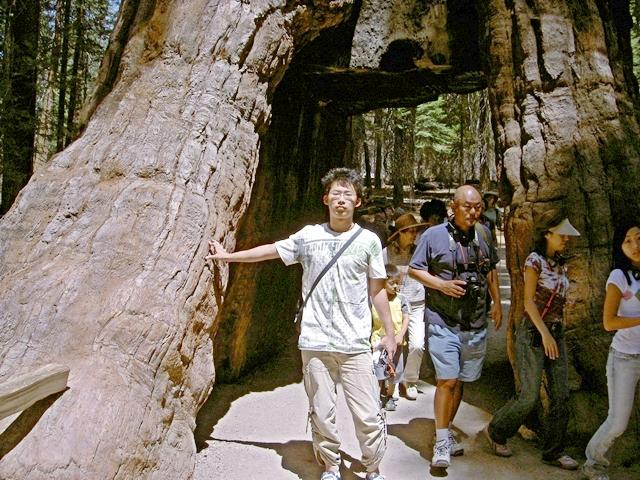}}		
	\hfil                                     
	\subfloat[SCL-LLE]{\includegraphics[width=25.5mm]{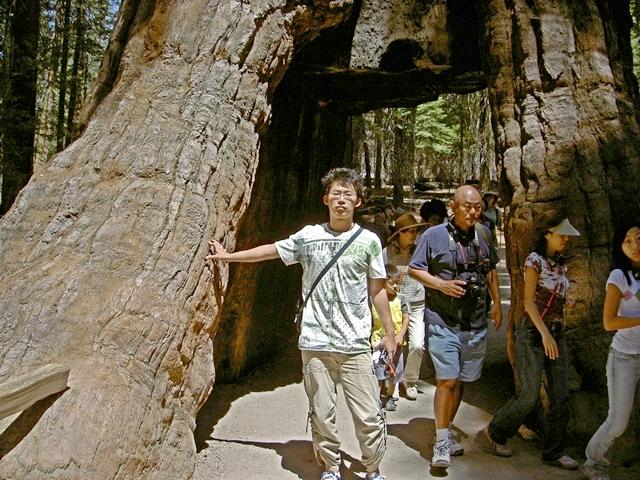}}
	\hfil	                                  
	\subfloat[RtxDIP]{\includegraphics[width=25.5mm]{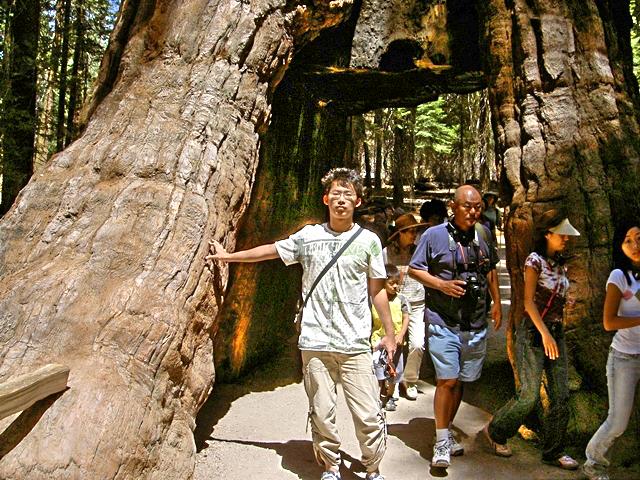}}
	\hfil	                                  
	\subfloat[Proposed]{\includegraphics[width=25.5mm]{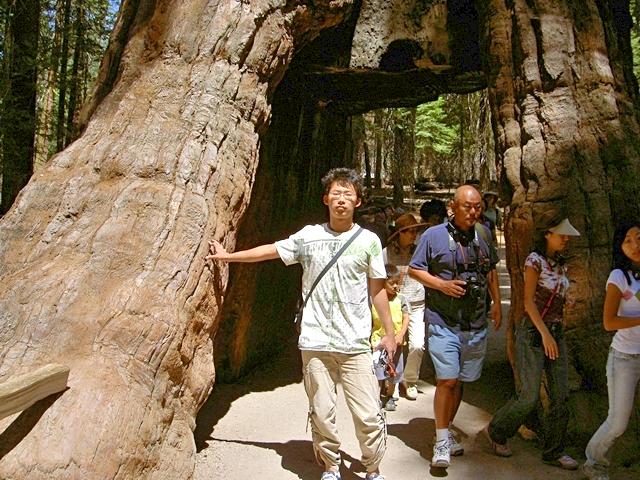}}	
	
	\caption{Comparison 2. ALSM and LIME achieve over-exposed results. There are evident artifacts in the results of KIND++, and RetinexDIP.  EnlightenGAN, NRMOE, ZeroDCE, and SCL-LLE produce results with varying degrees of color distortion. DFE, PnpRetinex, PLME, and the proposed method generate visually-pleasing results. Please zoom in to see details.}
	\label{Comparison_dcim02}
	\vspace{-2mm}
\end{figure*}

\begin{figure*}[h]
\vspace{-4mm}	
	\vspace{-4mm}
	
	\centering                                
	\subfloat[Input]{\includegraphics[width=25.5mm]{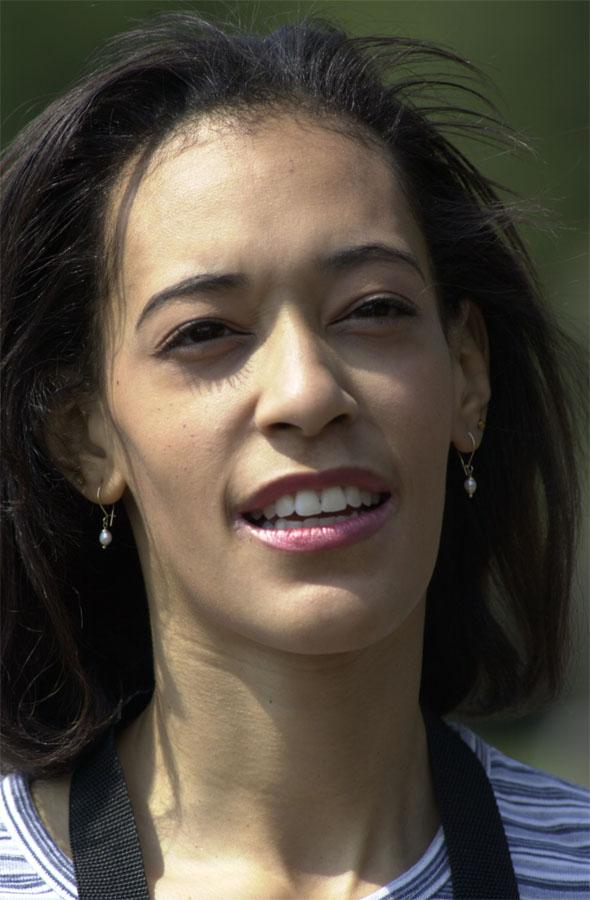}}		
	\hfil                                     
	\subfloat[DFE]{\includegraphics[width=25.5mm]{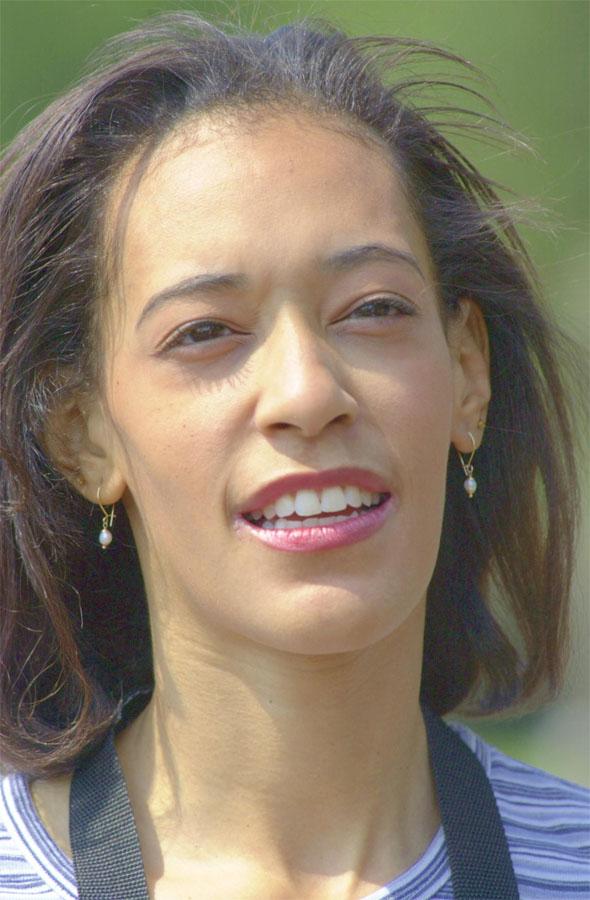}}		
	\hfil                                     
	\subfloat[ALSM]{\includegraphics[width=25.5mm]{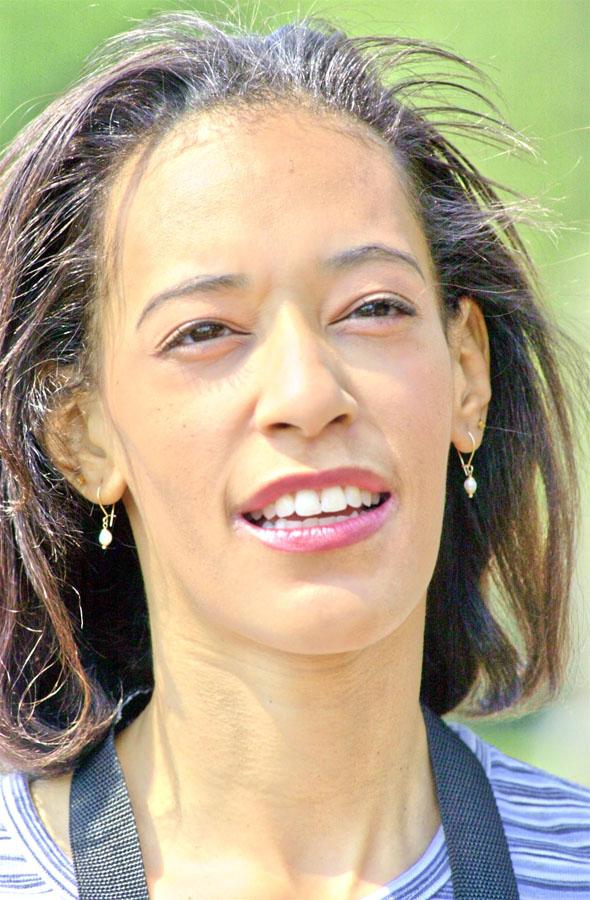}}	
	\hfil	                                  
	\subfloat[LIME]{\includegraphics[width=25.5mm]{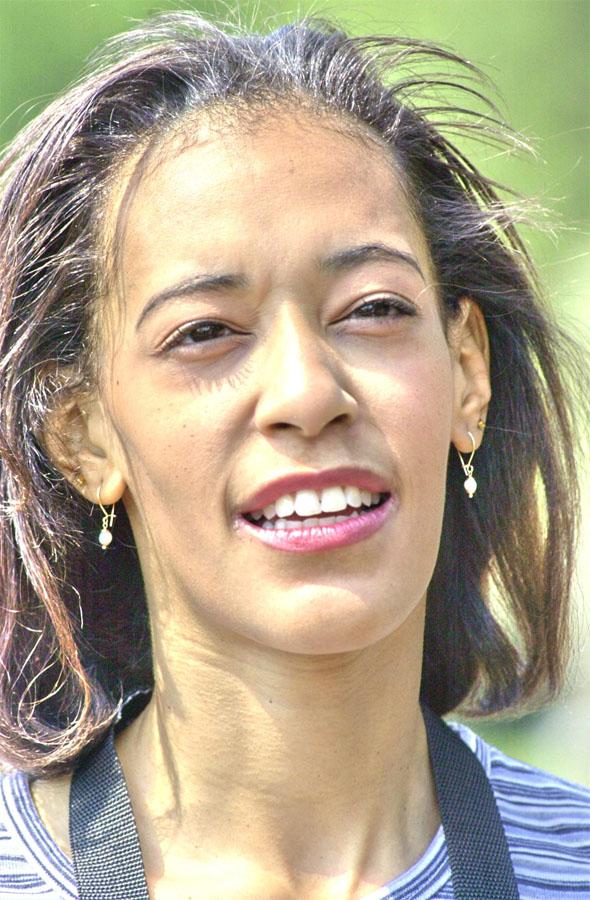}}		
	\hfil                                     
	\subfloat[MLLP]{\includegraphics[width=25.5mm]{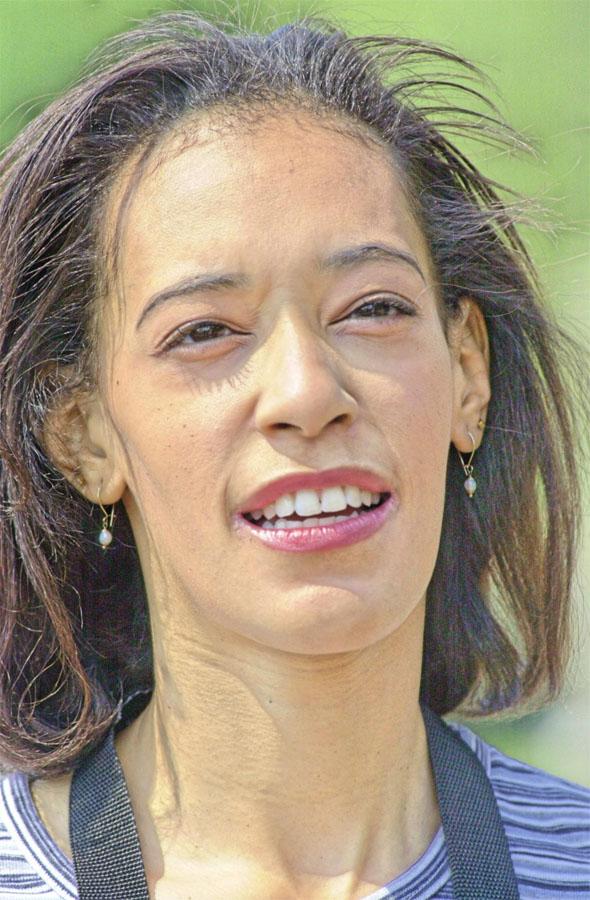}}
	\hfil	                                  
    \subfloat[PnpRtx]{\includegraphics[width=25.5mm]{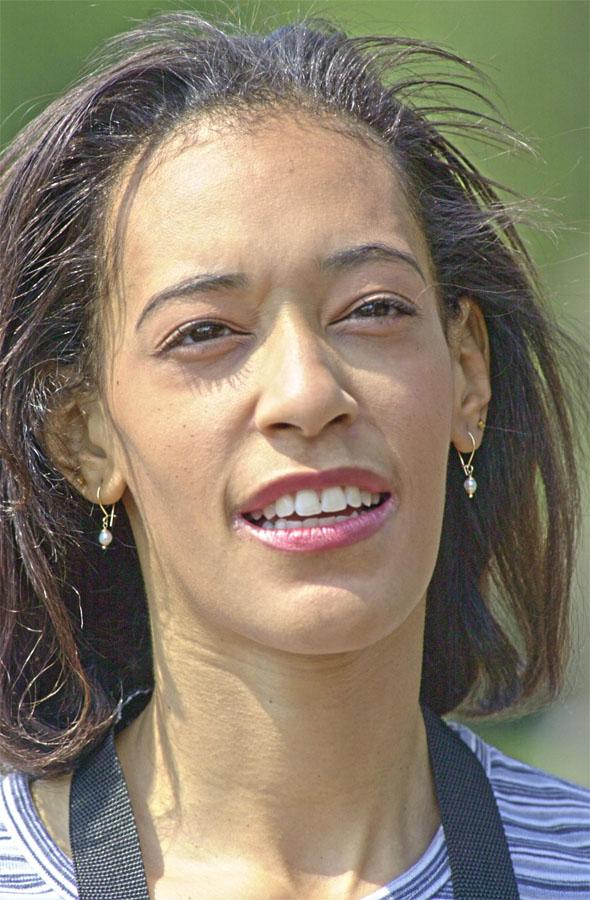}}
	\hfil	                                  
	\subfloat[NRMOE]{\includegraphics[width=25.5mm]{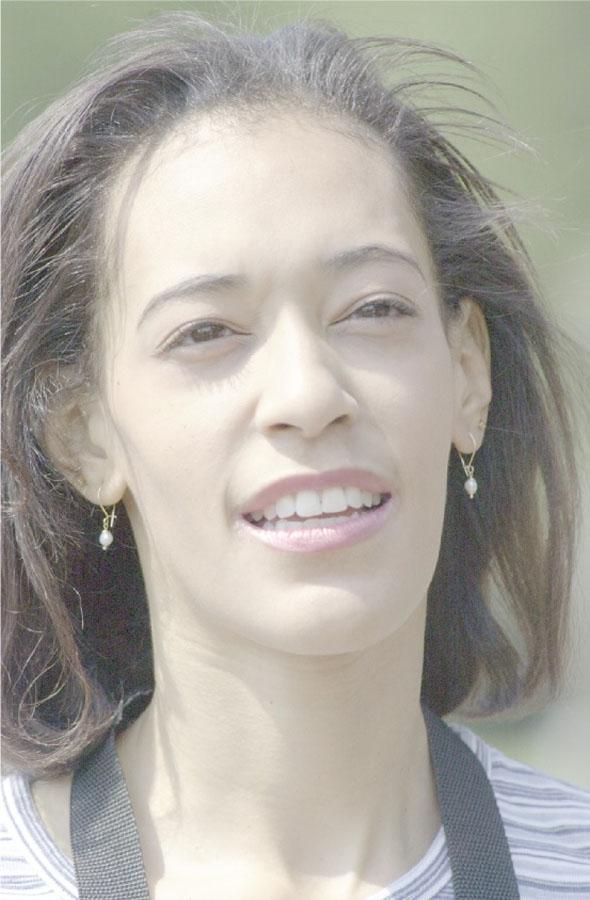}}		
	
	\vspace{-2.5mm}
	\centering                                
	\subfloat[PLME]{\includegraphics[width=25.5mm]{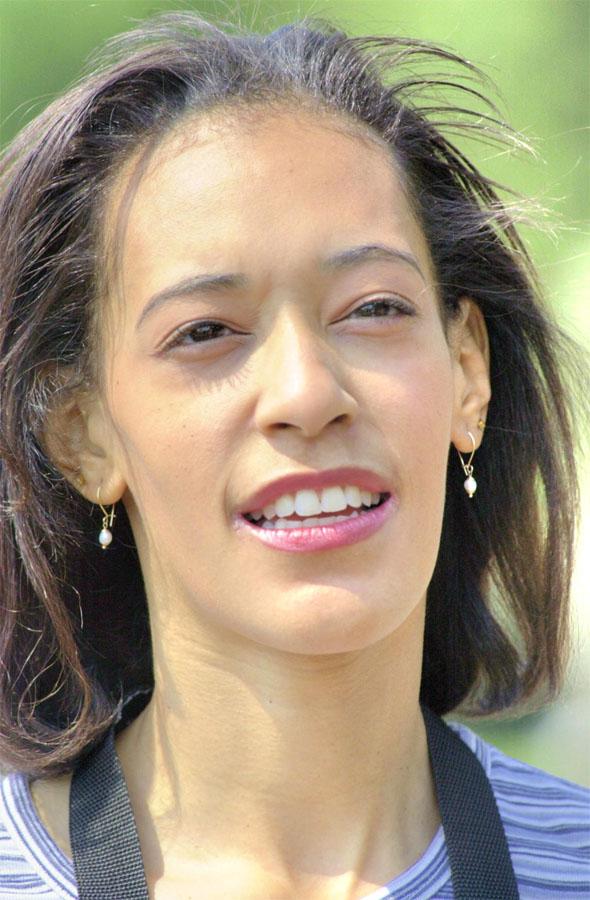}}
	\hfil                                     
	\subfloat[EGAN]{\includegraphics[width=25.5mm]{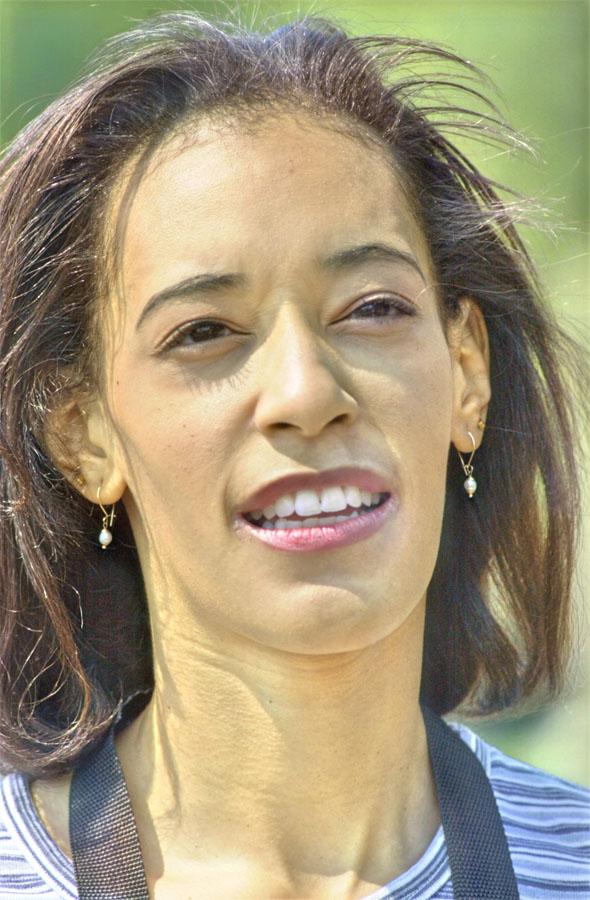}}
	\hfil                                     
	\subfloat[KIND++]{\includegraphics[width=25.5mm]{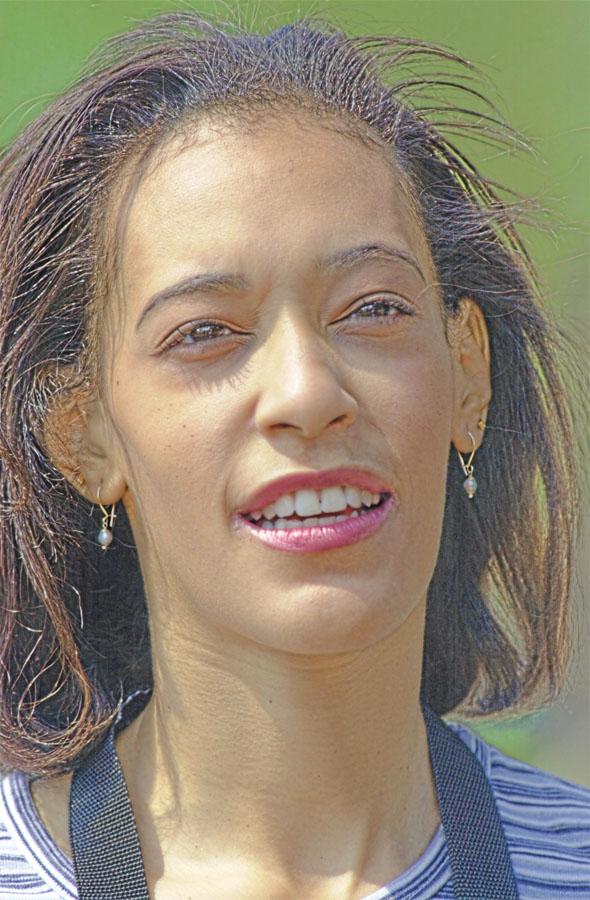}}
	\hfil	                                  
	\subfloat[ZeroDCE]{\includegraphics[width=25.5mm]{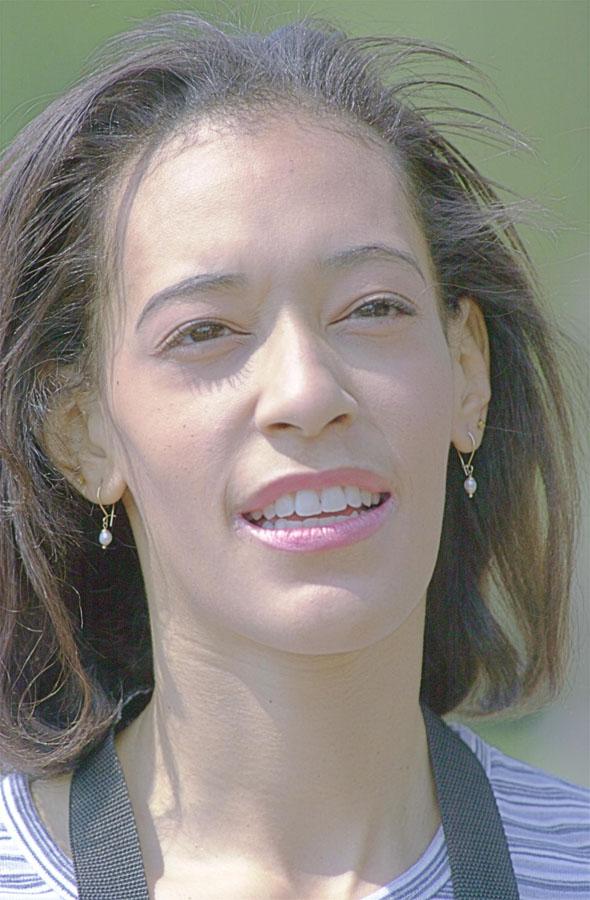}}		
	\hfil                                     
	\subfloat[SCL-LLE]{\includegraphics[width=25.5mm]{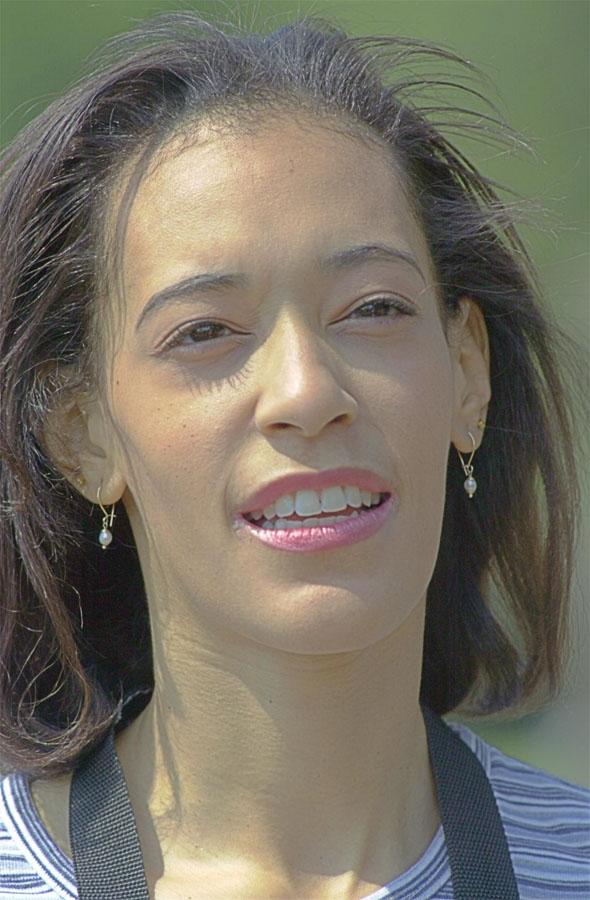}}
	\hfil	                                  
	\subfloat[RtxDIP]{\includegraphics[width=25.5mm]{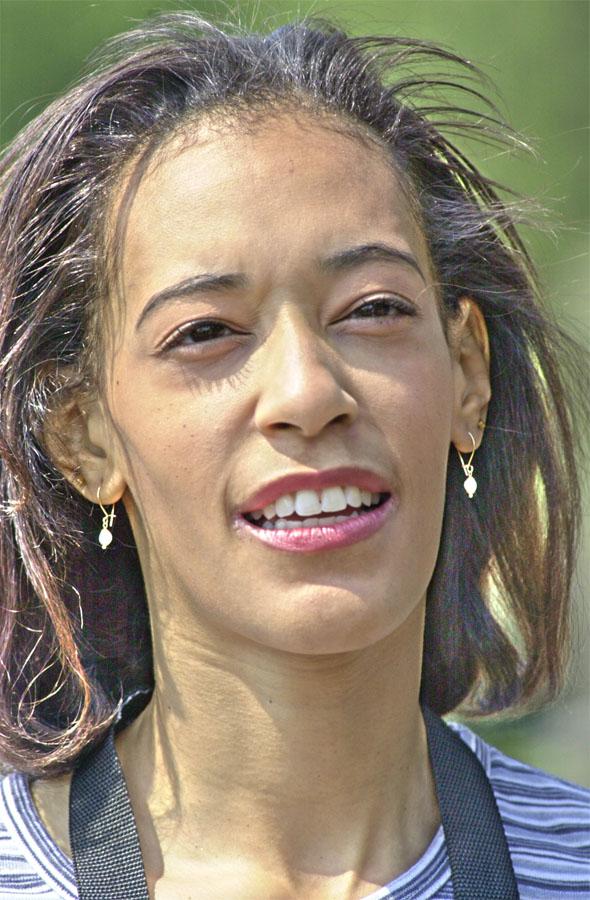}}
	\hfil	                                  
	\subfloat[Proposed]{\includegraphics[width=25.5mm]{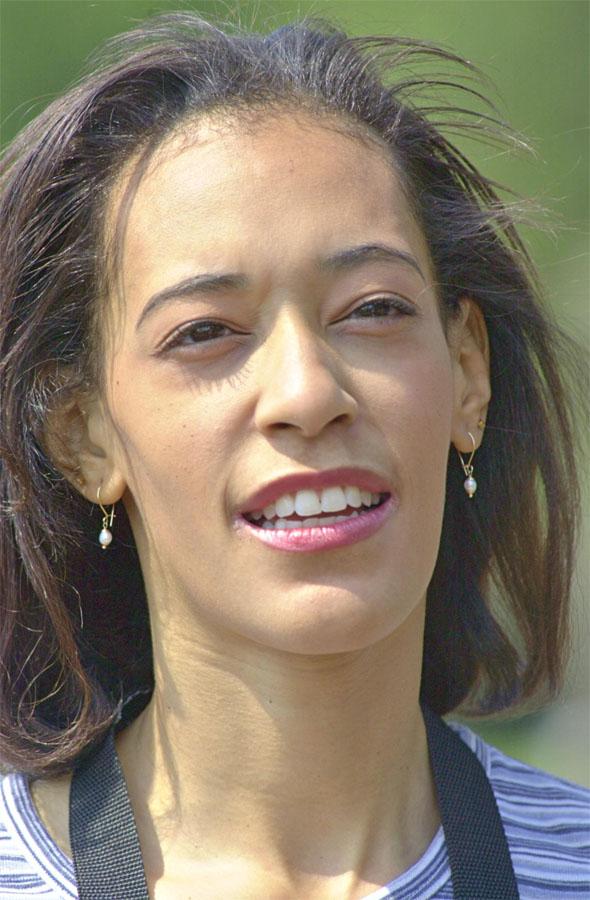}}

	\caption{Comparison 3. DFE, MLLP, PnpRetinex, PLME, RetinexDIP, and the proposed method produce better results than the other methods in terms of fidelity. Please zoom in to see details.}
	\label{Comparison_lady}
\vspace{-0mm}
\end{figure*}

\begin{figure*}[h]
\vspace{-4mm}	
	\centering                                
	\subfloat[Input]{\includegraphics[width=25.5mm]{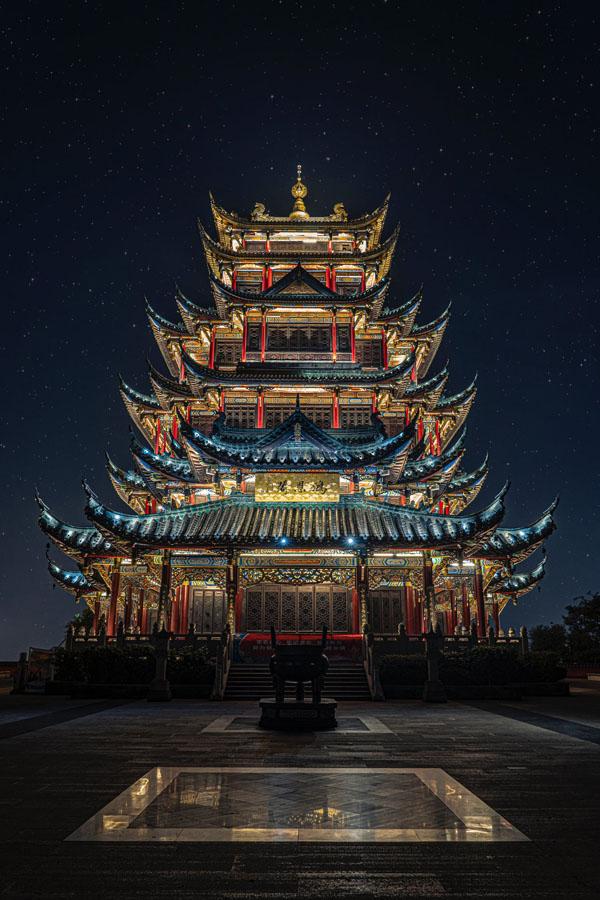}}		
	\hfil                                     
	\subfloat[DFE]{\includegraphics[width=25.5mm]{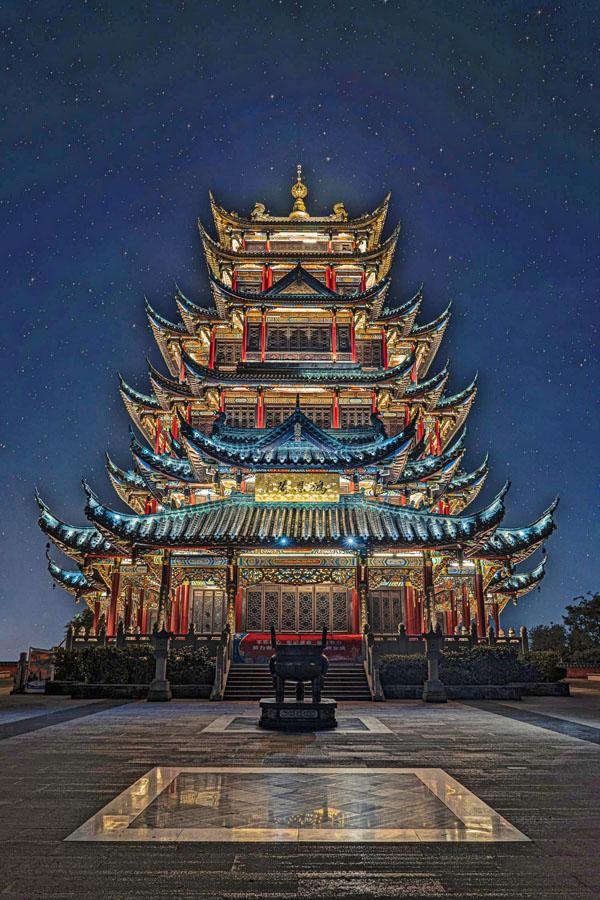}}		
	\hfil                                     
	\subfloat[ALSM]{\includegraphics[width=25.5mm]{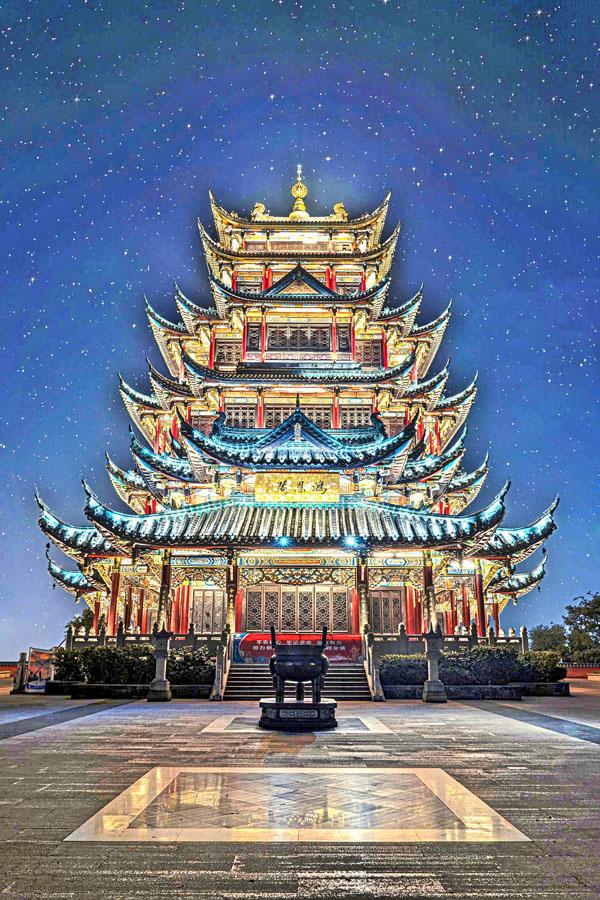}}
	\hfil	                                  
	\subfloat[LIME]{\includegraphics[width=25.5mm]{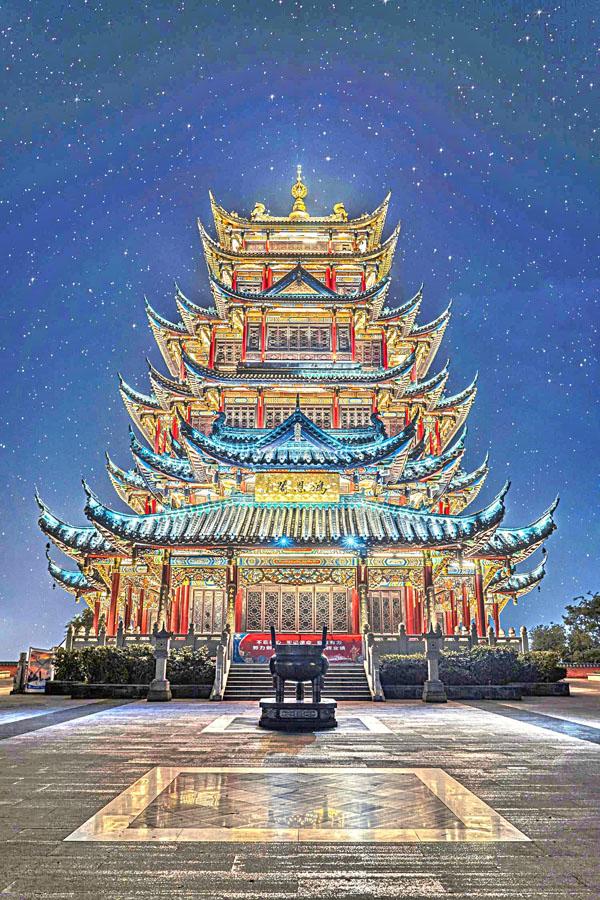}}
	\hfil                                     
	\subfloat[MLLP]{\includegraphics[width=25.5mm]{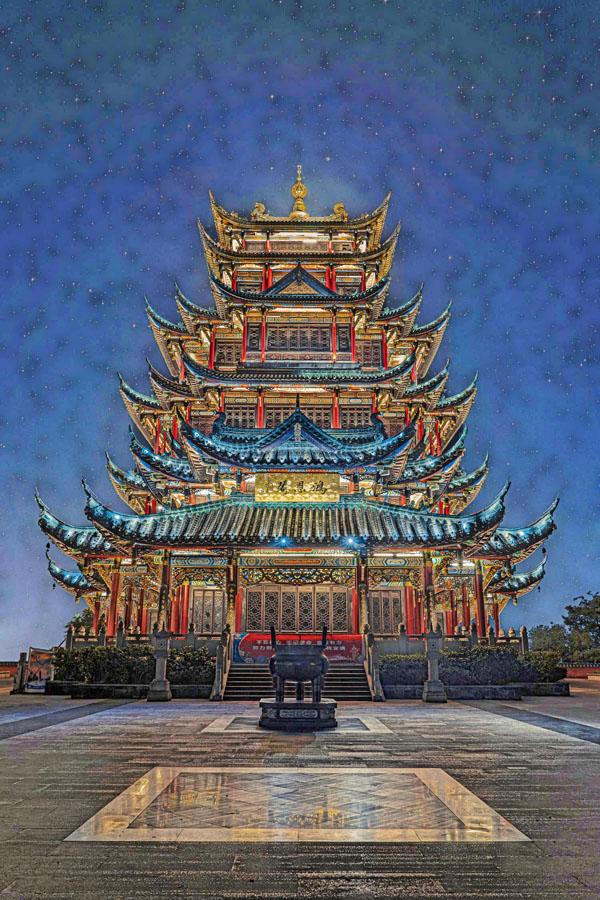}}
	\hfil	                                  
    \subfloat[PnpRtx]{\includegraphics[width=25.5mm]{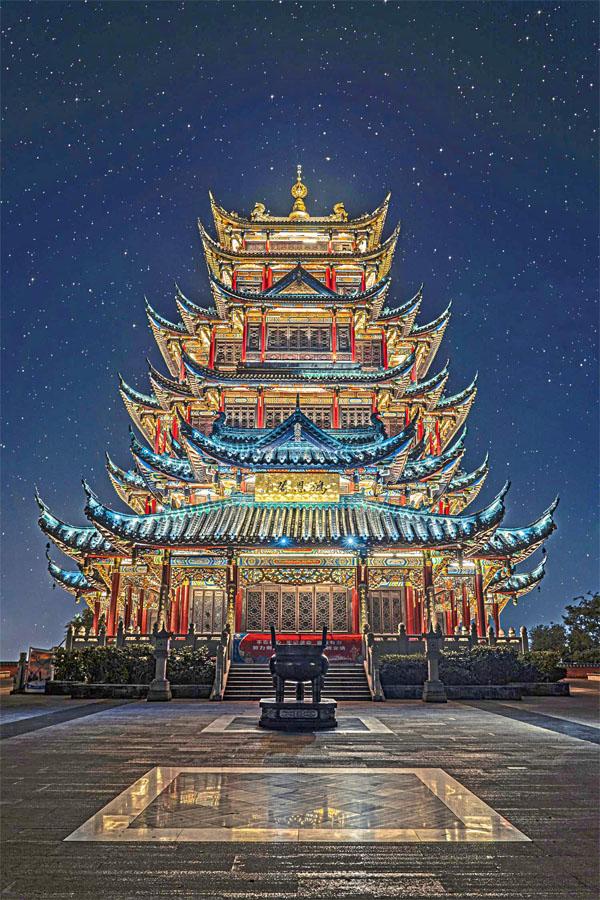}}
	\hfil	                                  
	\subfloat[NRMOE]{\includegraphics[width=25.5mm]{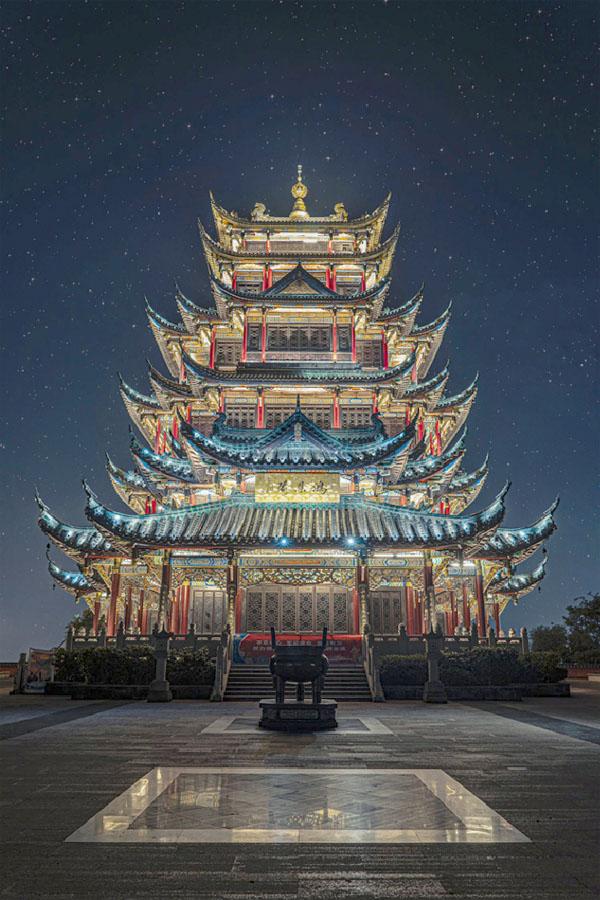}}		
	
	\vspace{-2.5mm}
	
	\centering                                
	\subfloat[PLME]{\includegraphics[width=25.5mm]{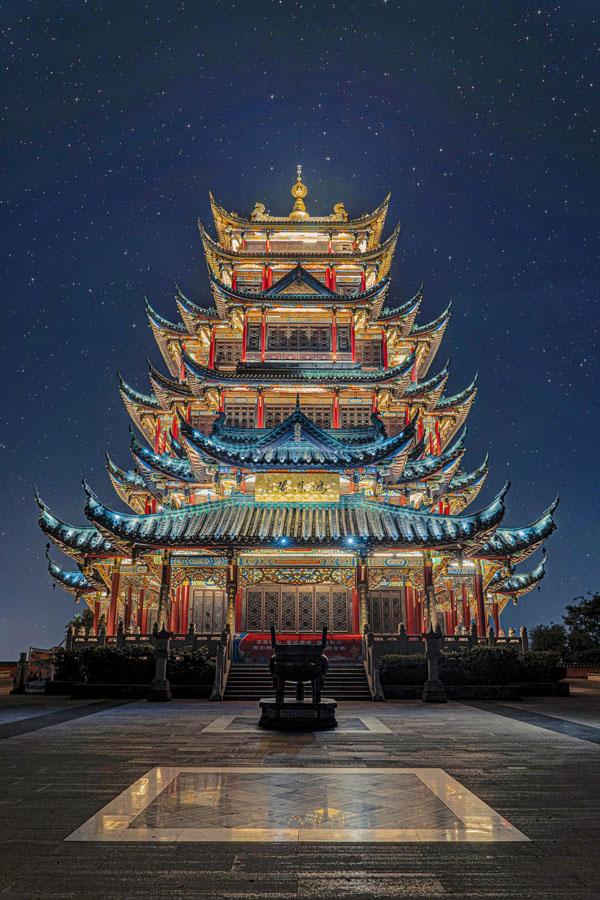}}
	\hfil                                     
	\subfloat[EGAN]{\includegraphics[width=25.5mm]{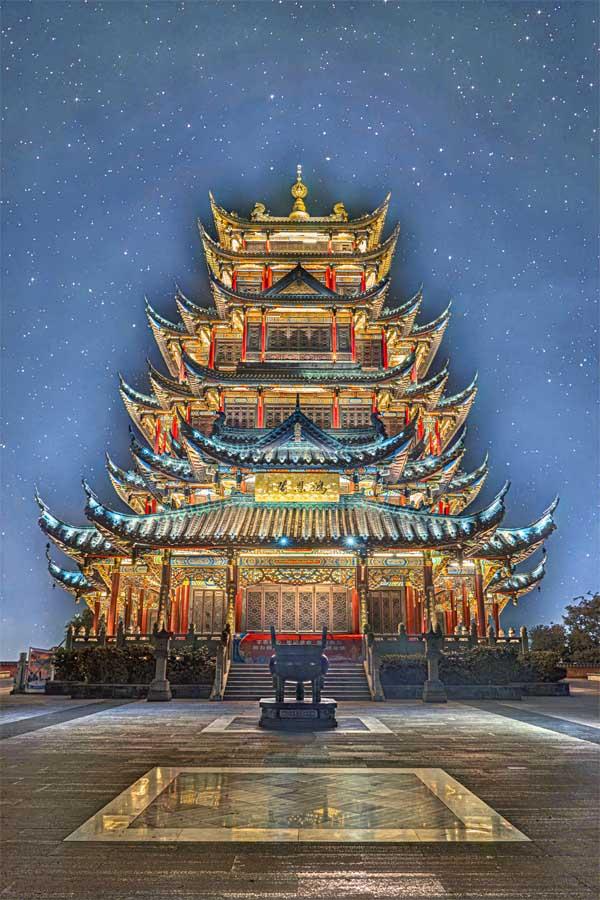}}
	\hfil                                     
	\subfloat[KIND++]{\includegraphics[width=25.5mm]{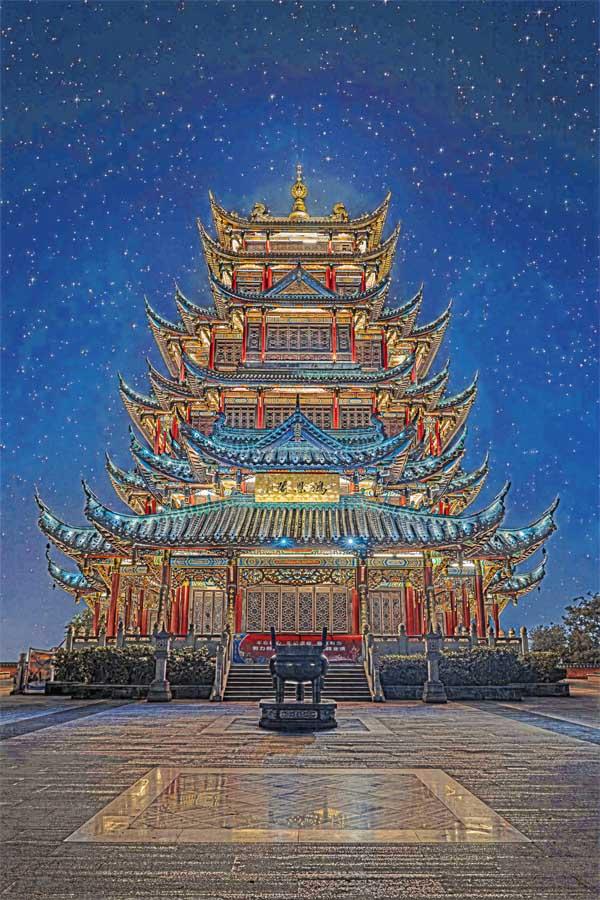}}
	\hfil	                                  
	\subfloat[ZeroDCE]{\includegraphics[width=25.5mm]{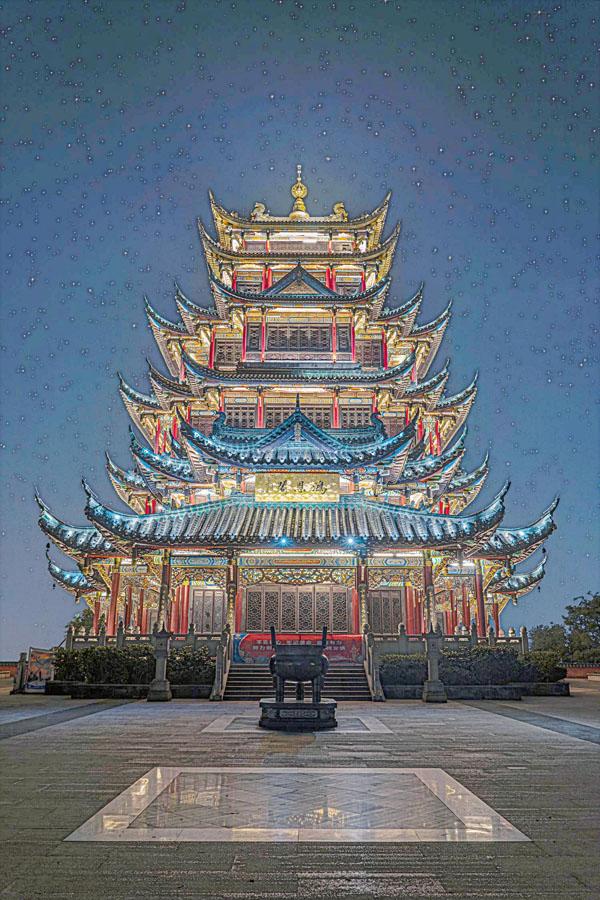}}		
	\hfil                                     
	\subfloat[SCL-LLE]{\includegraphics[width=25.5mm]{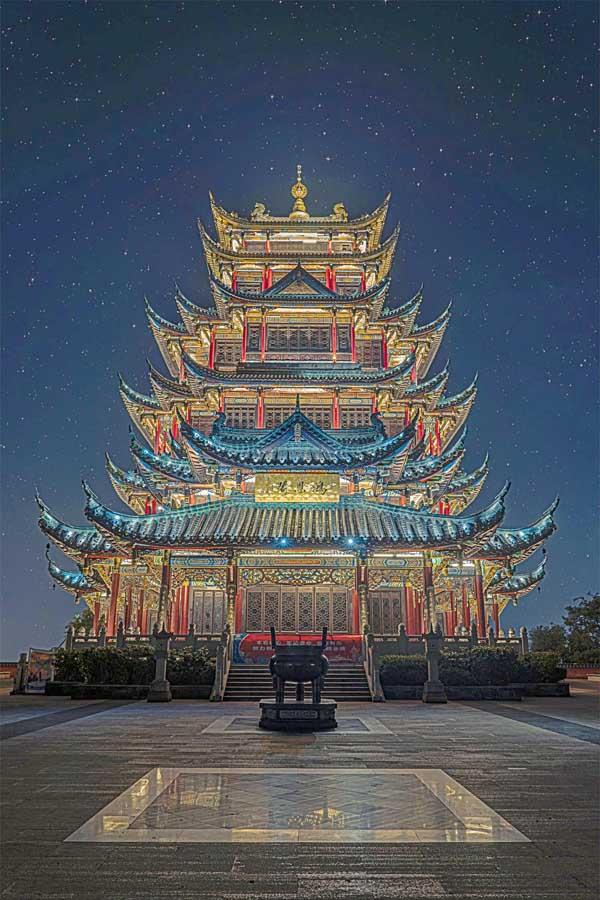}}
	\hfil	                                  
	\subfloat[RtxDIP]{\includegraphics[width=25.5mm]{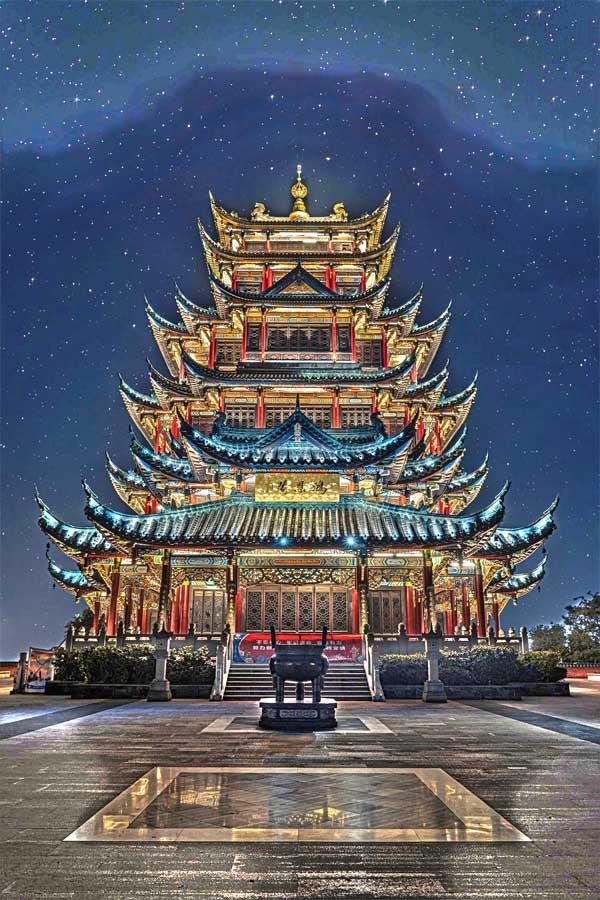}}
	\hfil	                                  
	\subfloat[Proposed]{\includegraphics[width=25.5mm]{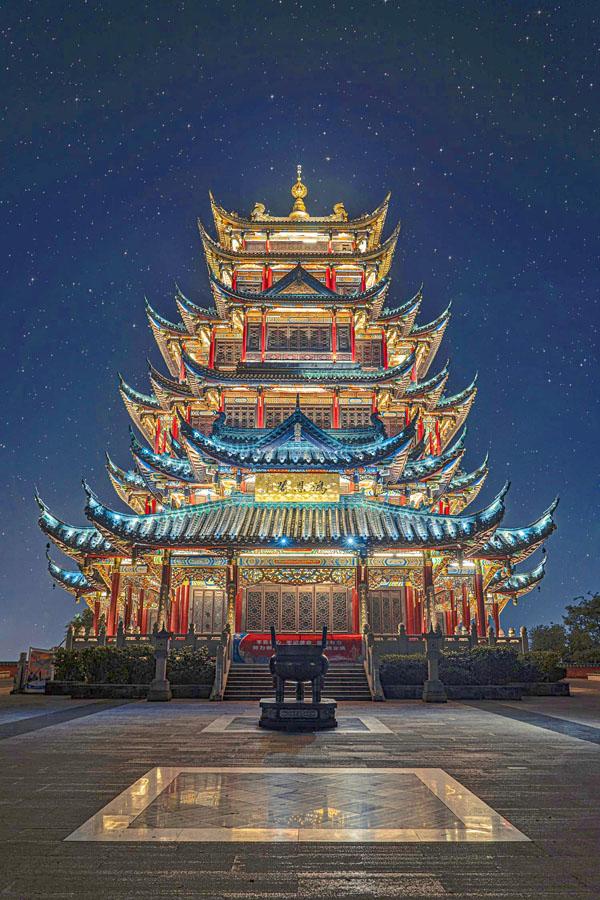}}
	
	\caption{Comparison 4. DFE, PnpRetinex, and the proposed method produce more visually-appealing results than the other methods. However, the result of DFE suffers from slight halo artifacts. Please zoom in to see details.}
	\label{Comparison_tower}
	\vspace{-4mm}
\end{figure*}

\vspace{2mm}
\noindent\textbf{1) Qualitative assessment}

\begin{figure*}[h]
\vspace{-8mm}
	
	\centering                                
	\subfloat[Input]{\includegraphics[width=25.5mm]{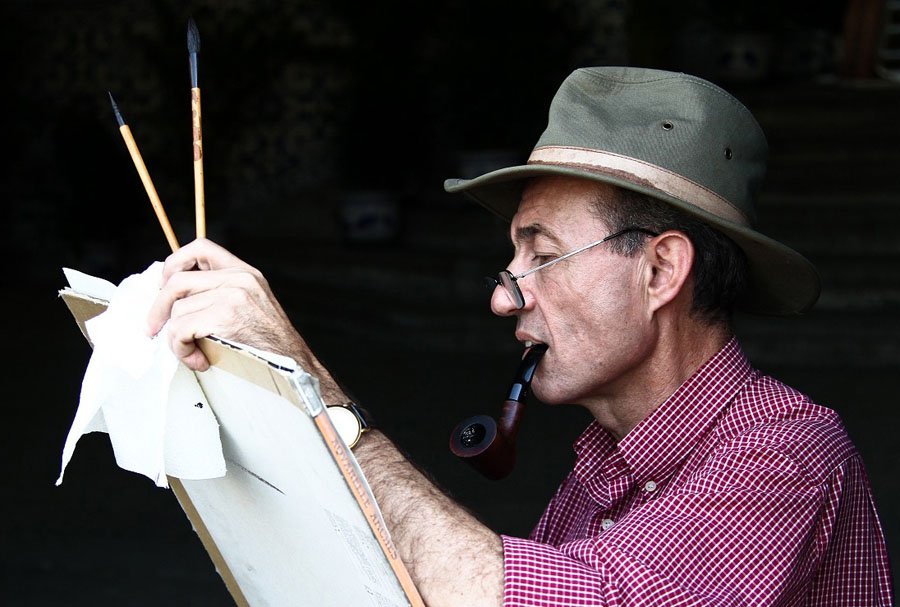}}		
	\hfil                                     
	\subfloat[DFE]{\includegraphics[width=25.5mm]{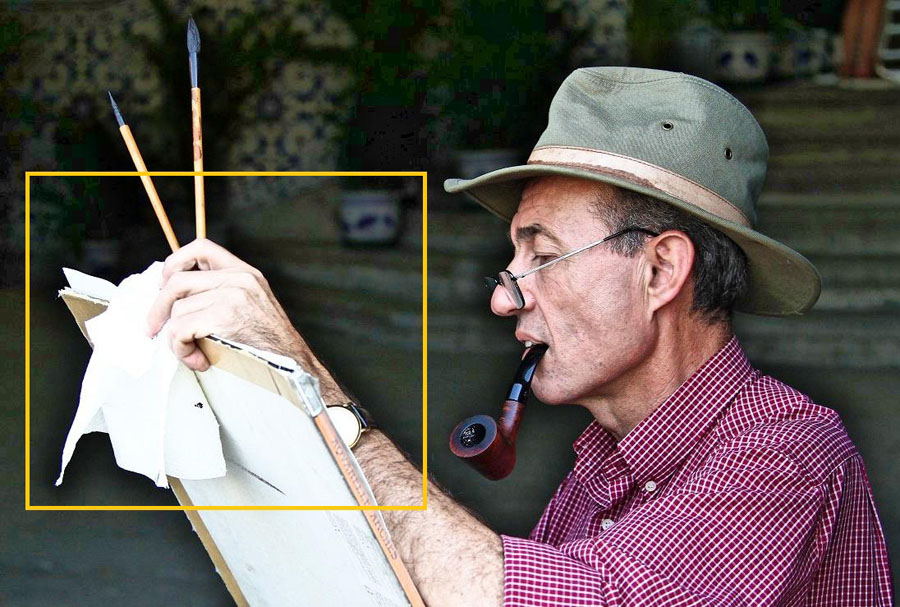}}		
	\hfil                                     
	\subfloat[ALSM]{\includegraphics[width=25.5mm]{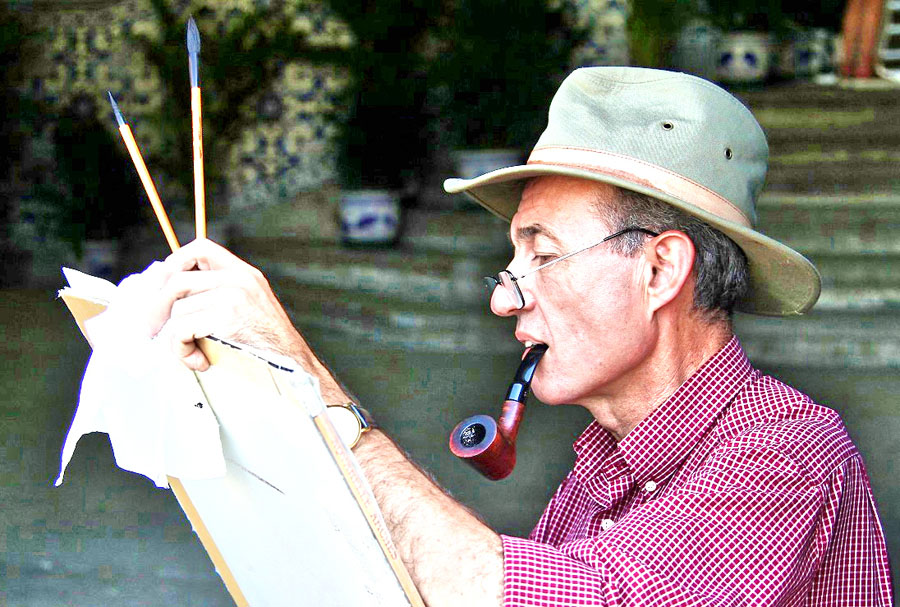}}
	\hfil	                                  
	\subfloat[LIME]{\includegraphics[width=25.5mm]{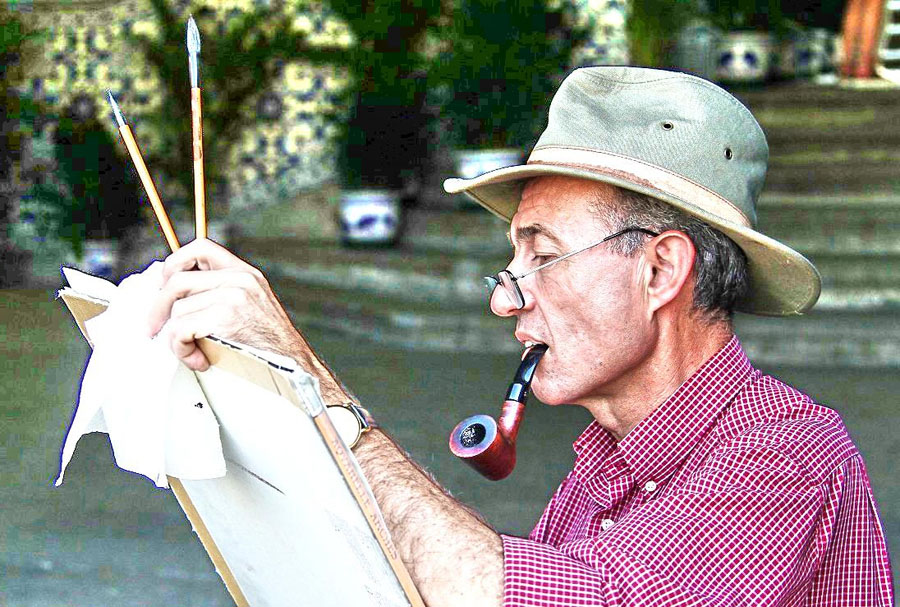}}
	\hfil                                     
	\subfloat[MLLP]{\includegraphics[width=25.5mm]{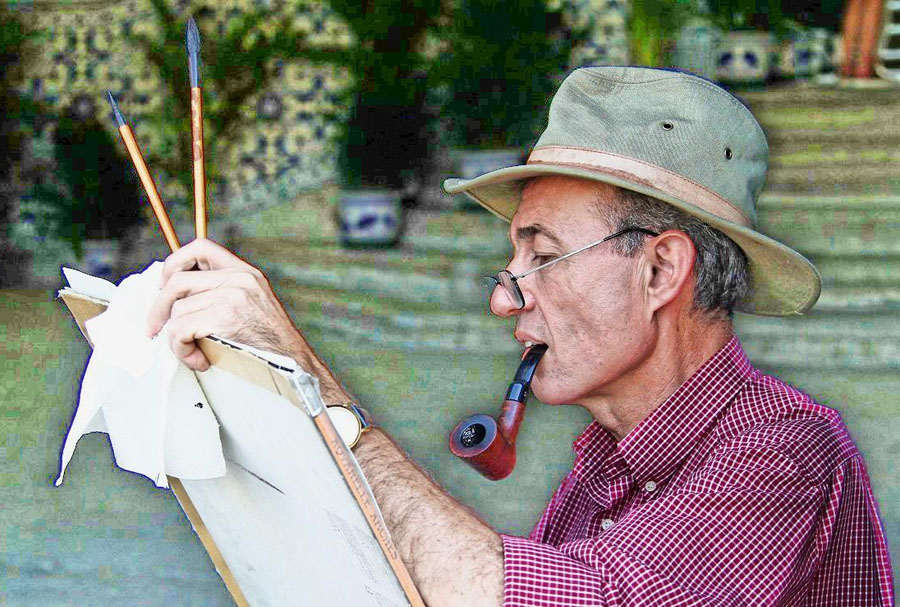}}
	\hfil	                                  
    \subfloat[PnpRtx]{\includegraphics[width=25.5mm]{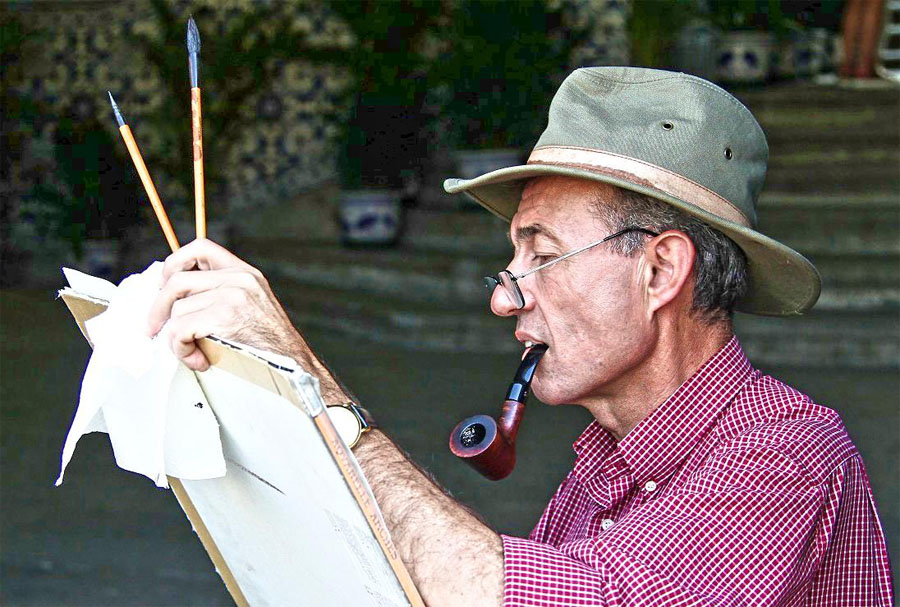}}
	\hfil	                                  
	\subfloat[NRMOE]{\includegraphics[width=25.5mm]{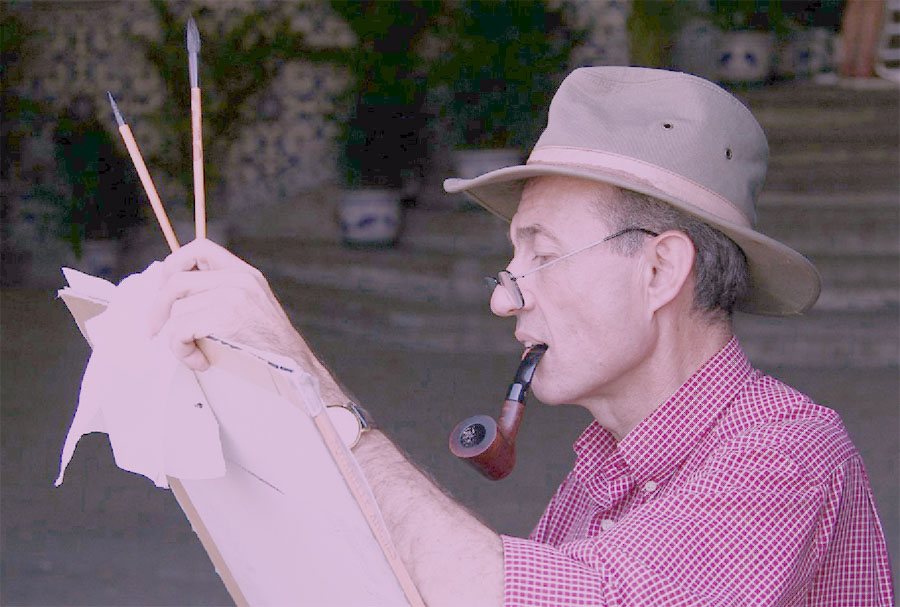}}		
	
	\vspace{-2.5mm}
	
	\centering                                
	\subfloat[PLME]{\includegraphics[width=25.5mm]{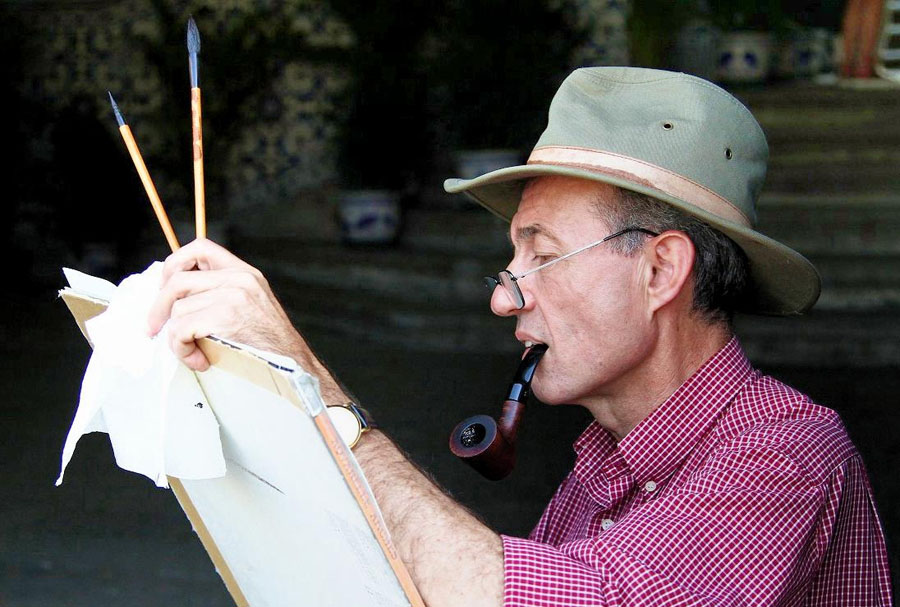}}
	\hfil                                     
	\subfloat[EGAN]{\includegraphics[width=25.5mm]{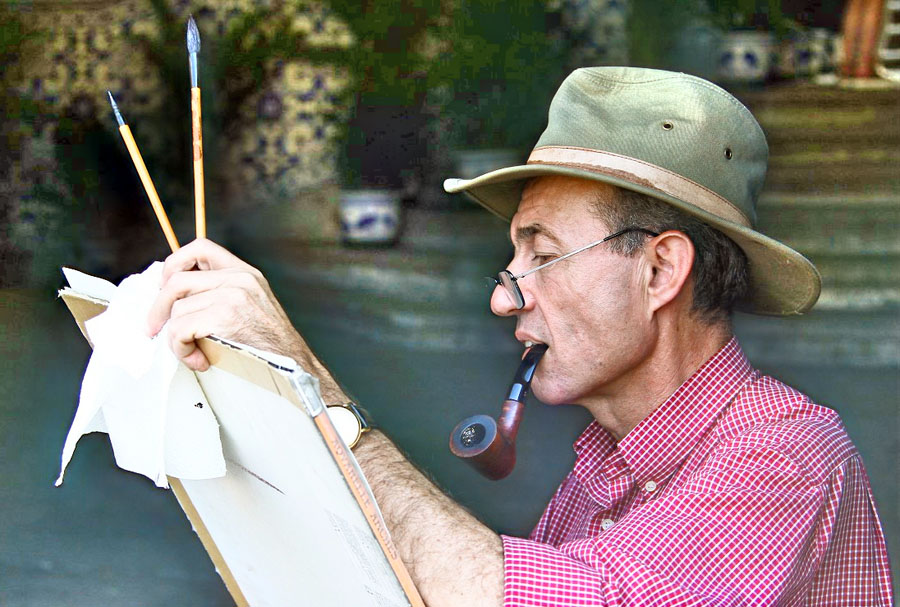}}	
	\hfil                                     
	\subfloat[KIND++]{\includegraphics[width=25.5mm]{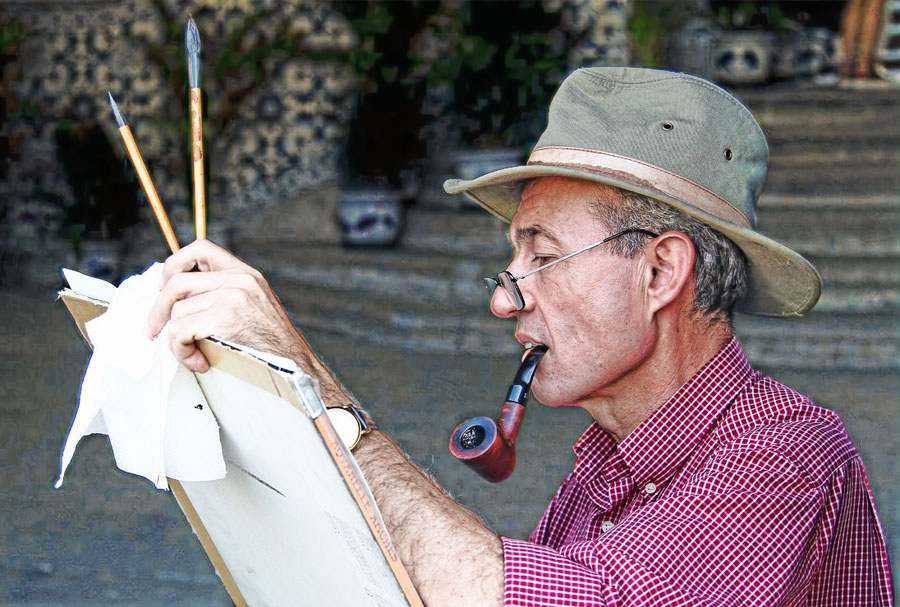}}
	\hfil	                                  
	\subfloat[ZeroDCE]{\includegraphics[width=25.5mm]{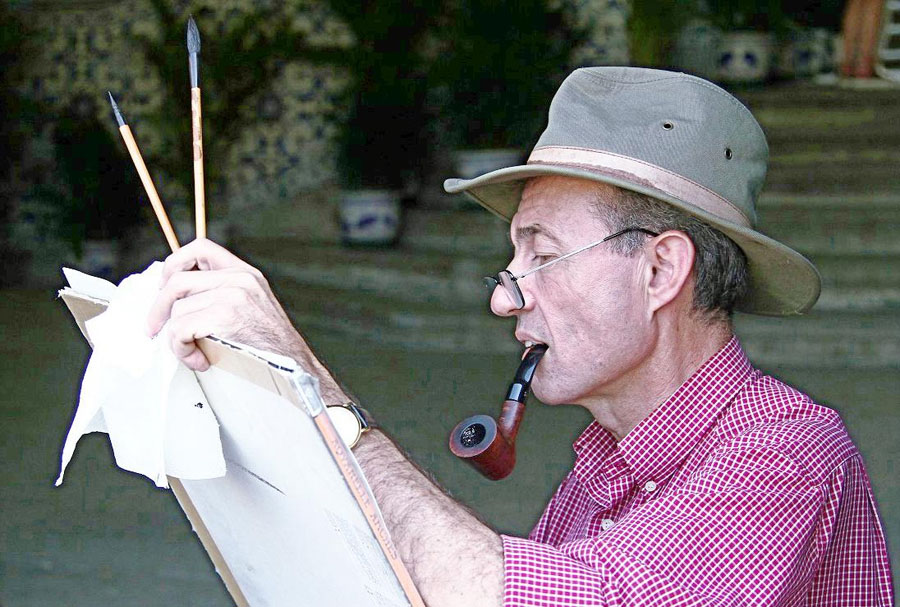}}		
	\hfil                                     
	\subfloat[SCL-LLE]{\includegraphics[width=25.5mm]{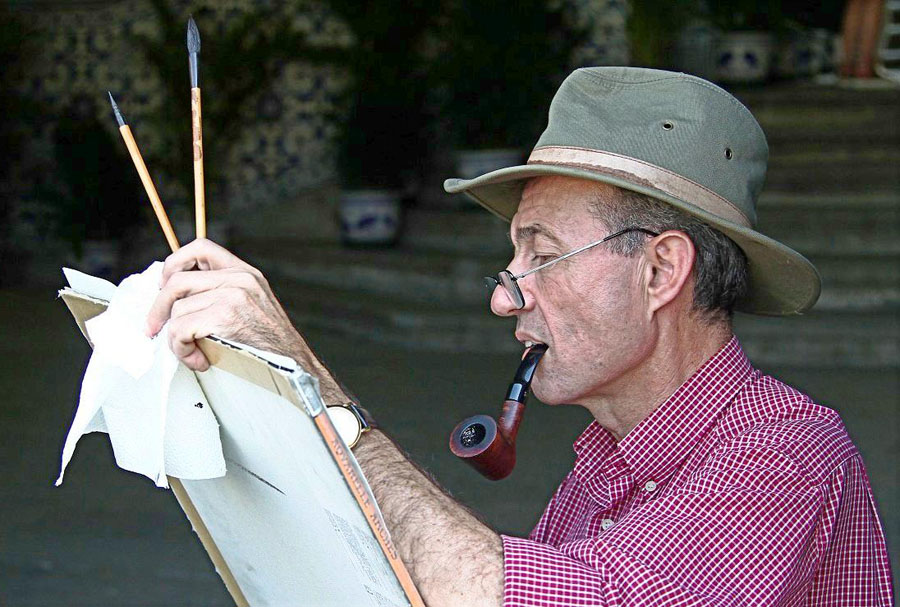}}
	\hfil	                                  
	\subfloat[RtxDIP]{\includegraphics[width=25.5mm]{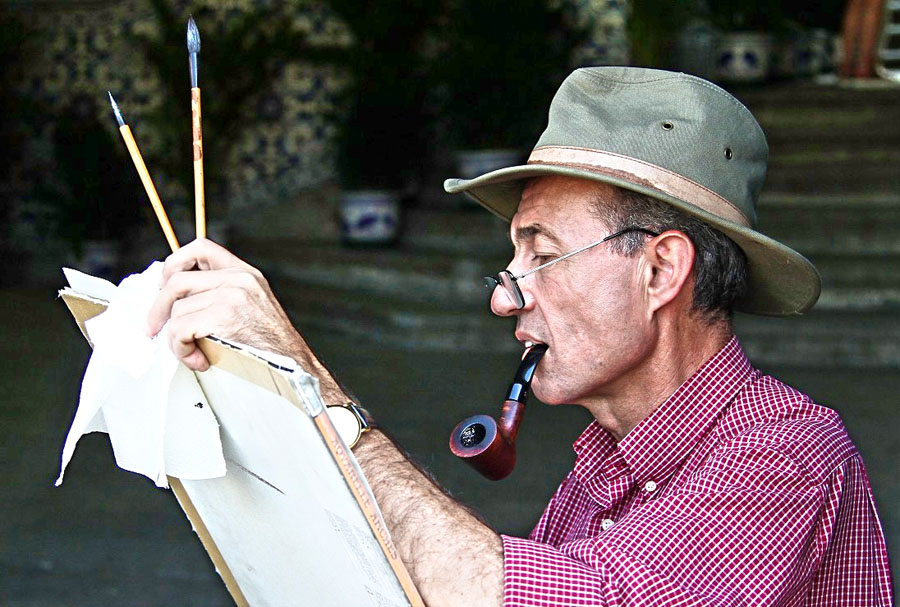}}
	\hfil	                                  
	\subfloat[Proposed]{\includegraphics[width=25.5mm]{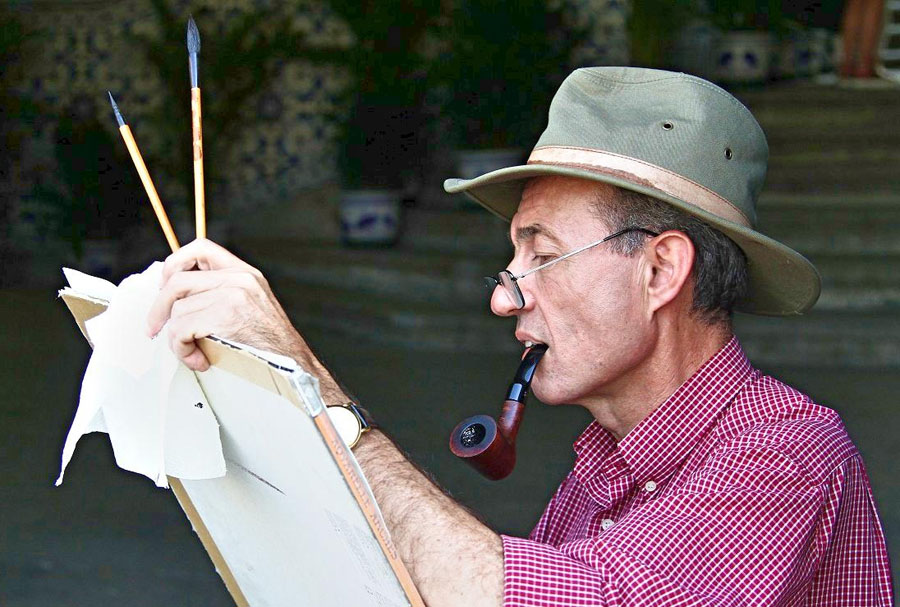}}
	
	\caption{Comparison 5. DFE produces noticeable halos around abrupt bright edges, see the marked region. PnpRetinex, PLME, SCL-LLE, RetinexDIP, and the proposed method produce more natural-looking results than other methods. }
	\label{Comparison_artist}
	
	\vspace{-2mm}
\end{figure*}

\begin{figure*}[h]
\vspace{-4mm}		
	\centering
	\subfloat[Input]{\includegraphics[width=25.5mm]{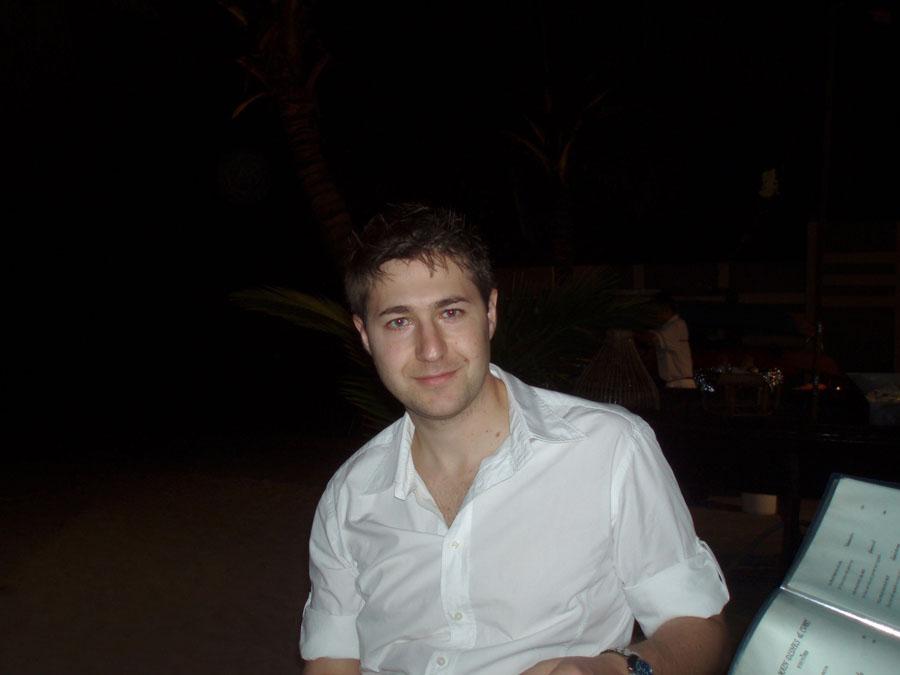}}		
	\hfil
	\subfloat[DFE]{\includegraphics[width=25.5mm]{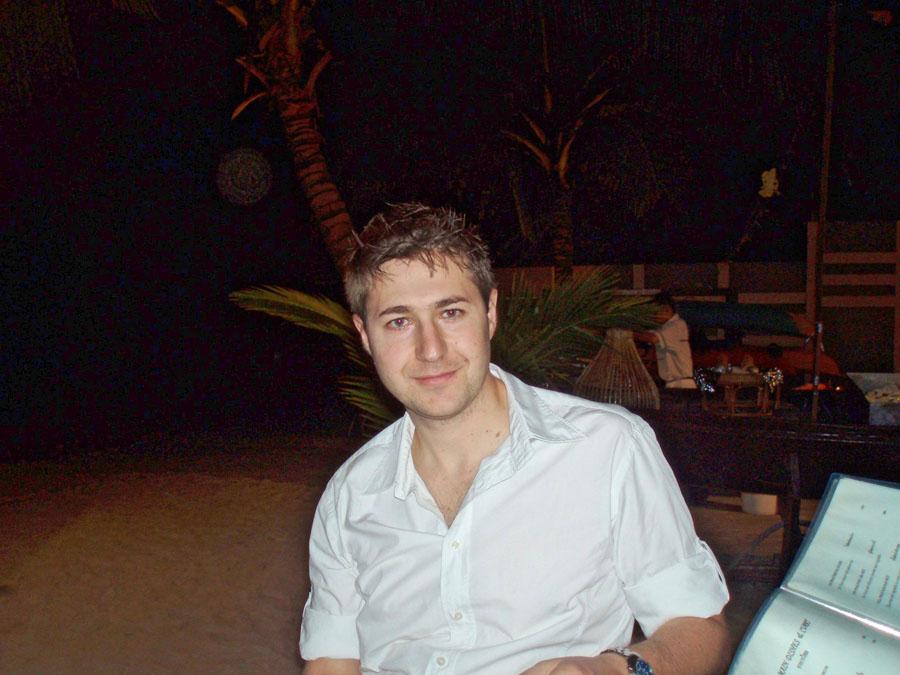}}		
	\hfil
	\subfloat[ALSM]{\includegraphics[width=25.5mm]{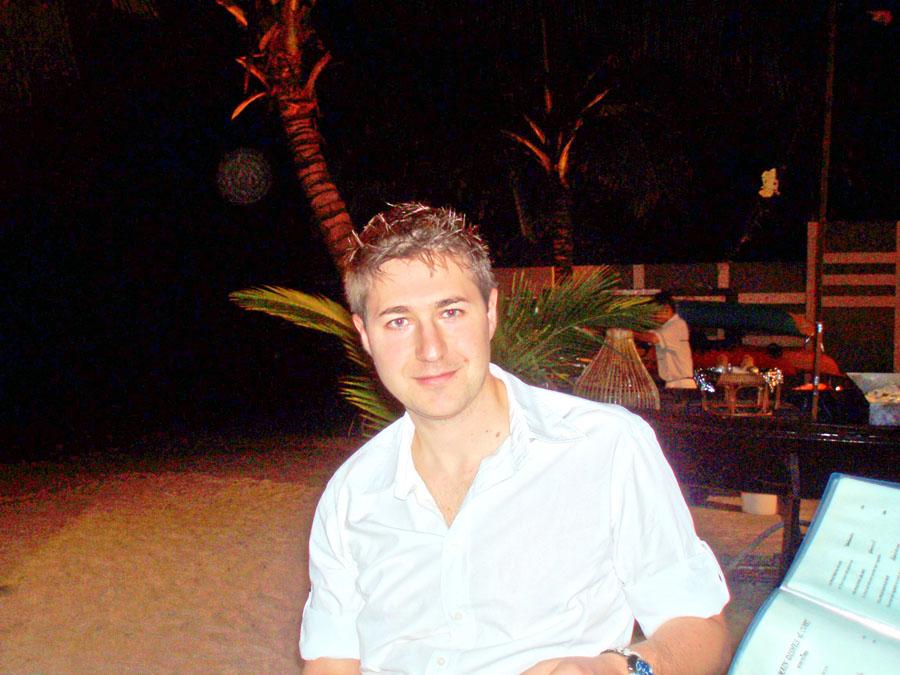}}
	\hfil	
	\subfloat[LIME]{\includegraphics[width=25.5mm]{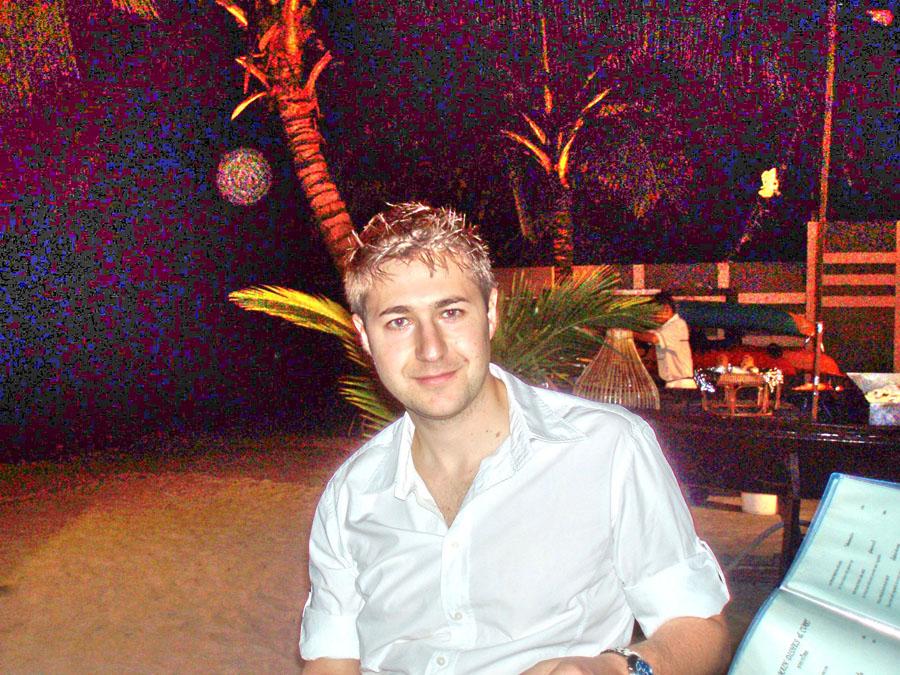}}
	\hfil
	\subfloat[MLLP]{\includegraphics[width=25.5mm]{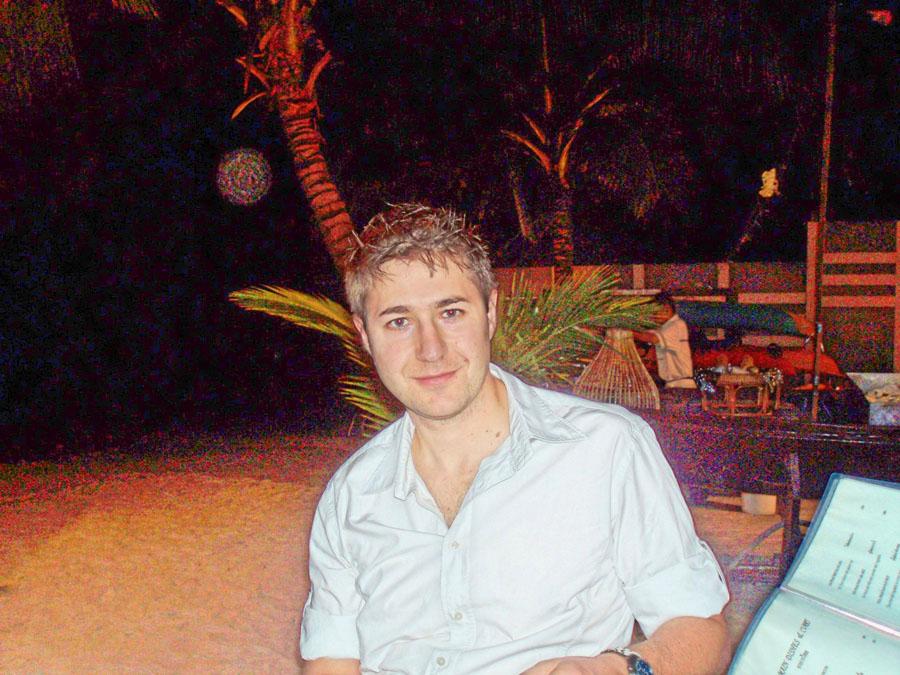}}
	\hfil	
    \subfloat[PnpRtx]{\includegraphics[width=25.5mm]{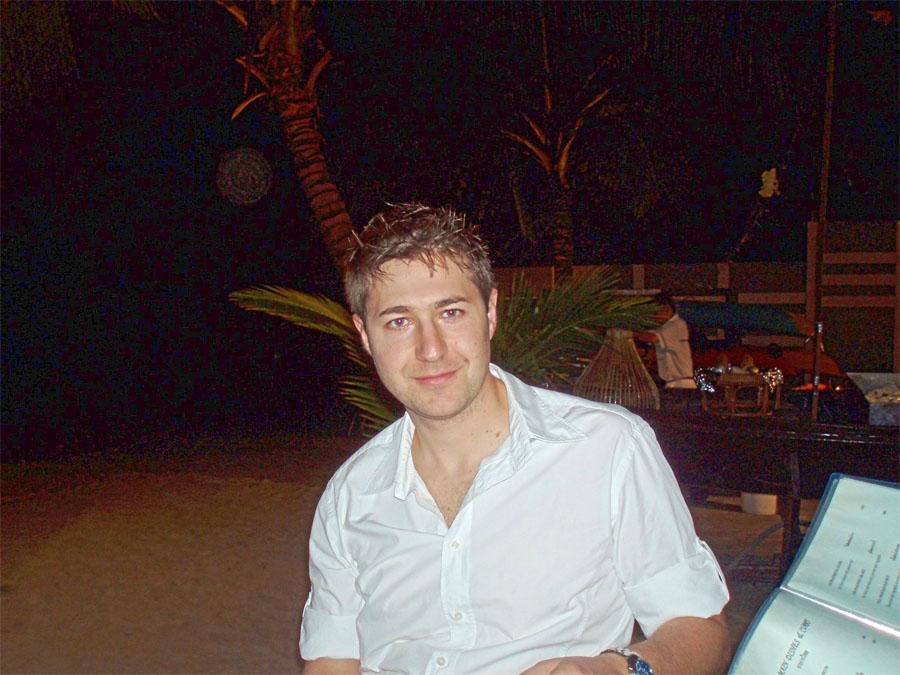}}
	\hfil	
	\subfloat[NRMOE]{\includegraphics[width=25.5mm]{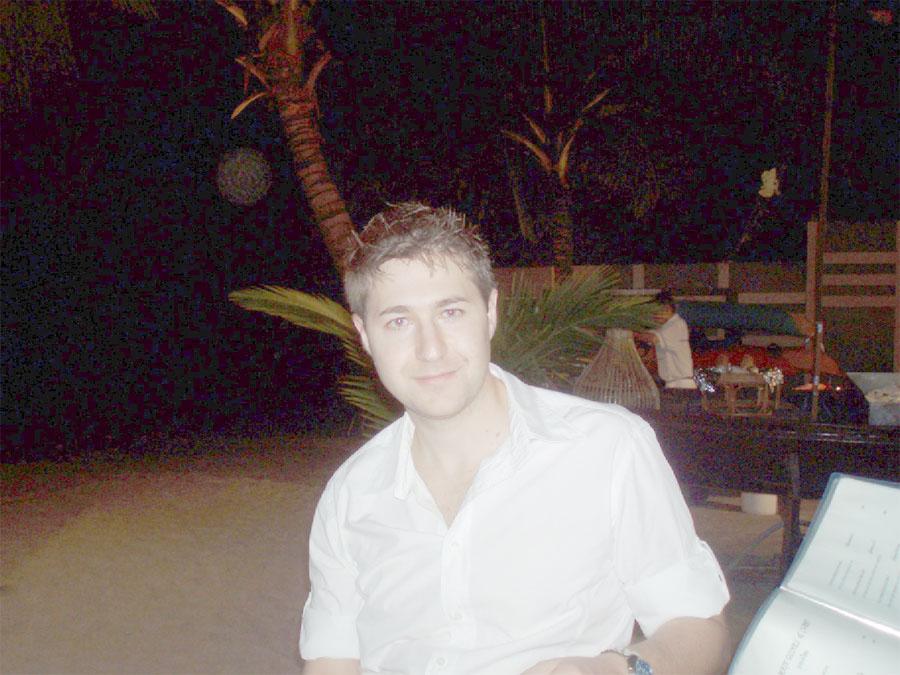}}		
	\vspace{-2.5mm}		
	
	\centering
	\subfloat[PLME]{\includegraphics[width=25.5mm]{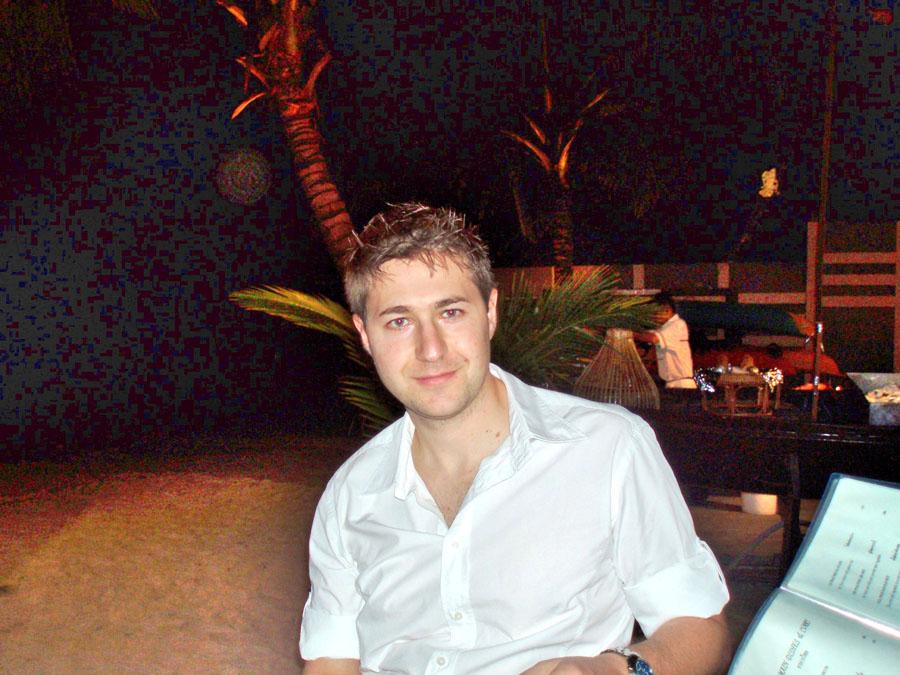}}
	\hfil
	\subfloat[EGAN]{\includegraphics[width=25.5mm]{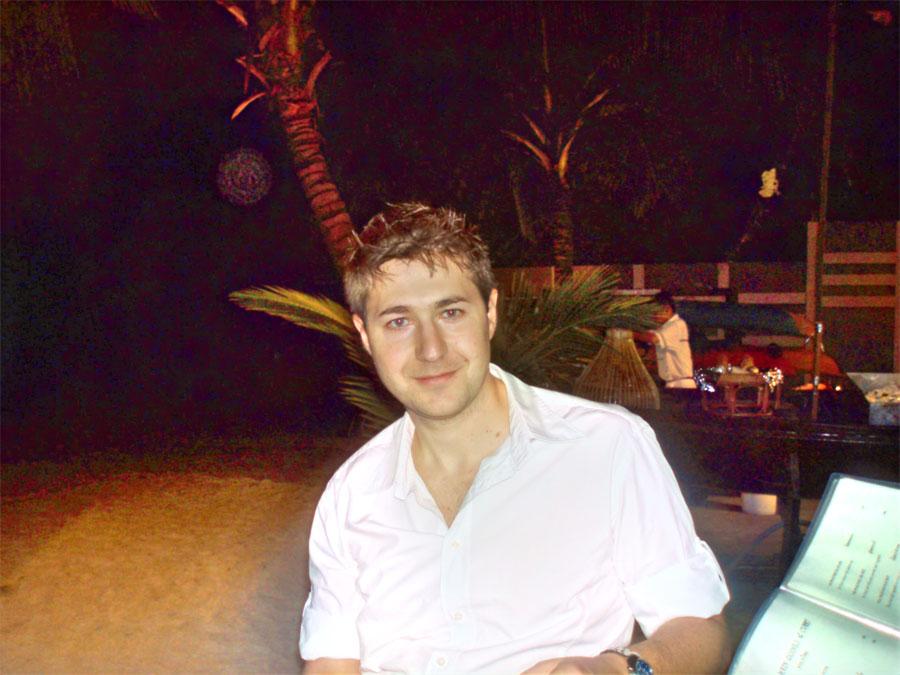}}
	\hfil		
	\subfloat[KIND++]{\includegraphics[width=25.5mm]{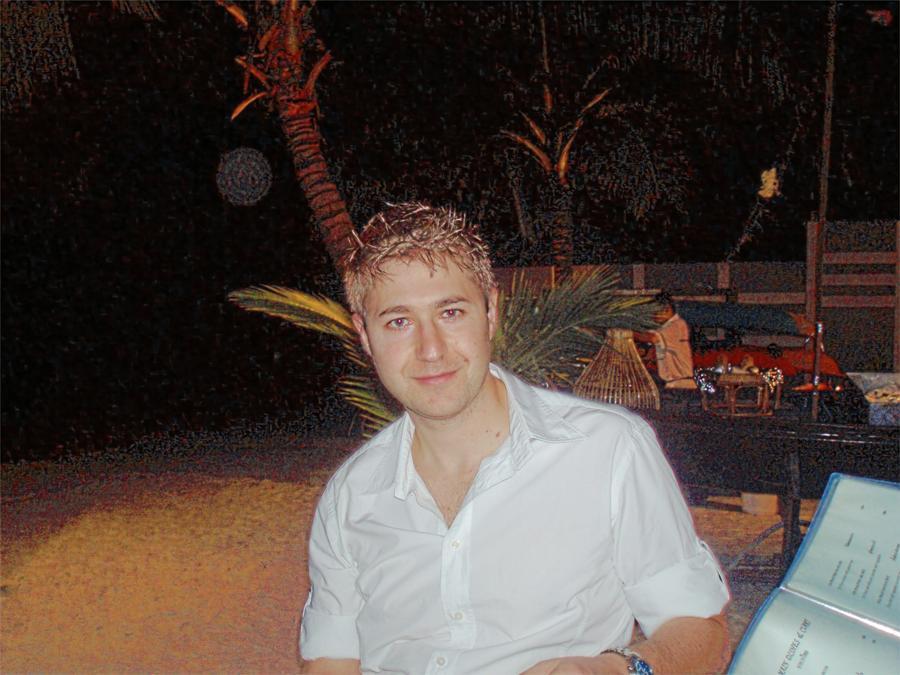}}
	\hfil	
	\subfloat[ZeroDCE]{\includegraphics[width=25.5mm]{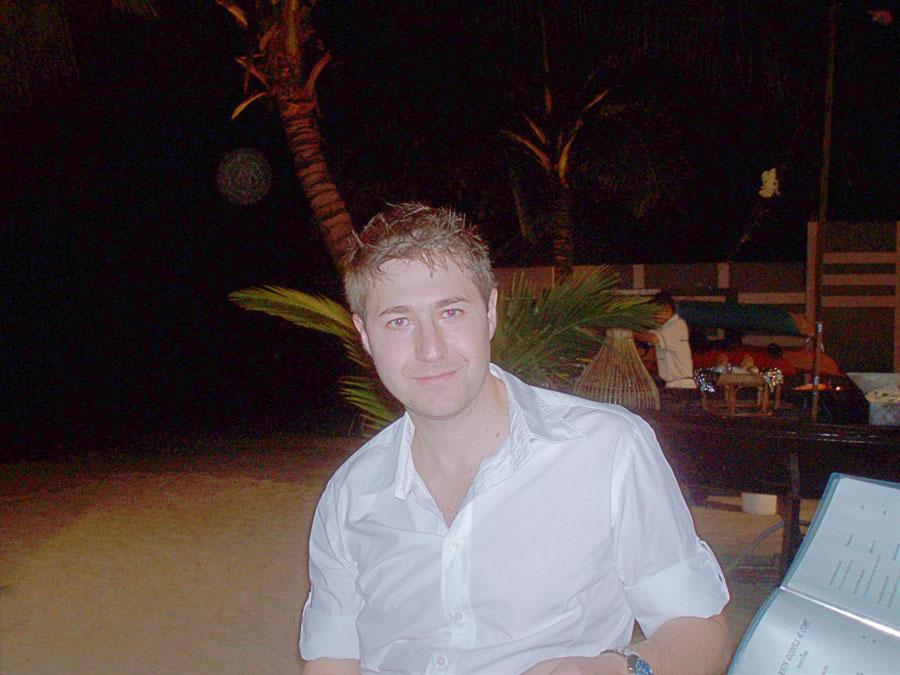}}		
	\hfil	
	\subfloat[SCL-LLE]{\includegraphics[width=25.5mm]{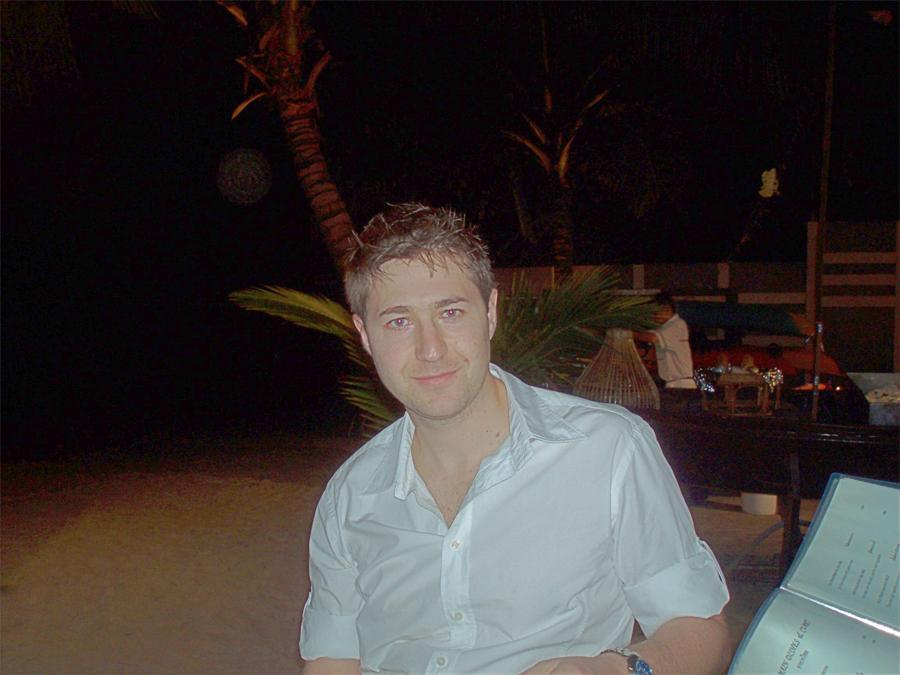}}
	\hfil
	\subfloat[RtxDIP]{\includegraphics[width=25.5mm]{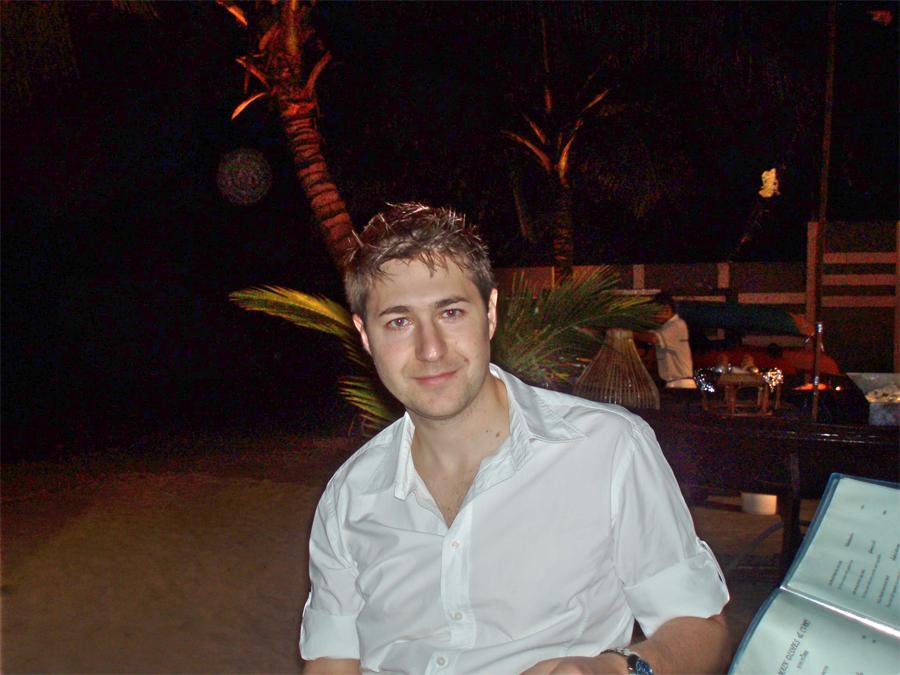}}
	\hfil	
	\subfloat[Proposed]{\includegraphics[width=25.5mm]{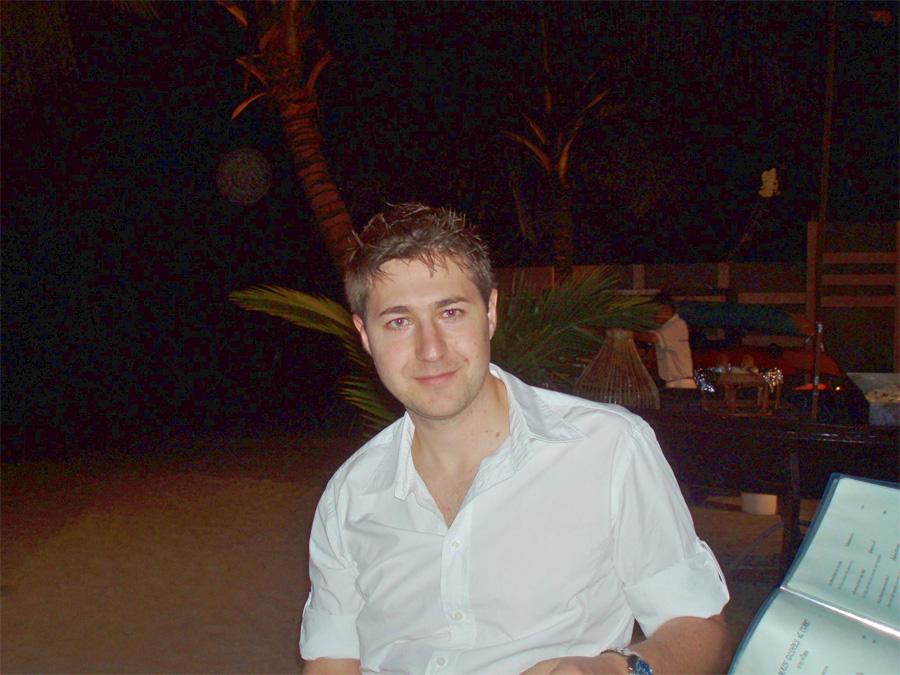}}
	
	\caption{Comparison 6. The results of ALSM, LIME, and PLME suffer from varying degrees of overexposure. ZeroDCE,SCL-LLE, and NRMOE produce rather pale images. The results of MLLP, EnlightenGAN and KIND++ contain noticeable artifacts. DFE, PnpRetinex, RetinexDIP, and the proposed method yield similar results. }
	\label{Comparison_vvman}	
	\vspace{-2mm}
\end{figure*}

Figs. \ref{Comparison_night} to \ref{Comparison_vvman} are representative comparisons. Referring to these results, we make the following comments: 

\begin {enumerate}
\item 
Both ALSM and LIME can significantly improve contrast, but tend to over-enhance images, resulting in unnatural appearance, see Figs.\ref{Comparison_dcim02}(c), and (d) to \ref{Comparison_vvman}(c), and (d).

\item 
MLLP is capable of a good degree of highlighting details, but may produce noticeable plaque artifacts (see Figs.\ref{Comparison_night}(e) and \ref{Comparison_tower}(e).

\item
NRMOE tends to introduce color distortion in the results, as shown in Figs.\ref{Comparison_lady}(g), \ref{Comparison_artist}(g), and \ref{Comparison_vvman}(g).
\item
DFE, PnpRetinex, and PLME are all effective in enhancing images with vivid color. Comparatively, DFE and PnpRetinex are better at bringing out details, as demonstrated in Figs.\ref{Comparison_tower}(b), (f), and (h), and Figs.\ref{Comparison_artist}(b), (f), and (g), respectively. However, DFE may give rise to prominent halos around abrupt bright edges (see the marked region in Fig.\ref{Comparison_artist}(b)). PnpRetinex tends to generate slightly over-sharpened edges, as seen in Figs.\ref{Comparison_dcim02}(f) and \ref{Comparison_lady}(f).

\item
All deep-learning based methods can significantly enhance contrast in shadow areas. However, the results frequently suffer from unstable image quality. EnlightenGAN, KIND++, and RetinexDIP easily generate unwanted artifacts when highlighting shadow areas (see Figs.\ref{Comparison_night}(i), (j), and (m) to \ref{Comparison_tower}(i), (j), and (m), respectively). Both ZeroDCE and SCL-LLE usually produce results with reduced color saturation. The reason may be that their limited training datasets are incapable of covering the various natural scenes.
\end {enumerate}

Compared to these enhancement methods, the proposed method offers a better balance between improving contrast and preserving naturalness without introducing undesired artifacts or color shifts.

\vspace{2mm}
\noindent\textbf{2) User study}
\vspace{2mm}

We conduct a user study with 13 observers to quantify the subjective assessment of the compared methods on Testset-1. The observers are trained from three aspects: 1) no severe artifacts such as over-, under-exposed regions, and halos are introduced, 2) the color rendition of the scene is perceptually natural, and 3) enhanced details are visually-pleasing \cite{WangNaturalness}.

\begin{table}[t]
\caption{Comparison of MOS on Testset-1. The best score is in red and the second best one is in blue.}
\vspace{-2mm}
\centering
\begin{tabular}{@{}lllllll@{}}
\toprule
         & \multicolumn{6}{c}{Testset-1}                                                                                                                                                     \\ \cmidrule(l){2-7}
Method   & NPE                         & MLLP                        & VV                          & DCIM                        & MEF                         & Average                        \\ \midrule
DFE      & 4.36                        & 4.23                        & 4.49                        & \textcolor{blue}{4.59}      & \textcolor{blue}{4.72}      & \textcolor{blue}{4.48}      \\ \midrule
LIME     & 4.34                        & 4.07                        & 4.20                        & 4.12                        & 4.29                        & 4.20                        \\ \midrule
ALSM     & 4.33                        & 3.89                        & 4.16                        & 4.03                        & 4.31                        & 4.14                        \\ \midrule
MLLP     & 4.29                        & 4.01                        & 3.83                        & 3.94                        & 4.06                        & 4.03                        \\ \midrule
NRMOE    & 3.63                        & 3.54                        & 2.67                        & 3.23                        & 3.53                        & 3.32                        \\ \midrule
PLME     & 4.20                        & 4.13                        & 4.49                        & 4.48                        & 4.67                        & 4.39                        \\ \midrule
PnpRetinex     & \textcolor{red}{4.45}       & \textcolor{blue}{4.28}      & \textcolor{blue}{4.54}      & \textcolor{red}{4.66}       & \textcolor{red}{4.79}       & \textcolor{red}{4.54}       \\ \midrule
EnlightenGAN     & 3.91                        & 3.92                        & 3.45                        & 4.15                        & 4.47                        & 3.98                        \\ \midrule
KIND++   & 3.57                        & 3.62                        & 3.31                        & 3.68                        & 4.04                        & 3.64                        \\ \midrule
ZeroDCE     & 4.29                        & 4.11                        & 4.21                        & 4.43                        & 4.71                        & 4.35                        \\ \midrule
SCL-LLE  & 4.13                        & 4.12                        & 4.32                        & 4.40                        & 4.58                        & 4.31                        \\ \midrule
RetinexDIP   & 3.74                        & 3.87                        & 3.75                        & 3.85                        & 4.22                        & 3.89                        \\ \midrule
Proposed & \textcolor{blue}{4.40}      & \textcolor{red}{4.31}       & \textcolor{red}{4.56}       & \textcolor{red}{4.66}       & \textcolor{red}{4.79}       & \textcolor{red}{4.54}       \\ \bottomrule
\end{tabular}
\label{mos}
\vspace{-6mm}
\end{table}

The user study is designed as follows: 1) an original image and one of its randomly ordered enhanced results are simultaneously displayed on the screen. The original is used as the reference in each trial. 2) The observers evaluate the enhanced images based on their understanding of the scene. The observers can switch enhanced images back and forth in each trial to make the final rating \cite{cao2021debiased}. 3) The quality score ranges from 1 to 5 (worst to best quality) with a step-size of one. Results achieved by different methods may have the same score, which means that the visual differences between the results are not enough for the observer to make a preferred rating. These scores are then averaged to yield the Mean Opinion Score (MOS) value for each method.

The MOS values for each image set are listed in Table \ref{mos}. The proposed method obtains the second highest score in NPE dataset, and the highest scores in the other four datasets. It also achieves the highest average MOS value aross all test images, tied for first place with PnpRetinex. This small-scale user study provides additional support that the proposed method outperforms other compared methods in terms of visual quality.

\vspace{2mm}
\noindent\textbf{3) Quantitative assessment}

Table \ref{quancomp_nr} shows the evaluation results on Testset-1. The proposed method obtains the lowest BIQI values on all datasets. Therefore, our method yields results with better naturalness perception than other compared methods. In terms of the NIQE metric, the proposed method achieves the lowest average values on NPE, MLLP, and DCIM, and the second lowest values on VV and MEF, slightly higher than DFE and EnlightenGAN, respectively. This means that our method can compete with state-of-the-art methods in producing results similar to natural images. Our method ranks behind PnpRetinex on NPE, MLLP, DCIM, and VV in terms of average NFERM values, and behind PLME and PnpRetinex on MEF. This indicates that our method is comparable to the state-of-the-art methods in terms of image quality based on the HVS-aware features of natural scenes.

\begin{table*}[]
\caption{ Comparison of NIQE(G), BIQI(B) and NFERM(F) values for Testset-1. The scores for the first, second, and third places are marked in red, blue, and green, respectively.}
\vspace{-2mm}
\centering
\begin{tabular}{@{}llllllllllllllll@{}}
\toprule
             & \multicolumn{3}{c}{NPE}                                                                   & \multicolumn{3}{c}{MLLP}                                                                  & \multicolumn{3}{c}{DCIM}                                                                  & \multicolumn{3}{c}{VV}                                                                    & \multicolumn{3}{c}{MEF}                                                                   \\ \cmidrule(l){2-16} 
Method       & \multicolumn{1}{c}{N}       & \multicolumn{1}{c}{B}        & \multicolumn{1}{c}{F}        & \multicolumn{1}{c}{N}       & \multicolumn{1}{c}{B}        & \multicolumn{1}{c}{F}        & \multicolumn{1}{c}{N}       & \multicolumn{1}{c}{B}        & \multicolumn{1}{c}{F}        & \multicolumn{1}{c}{N}       & \multicolumn{1}{c}{B}        & \multicolumn{1}{c}{F}        & \multicolumn{1}{c}{N}       & \multicolumn{1}{c}{B}        & \multicolumn{1}{c}{F}        \\ \midrule
DFE          &\textcolor{blue}{2.84}       &\textcolor{blue}{30.13}       &\textcolor{green}{22.99}      &\textcolor{green}{2.82}      &\textcolor{green}{29.88}      &\textcolor{green}{15.97}      &\textcolor{blue}{2.81}       &\textcolor{green}{27.88}      &\textcolor{green}{14.84}      & \textcolor{red}{2.38}       &\textcolor{green}{30.82}      & 35.78                        &\textcolor{green}{2.70}      & 28.75                        & \textcolor{red}{17.14}       \\
LIME         & 3.16                        & 38.78                        & 23.21                        & 3.28                        & 40.87                        & 18.78                        & 3.02                        & 31.62                        & 20.38                        &\textcolor{green}{2.49}      & 32.89                        & 32.47                        & 3.52                        & 35.09                        & 24.05                        \\
MLLP         & 2.96                        & 34.07                        & 23.01                        & 2.93                        & 32.35                        & 17.79                        & 3.21                        & 28.44                        & 19.32                        & 2.56                        & 35.69                        & 40.03                        & 2.76                        & 29.87                        & 19.08                        \\
ALSM         & 3.04                        & 36.37                        & 25.66                        & 3.09                        & 35.59                        & 16.67                        & 2.95                        & 29.52                        & 20.05                        & 2.50                        & 31.47                        & 34.90                        & 2.71                        & 30.50                        & 19.07                        \\
NRMOE        & 3.51                        & 36.19                        & 25.71                        & 3.35                        & 34.45                        & 20.29                        & 3.08                        & 29.08                        & 21.81                        & 2.66                        & 35.2                         & 39.76                        & 3.17                        & 29.79                        & 20.22                        \\
PLME         & 2.91                        & 32.28                        & 23.79                        & \textcolor{blue}{2.78}      &\textcolor{blue}{29.85}       & 16.53                        & 2.93                        & 28.32                        & \textcolor{red}{12.01}       & 2.51                        & 31.48                        & 34.61                        & 2.93                        &\textcolor{blue}{28.64}       &\textcolor{blue}{17.47}       \\
PnpRetinex   &\textcolor{green}{2.85}      &\textcolor{green}{30.86}      & \textcolor{red}{21.50}       & 2.97                        & 34.52                        & \textcolor{red}{15.01}       & 2.85                        &\textcolor{blue}{27.85}       & 15.07                        & 2.51                        &\textcolor{blue}{30.97}       & \textcolor{red}{22.47}       & 2.78                        & 32.07                        & 18.23                        \\
RetinexDIP   & 3.31                        & 35.09                        & 25.21                        & 3.26                        & 29.93                        & 18.29                        & 3.35                        & 30.52                        & 16.99                        & 3.04                        & 31.08                        & 32.89                        & 3.06                        & 31.09                        & 24.10                        \\
EnlightenGAN & 3.35                        & 31.02                        & 28.17                        & 2.98                        & 26.10                        & 17.51                        & 3.56                        & 32.41                        & 18.84                        & 3.38                        & 37.65                        & 37.46                        & \textcolor{red}{2.59}       & 30.92                        & 18.95                        \\
KIND++       & 3.11                        & 33.53                        & 28.46                        & 3.32                        & 34.01                        & 17.16                        &\textcolor{green}{2.83}      & 31.50                        & 19.08                        & 3.25                        & 32.66                        & 44.07                        & 3.32                        & 34.52                        & 20.65                        \\
SCL-LLE      & 3.03                        & 31.83                        & 24.36                        & 2.97                        & 30.77                        & 16.69                        & 3.58                        & 28.08                        & 17.86                        & 2.61                        & 31.28                        & 37.09                        & 2.82                        &\textcolor{green}{28.73}      & 19.30                        \\
ZeroDCE     & 3.05                        & 32.54                        & 24.46                        & 3.14                        & 32.13                        & 17.25                        & 3.66                        & 28.25                        & 14.90                        & 2.68                        & 32.81                        &\textcolor{green}{31.39}      & 3.11                        & 29.39                        & 19.18                        \\
Proposed     & \textcolor{red}{2.79}       & \textcolor{red}{30.03}       &\textcolor{blue}{22.04}       & \textcolor{red}{2.77}      & \textcolor{red}{28.68}       &\textcolor{blue}{15.64}       &\textcolor{red}{2.66}        & \textcolor{red}{27.12}       &\textcolor{blue}{14.49}       &\textcolor{blue}{2.47}       & \textcolor{red}{30.54}       &\textcolor{blue}{27.02}       &\textcolor{blue}{2.65}       & \textcolor{red}{27.85}       &\textcolor{green}{17.93}      \\ \bottomrule
\end{tabular}
\label{quancomp_nr}
\vspace{-2mm}
\end{table*}

\begin{figure}[h]
	\vspace{-4mm}
	\centering
	\subfloat[Input]{\includegraphics[width=25.5mm]{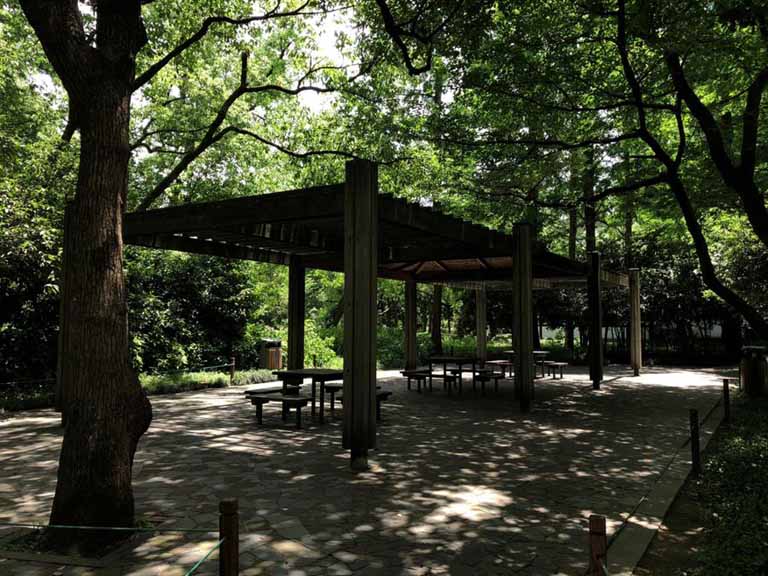}}		
	\hspace{0.05mm}
	\subfloat[Ref.Image]{\includegraphics[width=25.5mm]{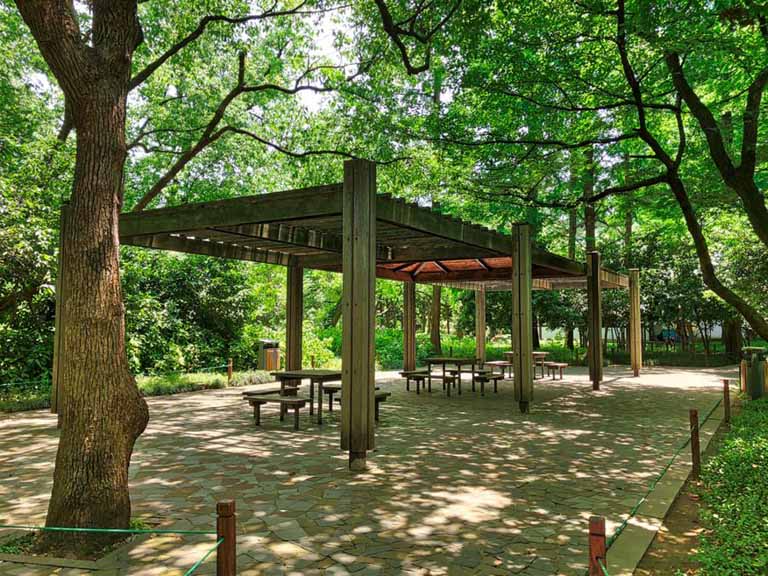}}		
	\hspace{0.05mm}
	\subfloat[DFE]{\includegraphics[width=25.5mm]{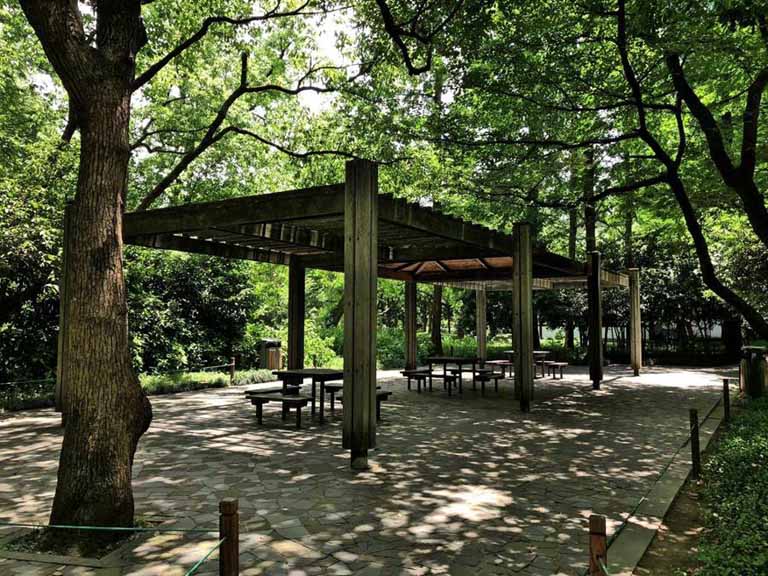}}		
	\hspace{0.05mm}
	\vspace{-2mm}
	
	\centering
	\subfloat[ALSM]{\includegraphics[width=25.5mm]{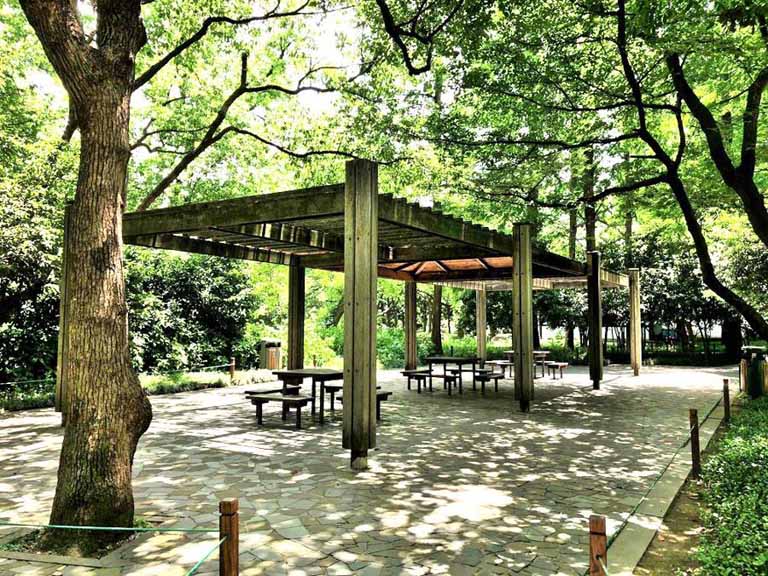}}	
	\hspace{0.05mm}	
	\subfloat[LIME]{\includegraphics[width=25.5mm]{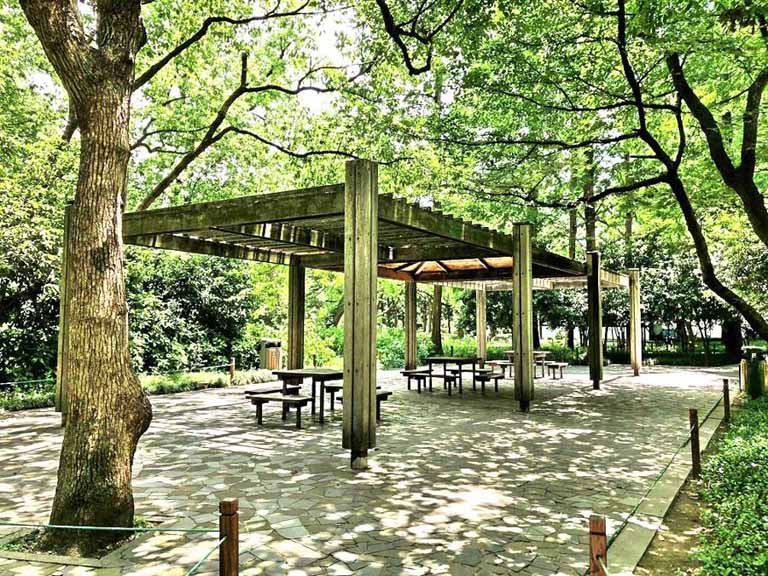}}		
	\hspace{0.05mm}
	\subfloat[MLLP]{\includegraphics[width=25.5mm]{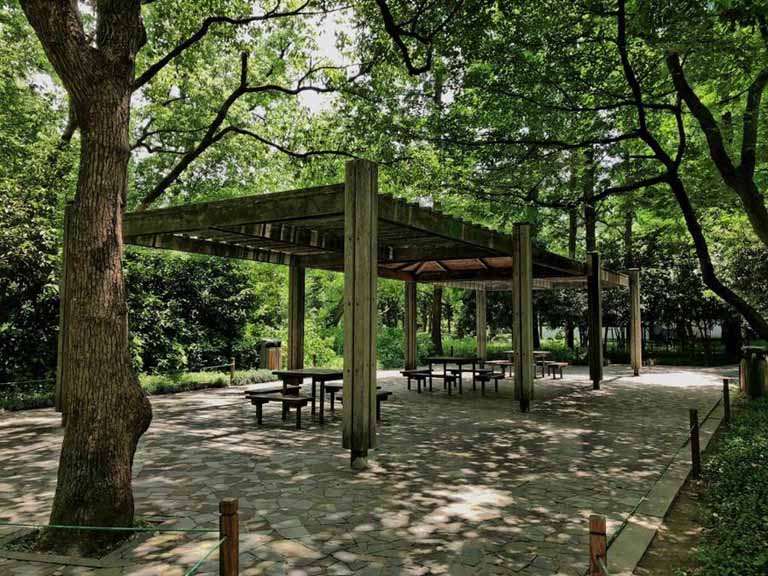}}
	\hspace{0.05mm}	   
    \vspace{-2mm}
    
    \centering    
    \subfloat[PnpRtx]{\includegraphics[width=25.5mm]{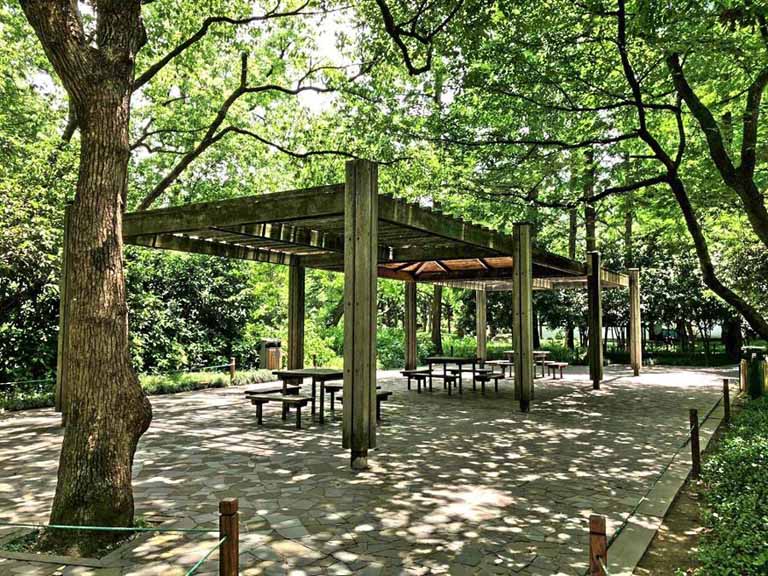}}
	\hspace{0.05mm} 	
	\subfloat[NRMOE]{\includegraphics[width=25.5mm]{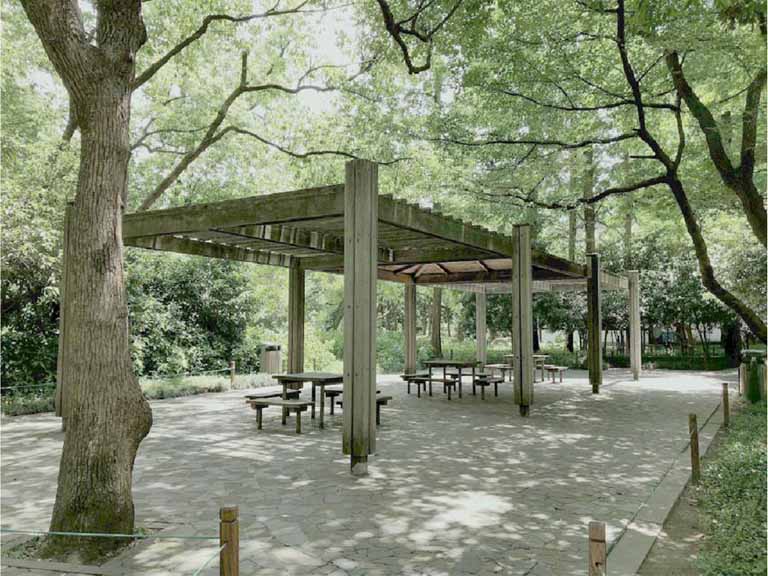}}		
	\hspace{0.05mm}
	\subfloat[PLME]{\includegraphics[width=25.5mm]{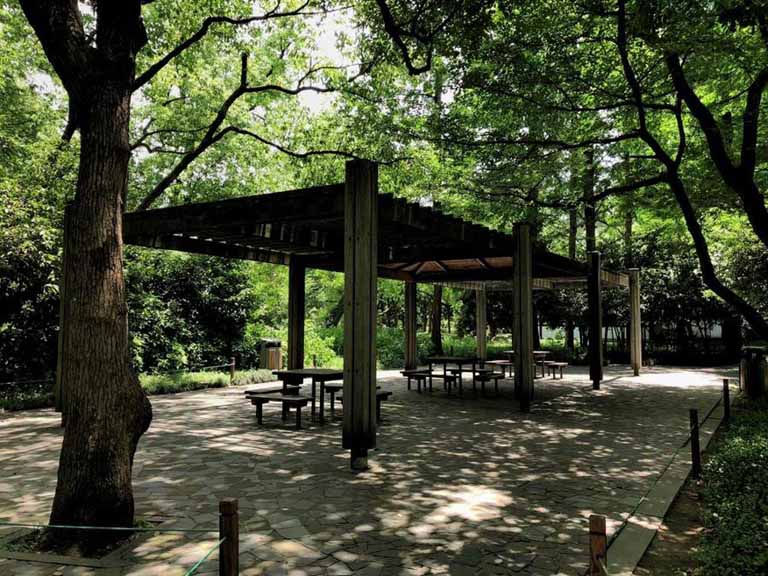}}
	\hspace{0.05mm}	
	\vspace{-2mm}
	
	\centering
	\subfloat[EGAN]{\includegraphics[width=25.5mm]{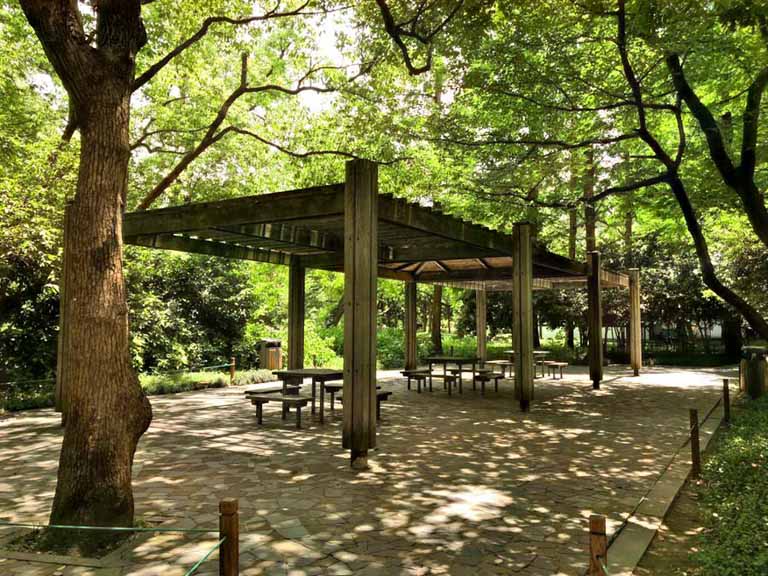}}
	\hspace{0.05mm}
	\subfloat[KIND++]{\includegraphics[width=25.5mm]{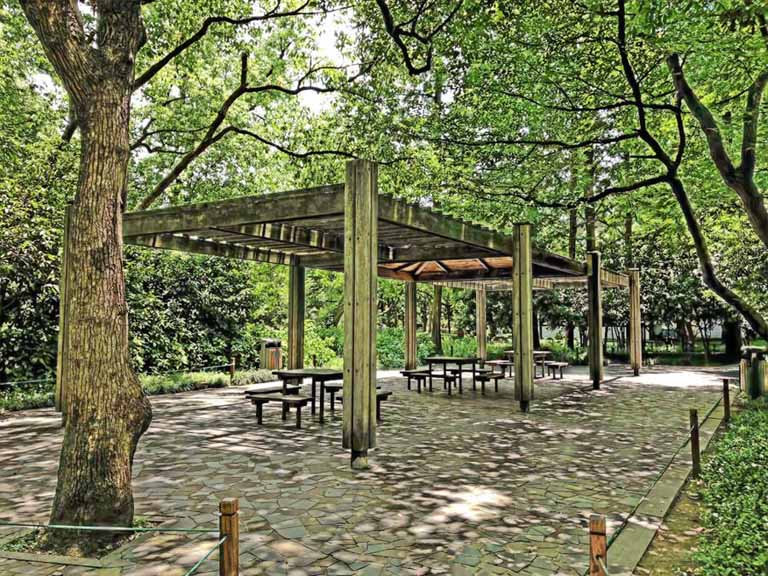}}
	\hspace{0.05mm}	
	\subfloat[ZeroDCE]{\includegraphics[width=25.5mm]{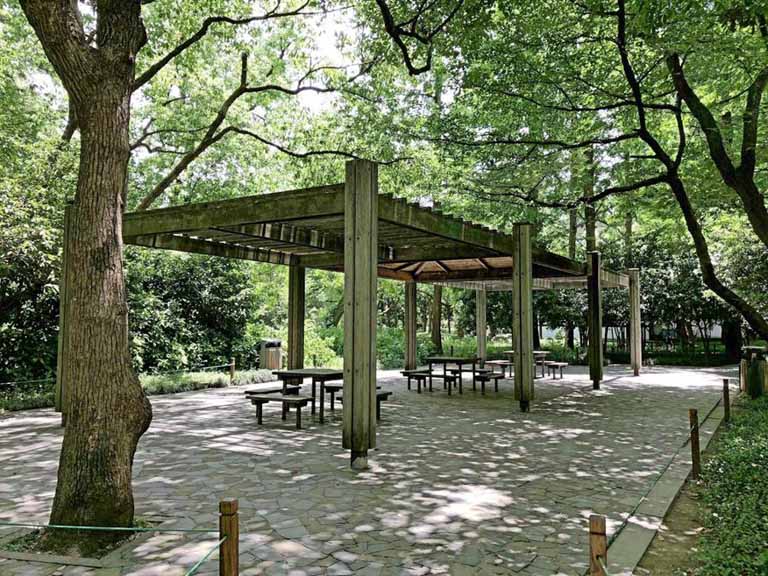}}		
	\vspace{-2mm}
	
	\centering
	\subfloat[SCL-LLE]{\includegraphics[width=25.5mm]{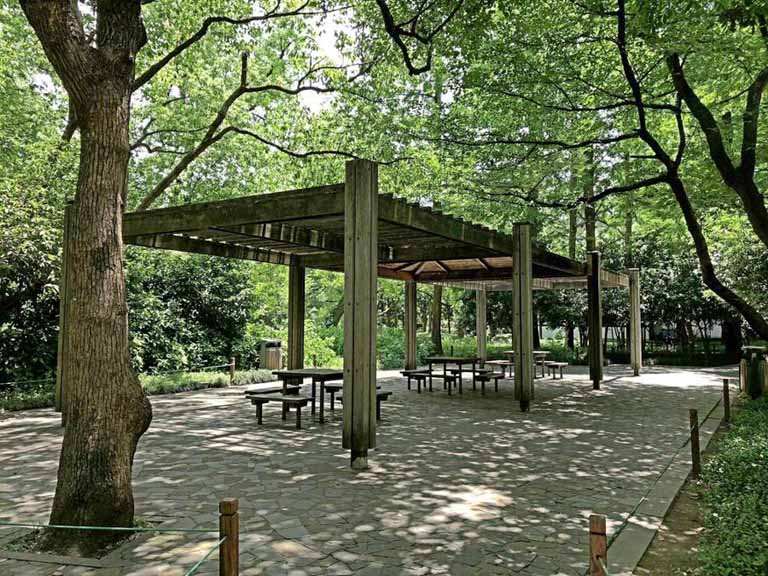}}
	\hspace{0.05mm}	
	\subfloat[RtxDIP]{\includegraphics[width=25.5mm]{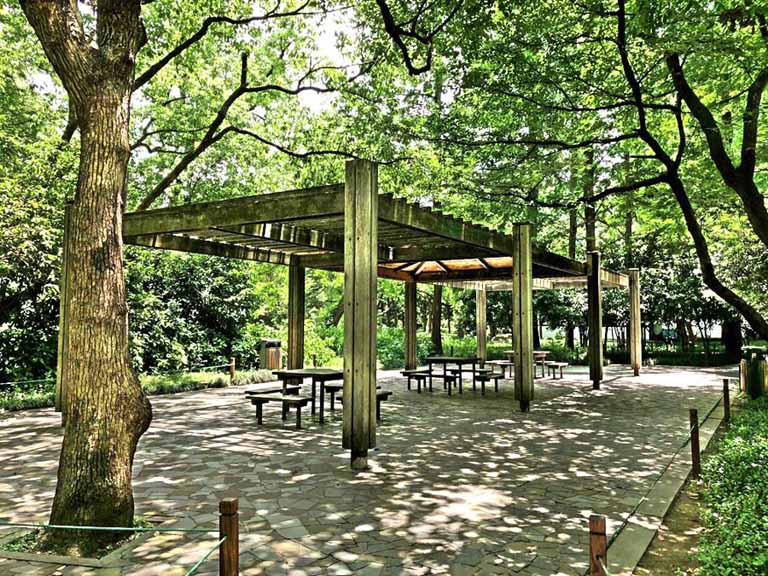}}
	\hspace{0.05mm}
	\subfloat[Proposed]{\includegraphics[width=25.5mm]{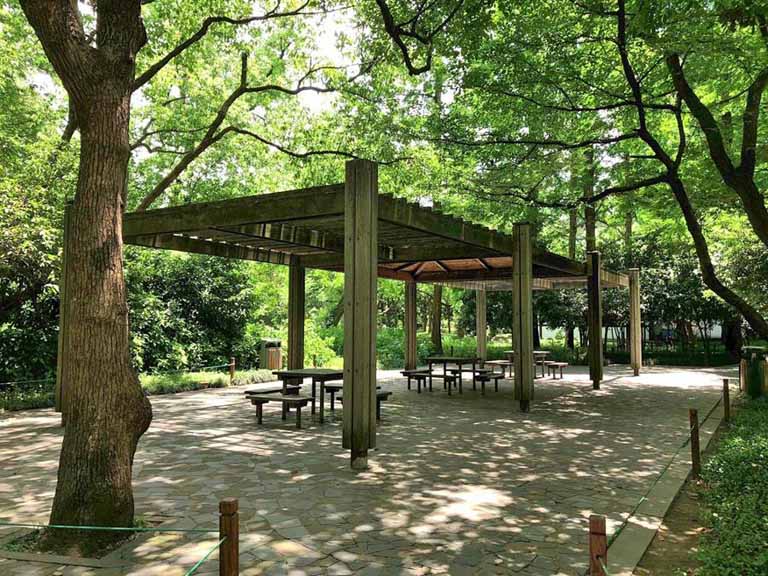}}
	
	\caption{Comparison 1 of results from SICE.}
	\label{Comparison_sice1}
	\vspace{-2mm}
\end{figure}

\begin{table}[h]
\caption{ Quantitative assessment on Testset-2. The scores for the first, second, and third places are marked in red, blue, and green, respectively.}
\vspace{-2mm}
\centering
\begin{tabular}{lllllll}
\hline
Method       & NIQE                        & BIQI                         & NFERM                        & PSNR                         & SSIM                         & LOE                          \\ \hline
DFE          & \textcolor{blue}{2.55}     & 30.67                        & 20.46                        & 16.57                        & 0.651                        & 383.3                        \\
LIME         & 2.72                        & 35.40                        & 22.55                        & 14.52                        & 0.623                        & 417.9                        \\
MLLP         & \textcolor{red}{2.45}       & \textcolor{green}{30.46}     & 23.50                        & 16.68                        & 0.647                        & 528.8                        \\
ALSM         & 2.57                        & 33.55                        & 23.86                        & 14.55                        & 0.635                        & 389.7                        \\
NRMOE        & 2.78                        & 31.24                        & 23.61                        & 15.72                        & 0.637                        & 416.6                        \\
PLME         & 2.63                        &\textcolor{blue}{30.35}       & \textcolor{red}{19.33}       & 16.82                        & 0.657                        & 425.9                        \\
PnpRetinex   & 2.58                        & 33.17                        &\textcolor{blue}{19.99}       & \textcolor{green}{17.09}     & 0.661                        & \textcolor{blue}{372.2}      \\
RetinexDIP   & 2.90                        & 31.90                        & 26.53                        & 15.62                        & 0.641                        & 544.9                        \\
EnlightenGAN & 2.68                        & 30.76                        & 21.55                        & 16.47                        & \textcolor{blue}{0.683}      & 528.9                        \\
KIND++       & 2.93                        & 32.67                        & 25.76                        & 16.64                        & 0.658                        & 586.5                        \\
SCL-LLE      & 2.75                        & 30.71                        & 26.28                        & 16.93                        & 0.650                        & \textcolor{green}{378.2}     \\
ZeroDCE     & 2.70                        & 30.94                        & 23.26                        & \textcolor{red}{18.82}       & \textcolor{red}{0.693}       & 408.9                        \\
Proposed     &\textcolor{blue}{2.55}       & \textcolor{red}{29.55}       & \textcolor{green}{20.20}     & \textcolor{blue}{17.33}      & \textcolor{green}{0.672}     & \textcolor{red}{359.0}       \\ \bottomrule
\end{tabular}
\label{quancomp_fr}
\end{table}

\begin{figure}[h]
	\vspace{-4mm}
	\centering
	\subfloat[Input]{\includegraphics[height=30mm]{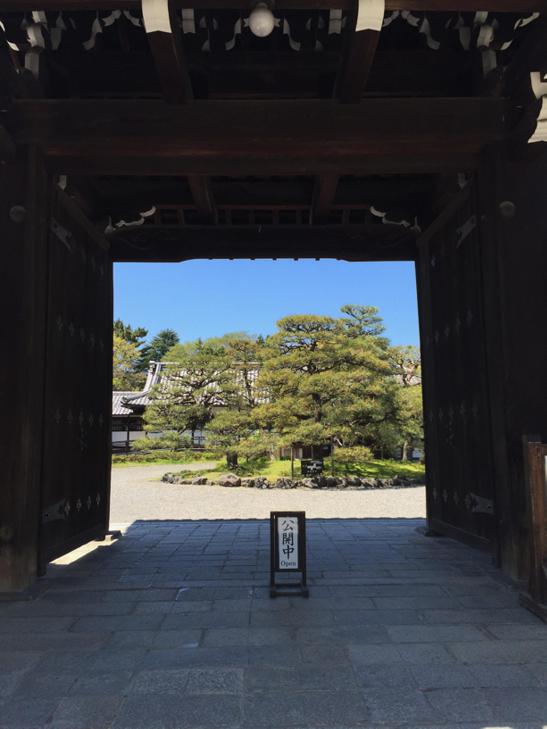}}
	\hspace{0.05mm}
	\subfloat[Ref.Image]{\includegraphics[height=30mm]{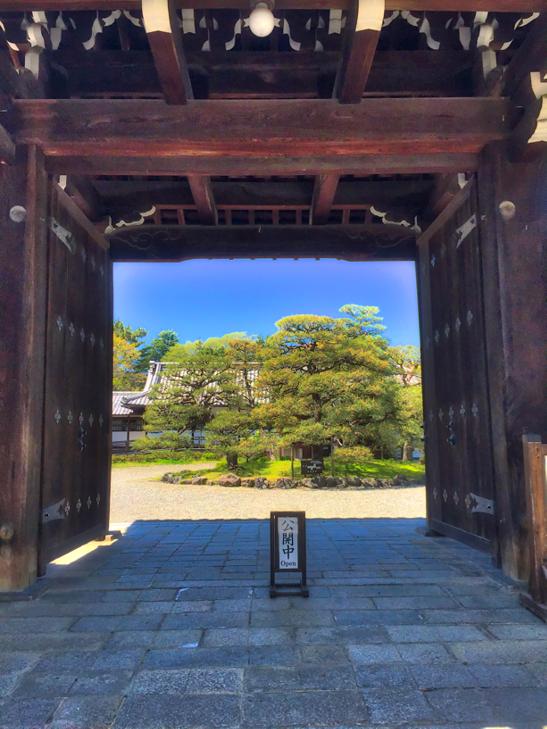}}		
	\hspace{0.05mm}
	\subfloat[DFE]{\includegraphics[height=30mm]{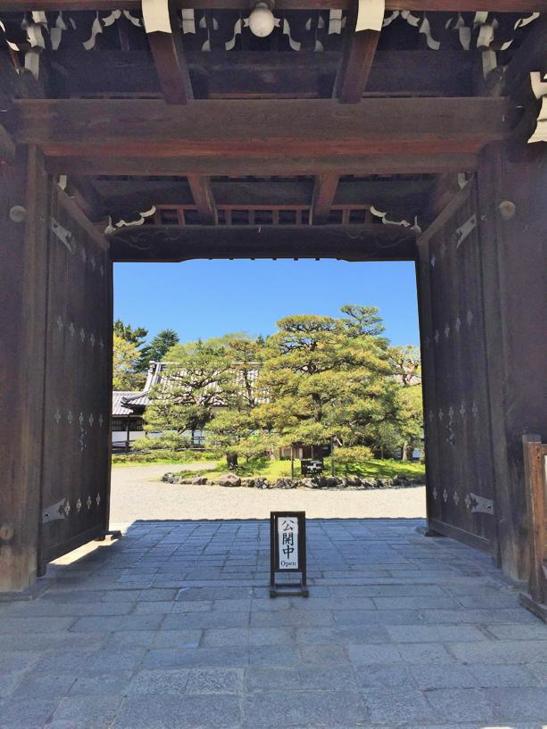}}		
	\hspace{0.05mm}
	\vspace{-2mm}
	
	\centering
	\subfloat[ALSM]{\includegraphics[height=30mm]{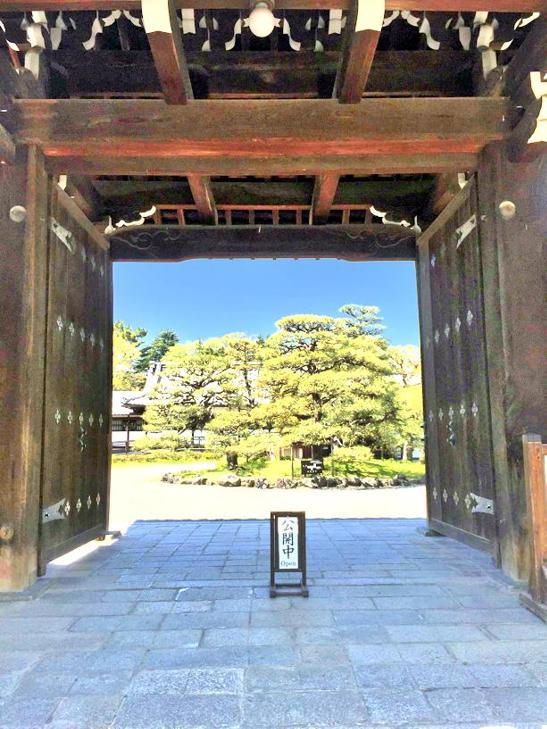}}	
	\hspace{0.05mm}	
	\subfloat[LIME]{\includegraphics[height=30mm]{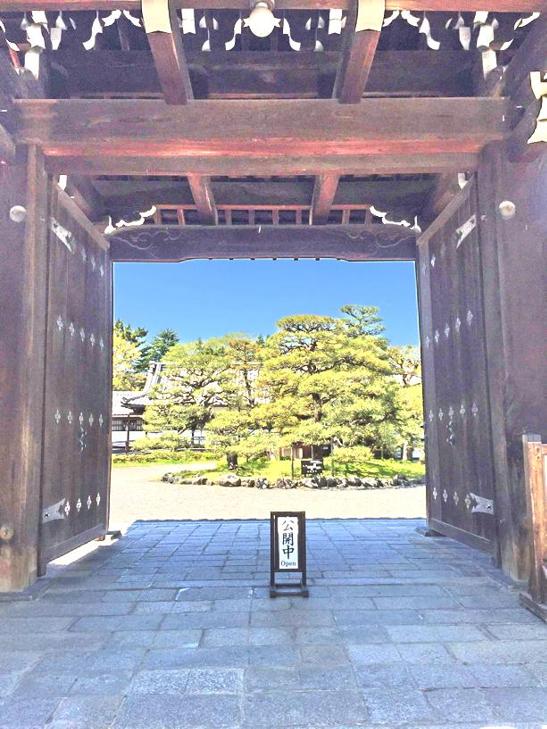}}		
	\hspace{0.05mm}
	\subfloat[MLLP]{\includegraphics[height=30mm]{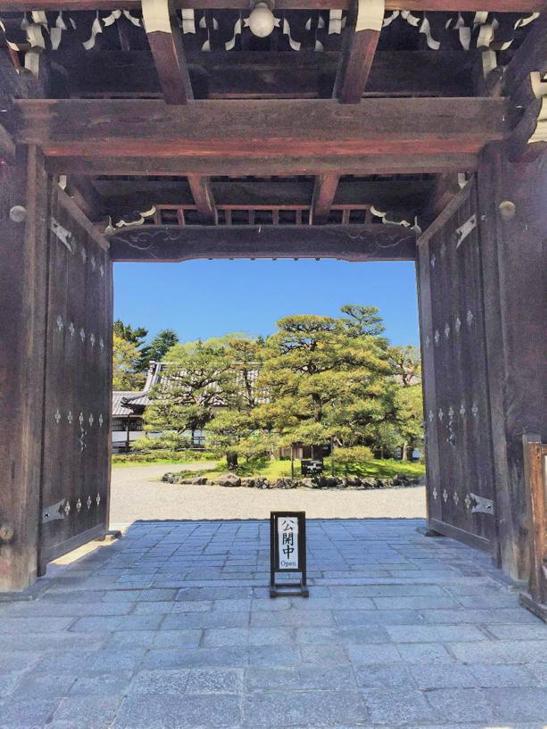}}
	\hspace{0.05mm}	
	\vspace{-2mm}
	
	\centering   
    \subfloat[PnpRtx]{\includegraphics[height=30mm]{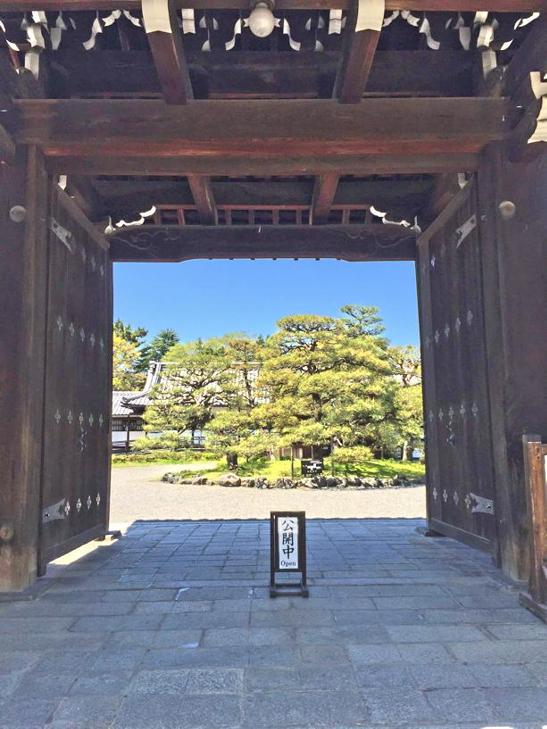}}
	\hspace{0.05mm} 	
	\subfloat[NRMOE]{\includegraphics[height=30mm]{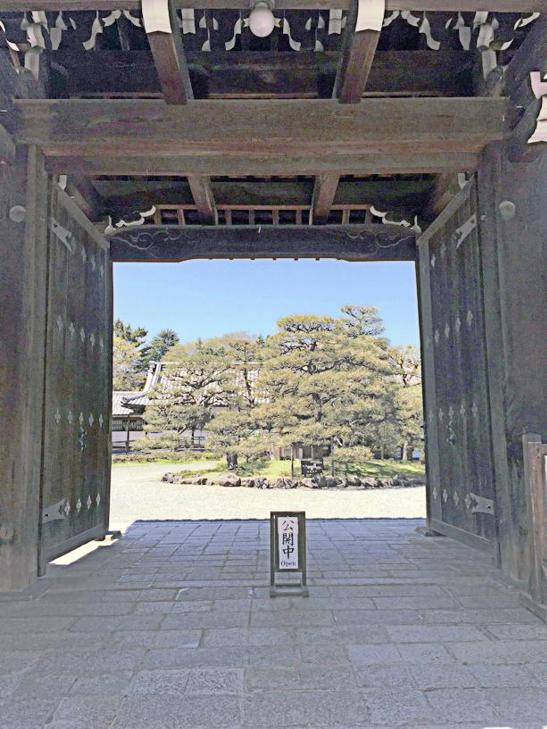}}		
	\hspace{0.05mm}
	\subfloat[PLME]{\includegraphics[height=30mm]{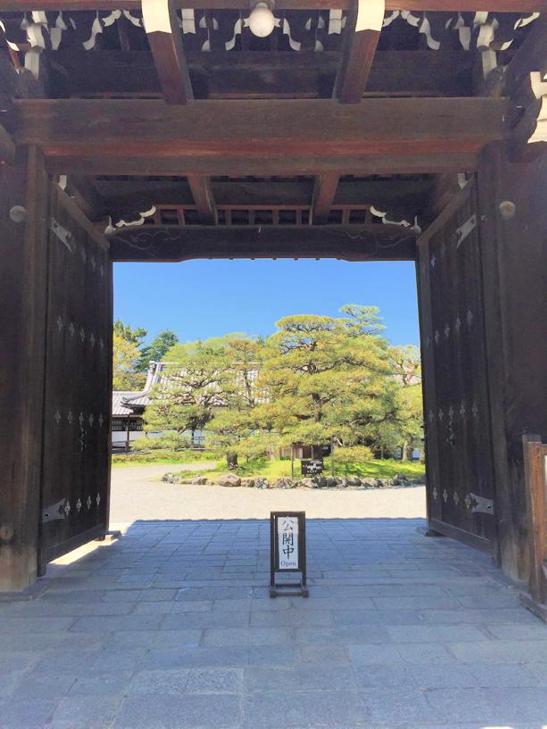}}
	\hspace{0.05mm}	
	\vspace{-2mm}
	
	\centering
	\subfloat[EGAN]{\includegraphics[height=30mm]{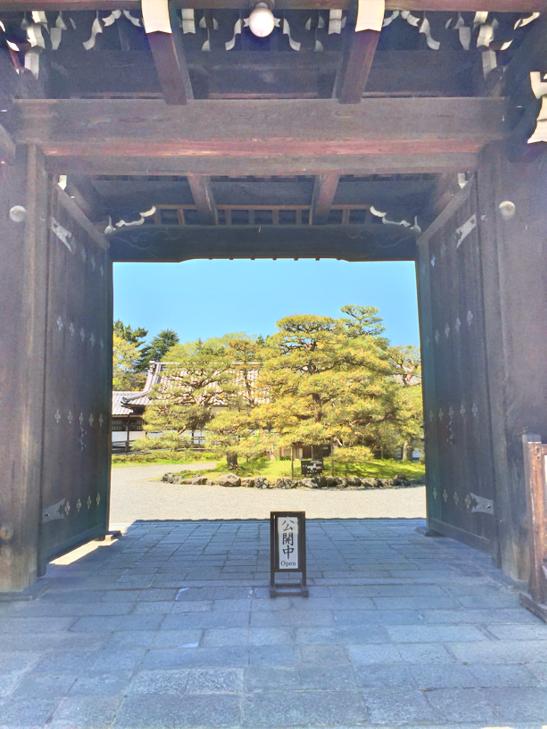}}
	\hspace{0.05mm}			
	\subfloat[KIND++]{\includegraphics[height=30mm]{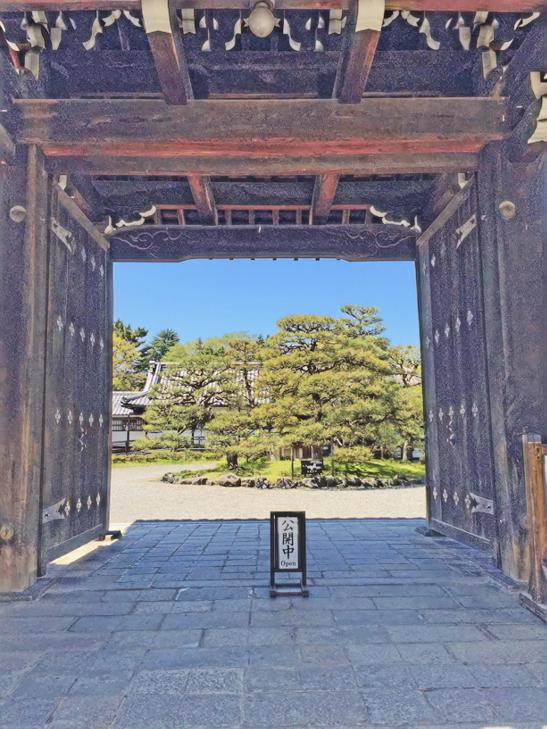}}
	\hspace{0.05mm}	
	\subfloat[ZeroDCE]{\includegraphics[height=30mm]{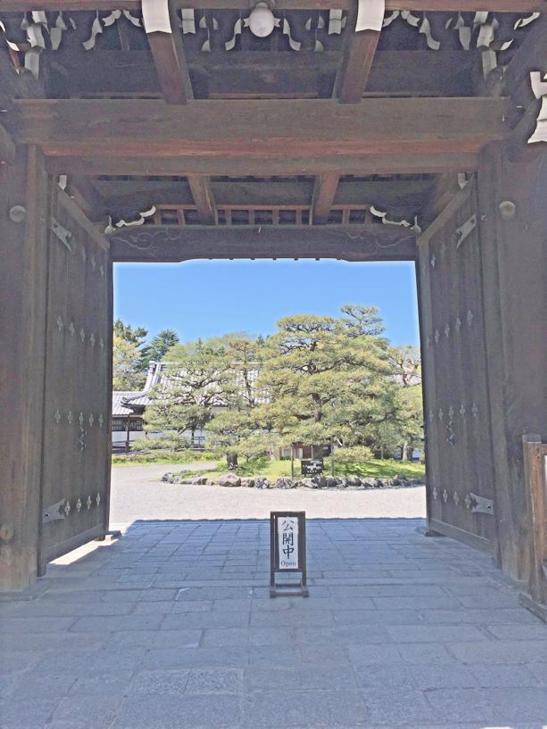}}		
	\hspace{0.05mm}
	\vspace{-2mm}
	
	\centering	
	\subfloat[SCL-LLE]{\includegraphics[height=30mm]{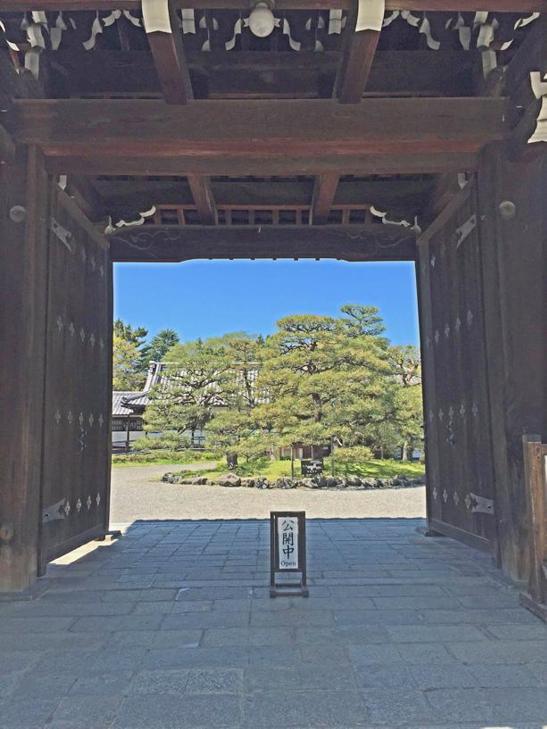}}
	\hspace{0.05mm}	
	\subfloat[RtxDIP]{\includegraphics[height=30mm]{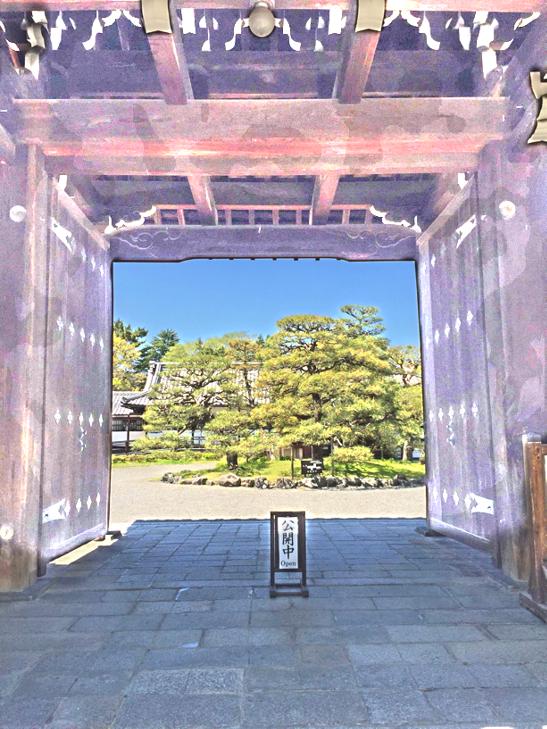}}
	\hspace{0.05mm}
	\subfloat[Proposed]{\includegraphics[height=30mm]{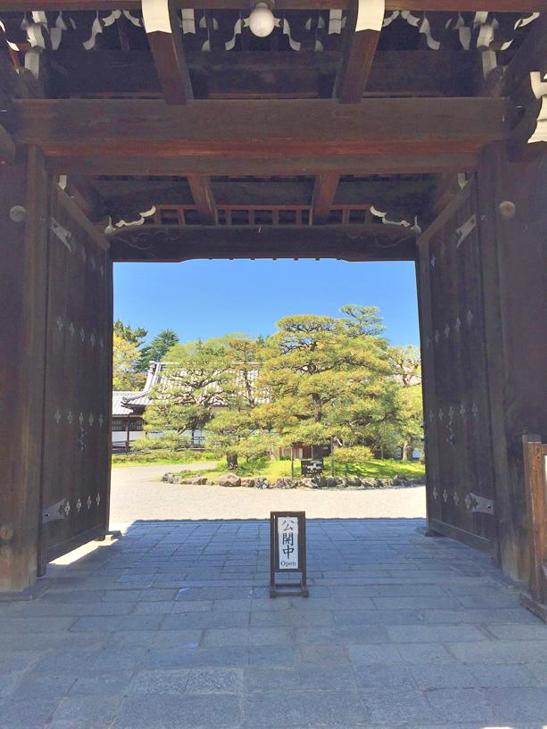}}
	
	\caption{Comparison 2 of results from SICE. }
	\label{Comparison_sice2}
	\vspace{-2mm}
\end{figure}

Table \ref{quancomp_fr} presents the quantitative assessment results on Testset-2. The proposed method achieves the best BIQI score, the second-best NIQE score, and the third-best NFERM score. This means that our method performs consistently in terms of the three no-reference IQA metrics. Our method also obtains the lowest LOE value for this dataset, indicating the best performance on naturalness preservation. ZeroDCE obtains the highest PSNR and SSIM values, and EnlightenGAN obtains the second highest SSIM value, while our method ranks second and third, respectively. The reason may be that ZeroDCE and EnlightenGAN are trained on the Part1 of SICE dataset and have learned how to enhance the test images with parameters trained from the training set. Figs.\ref{Comparison_sice1} and \ref{Comparison_sice2} present two examples from Testset-2 for visual comparison. We can observe that ZeroDCE and EnlightenGAN suffer from color deviation. The same drawback occurs in the results of SCL-LLE and NRMOE. KIND++ and RetinexDIP yield results with noticeable artifacts. We consider that the results obtained by DFE, PnpRetinex, MLLP, and our method are closer to the reference images than those by the other methods.

\noindent\textbf{4) Comparison on speed}

We compare the average computation time of all compared methods over 100 images with size of 1368*912. As shown in Table \ref{speed}, the proposed method is the fastest method on CPU platform and the averaged computation time of our method is 0.23s. The deep-learning based methods ZeroDCE and SCL-LLE are faster than our method, but they require expensive GPU resources.

\begin{table}[h]
\caption{Comparison of the average computation time (in seconds). The top three fastest methods are marked in red,blue, and green respectively.}
\vspace{-2mm}
\centering
\begin{tabular}{l|llll}
\hline
Platform & Method       & Time   & Method     & Time  \\ \hline
CPU      & DFE          & 8.65   & LIME       & 0.42  \\
         & ALSM         & 65.37  & MLLP       & 45.08 \\
         & NRMOE        & 3.56   & PLME       & 0.77  \\
         & PnpRetinex   & 13.24  & Proposed   & \textcolor{green}{0.23}  \\ \hline
GPU      & EnlightenGAN & 0.84  & KIND++     & 3.58  \\
         & SCL-LLE      & \textcolor{blue}{0.0069}   & RetinexDIP  & 42.37 \\
         & ZeroDCE     & \textcolor{red}{0.0041}    &              &       \\ \bottomrule
\end{tabular}
\label{speed}
\end{table}

\section{Discussion}\label{discussion}

\subsection{Comparison with Retinex decomposition}\label{rtx-cmp}
\begin{figure}[t]
\vspace{0mm}
	\captionsetup[subfloat]{labelsep=none,format=plain,labelformat=empty}
	\centering
	\subfloat[(a) Input]{\includegraphics[width=28mm]{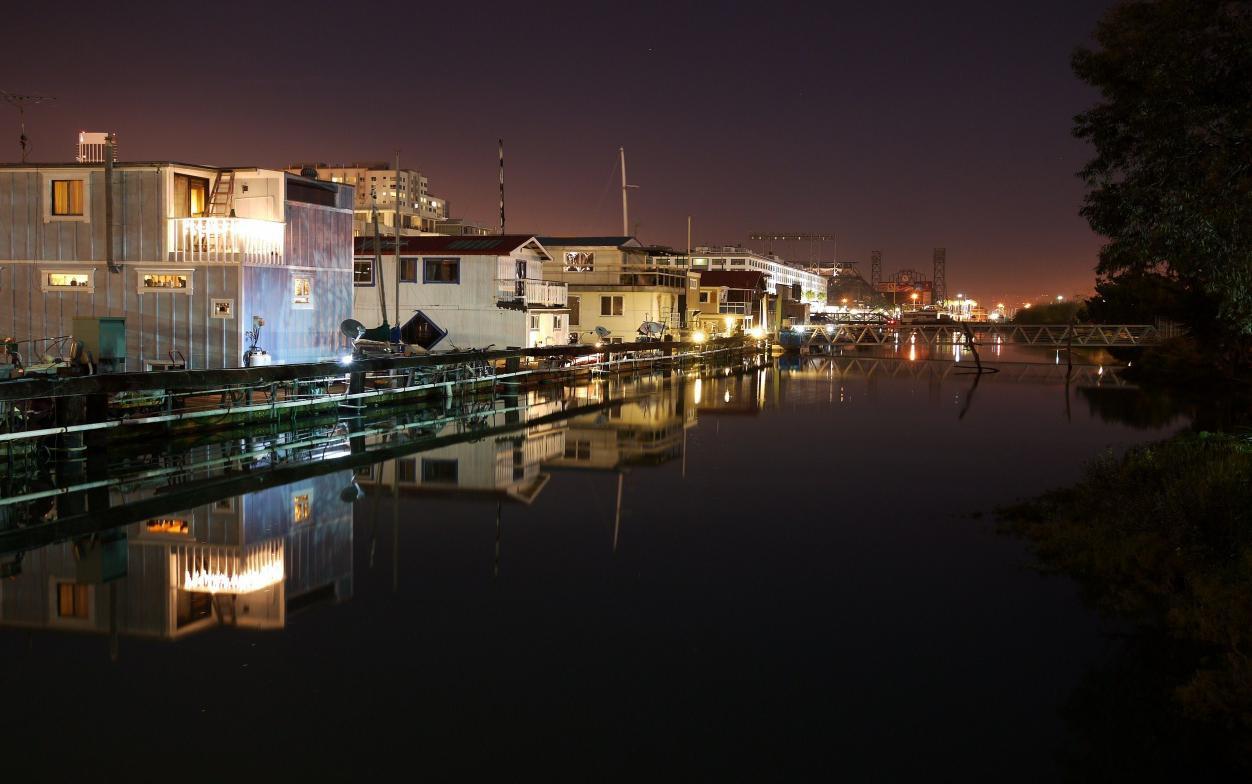}}
	
	\vspace{-2mm}
	\centering
	\subfloat{\includegraphics[width=28mm]{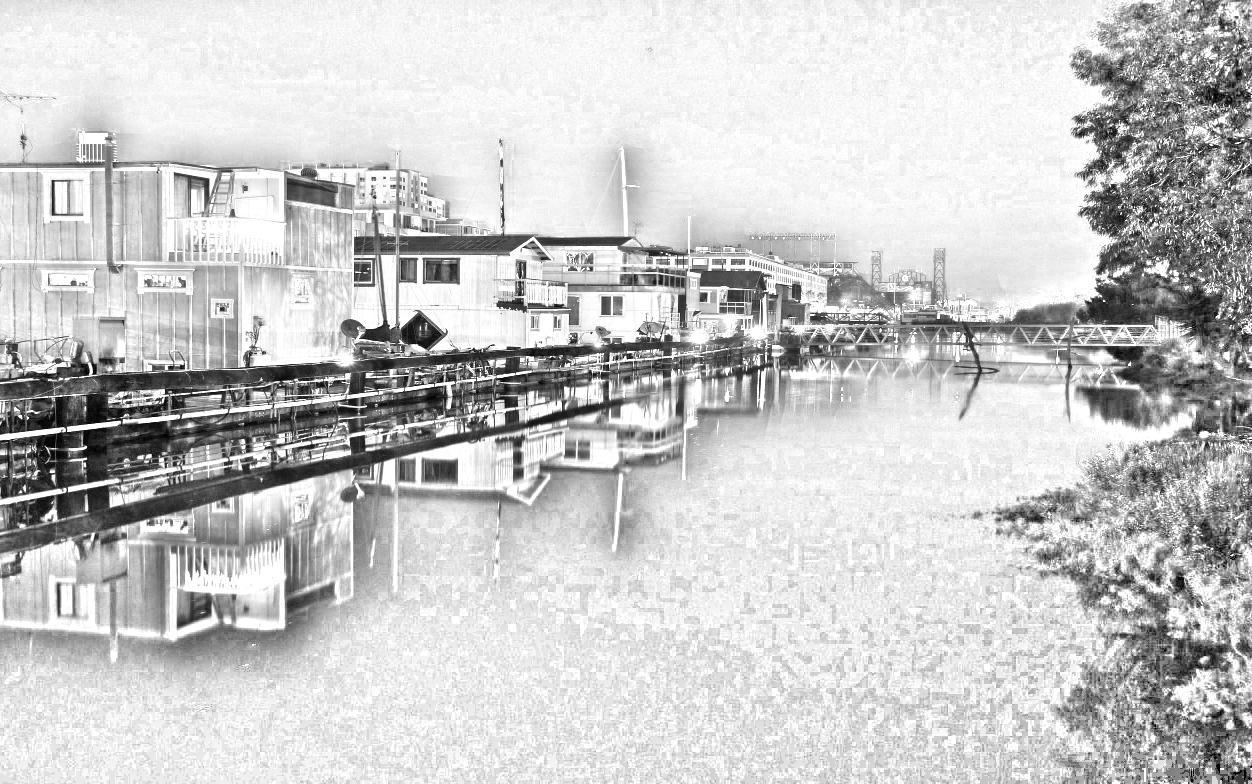}}
	\hspace{0.05mm}
	\subfloat[(b) DFE]{\includegraphics[width=28mm]{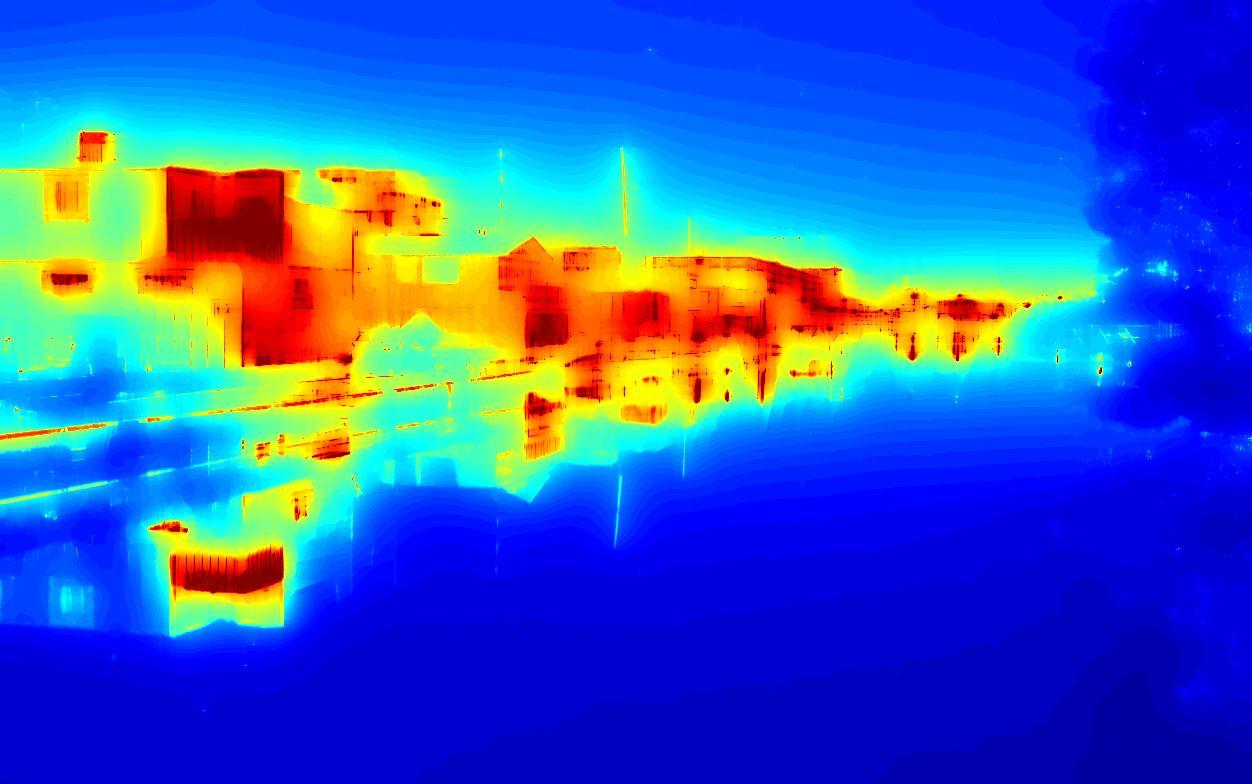}}
	\hspace{0.05mm}
	\subfloat{\includegraphics[width=28mm]{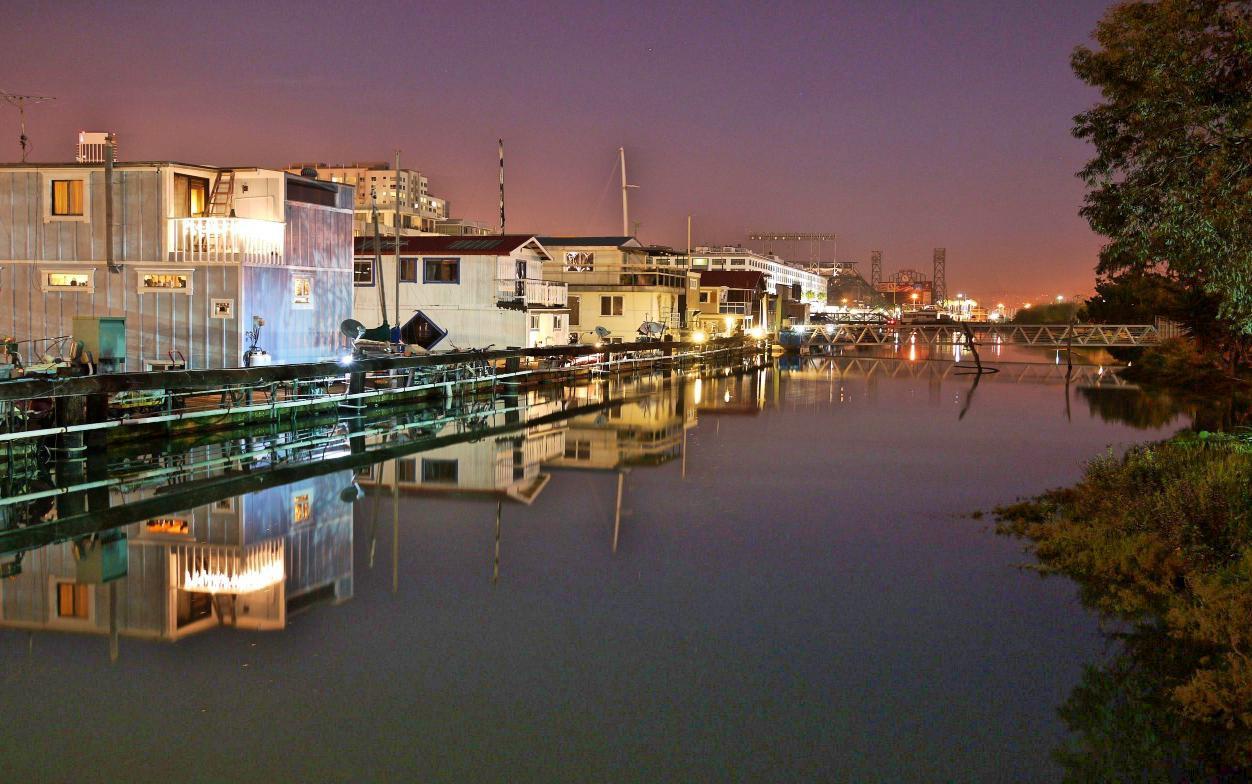}}
	
	\vspace{-2mm}	
	
	\centering	
	\subfloat{\includegraphics[width=28mm]{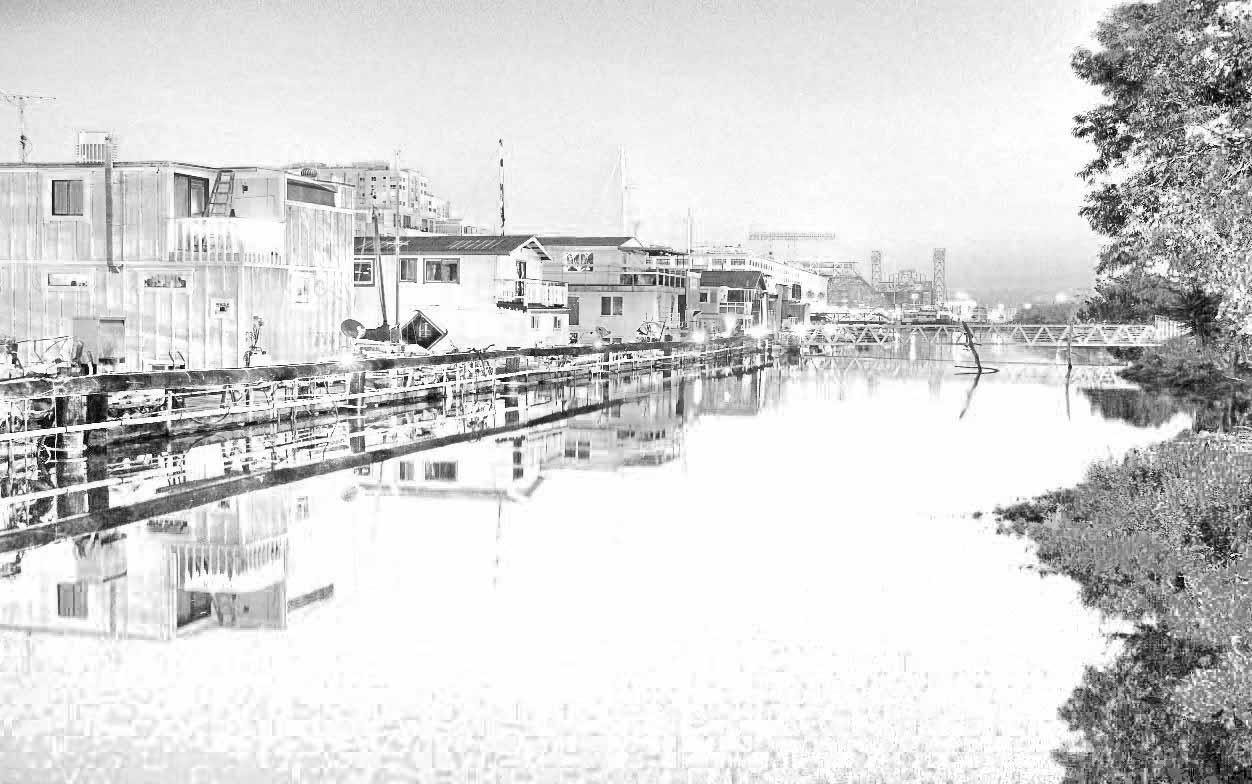}}
	\hspace{0.05mm}
	\subfloat[(c) LIME]{\includegraphics[width=28mm]{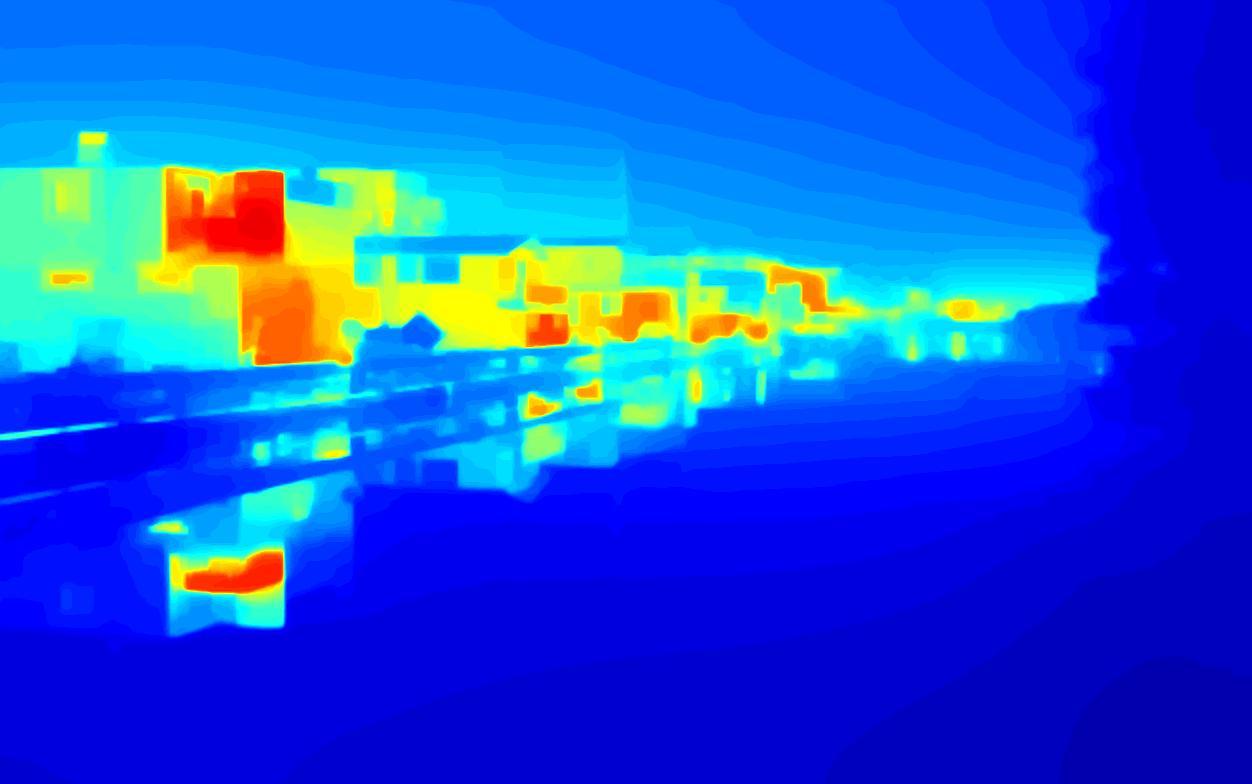}}
	\hspace{0.05mm}
	\subfloat{\includegraphics[width=28mm]{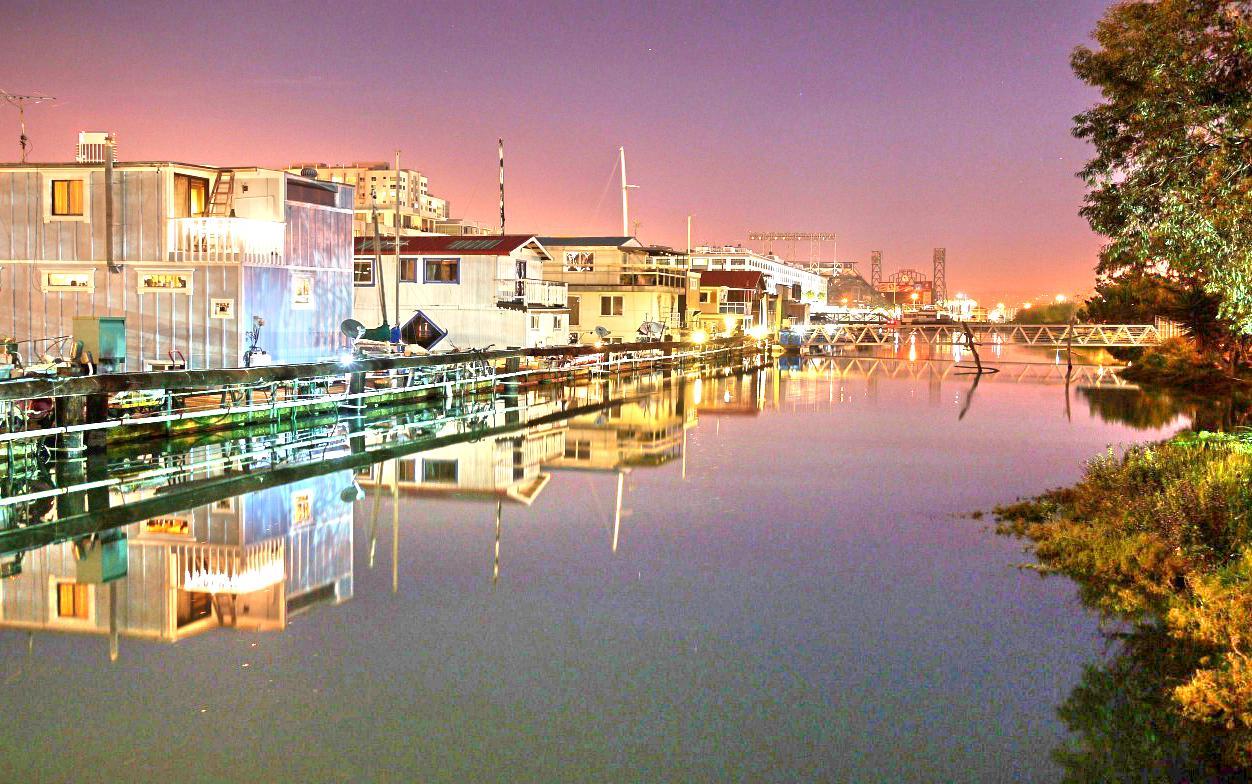}}
	
	\vspace{-2mm}
	\centering
	\subfloat{\includegraphics[width=28mm]{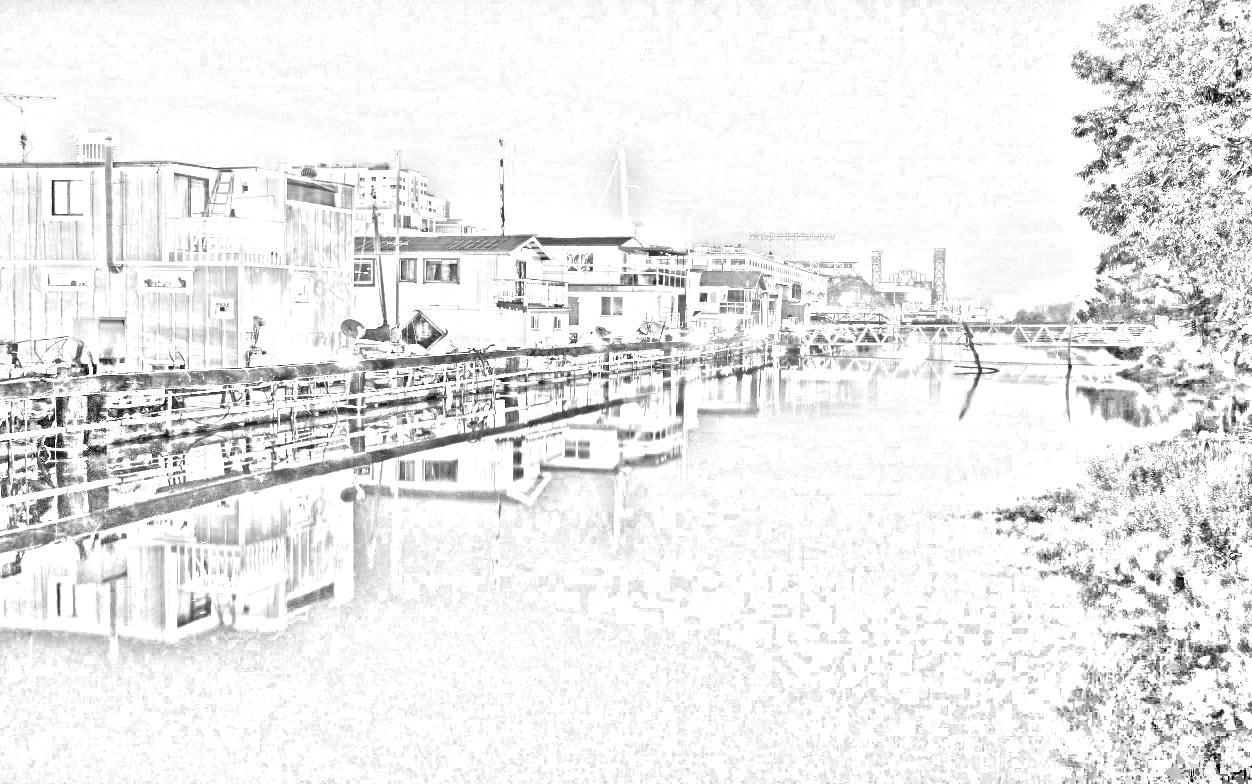}}
	\hspace{0.05mm}
	\subfloat[(d) PnpRetinex]{\includegraphics[width=28mm]{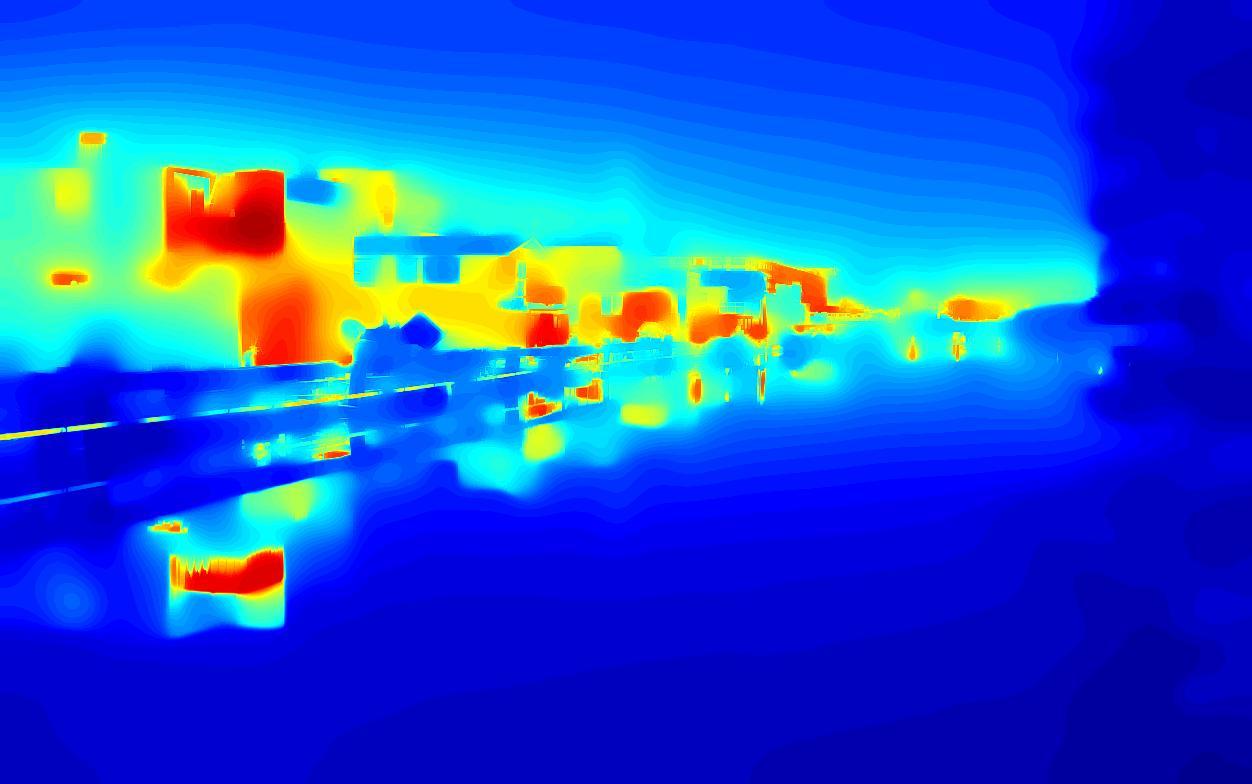}}
	\hspace{0.05mm}
	\subfloat{\includegraphics[width=28mm]{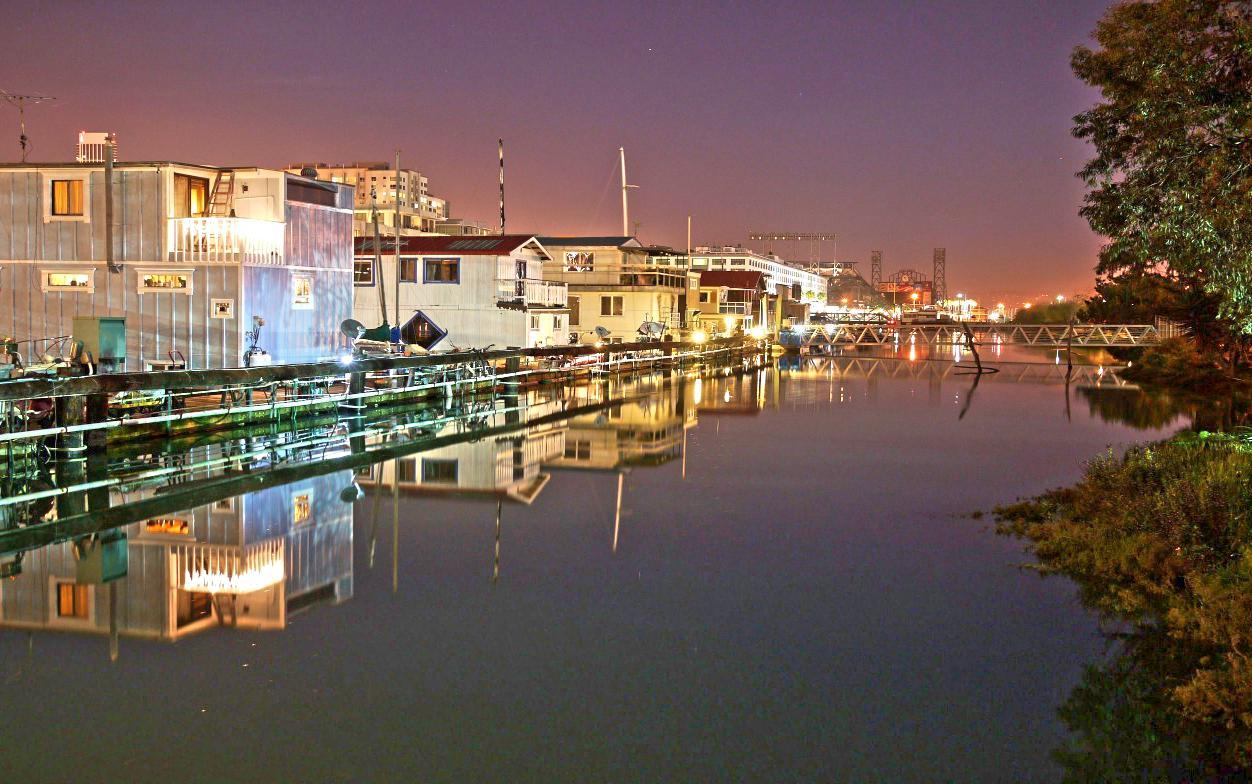}}

	\vspace{-2mm}
	\centering
	\subfloat{\includegraphics[width=28mm]{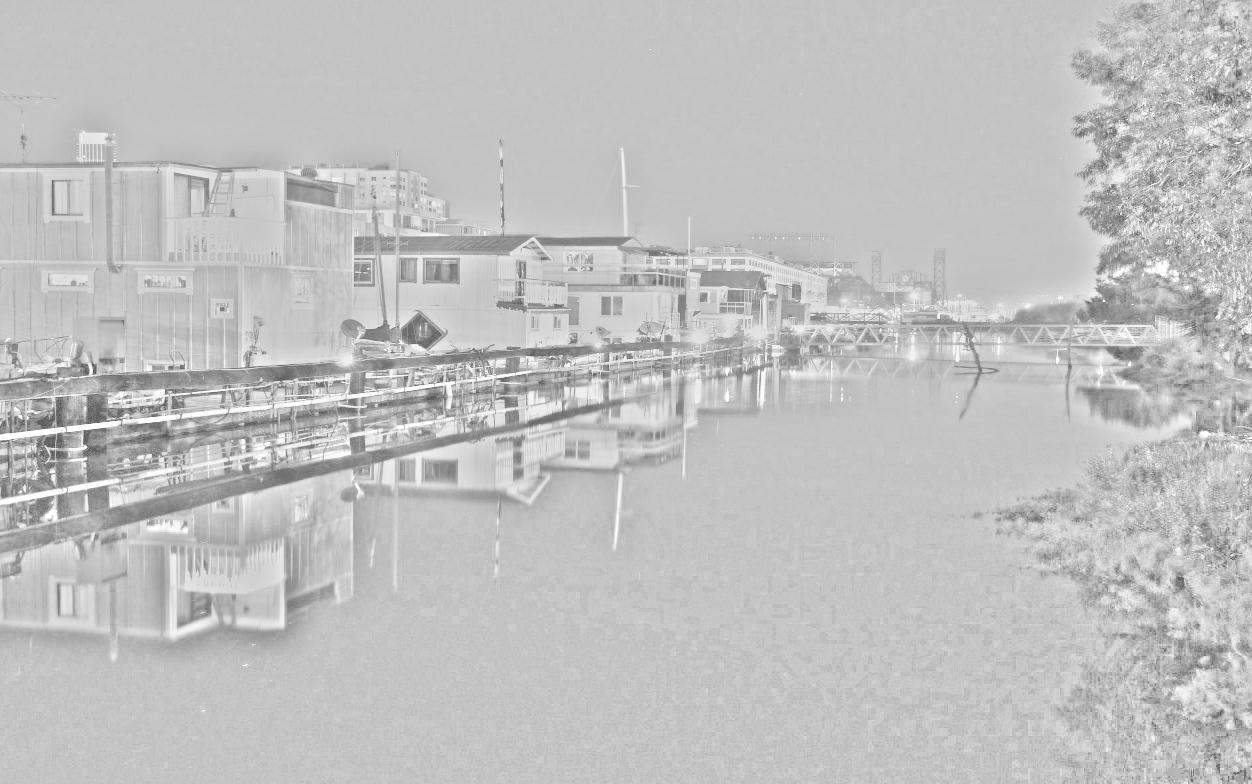}}
	\hspace{0.05mm}
	\subfloat[(e) Proposed]{\includegraphics[width=28mm]{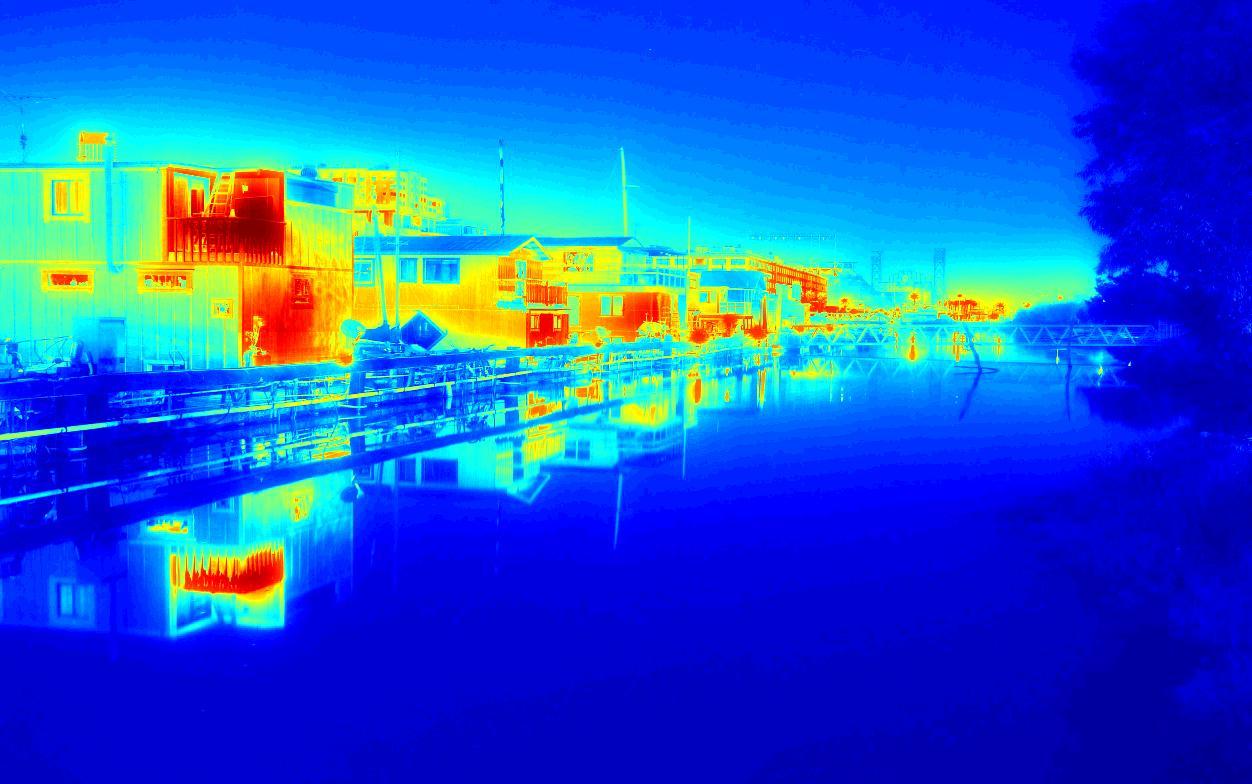}}
	\hspace{0.05mm}
	\subfloat{\includegraphics[width=28mm]{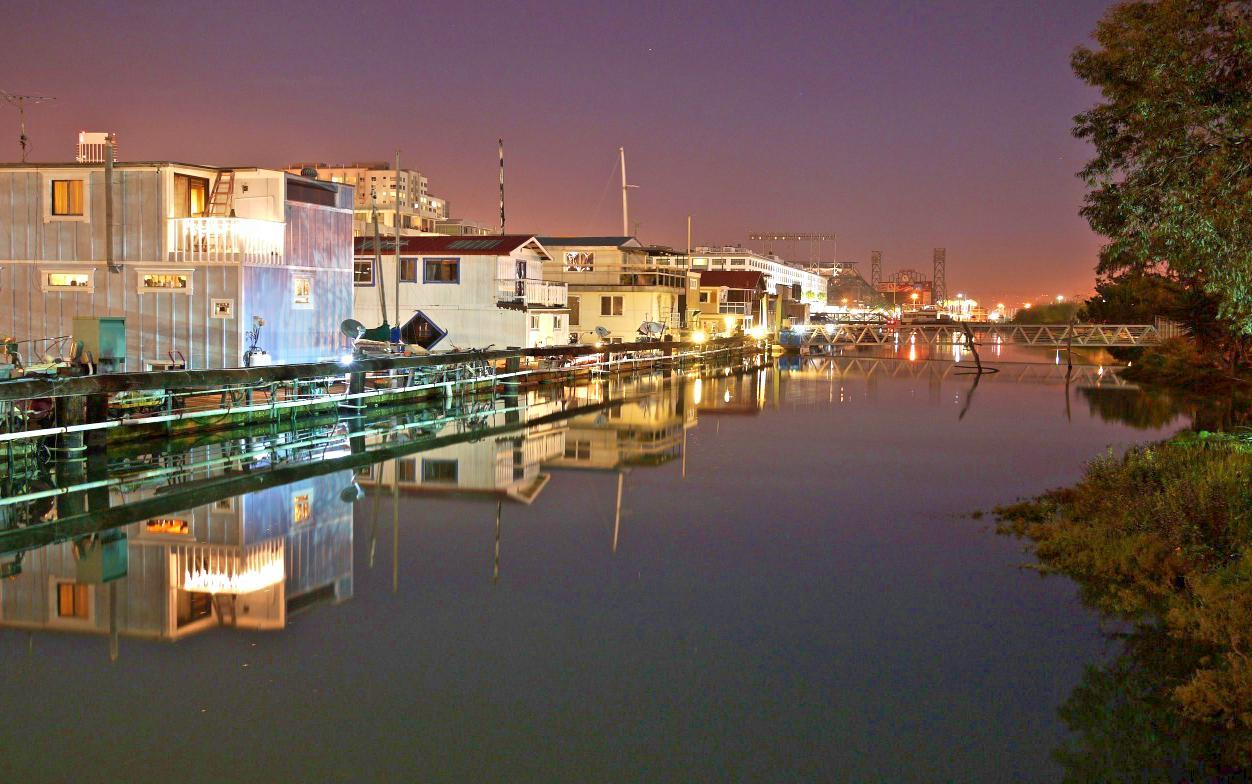}}

	
	\caption{Comparison with Retinex decomposition. The input image violates the assumptions made by Retinex theory. 
	}
	\label{fig-rtxcmp}
	\vspace{-4mm}
\end{figure}

Assuming that an image is the product of an illumination image and a reflectance image, Retinex-based enhancement methods deal with the challenge of decomposing a given image into the illumination image and the reflectance image. Computing such a decomposition for real scenes is an ill-posed problem. Consequently, any attempt to solve it must make some simplifying assumptions about the scene, such as the spatial smoothness of illumination, or the piece-wise consistency/similarity of albedo \cite{Guo2017LIME,lin2022low}. These assumptions about the scene imaging process are easily violated in real scenes. An example is shown in Fig.\ref{fig-rtxcmp}(a). The lower half of (a) is the reflection image from a highly reflective surface (e.g., a mirror or a still pond). The nearly identical reflectance of the pond surface suggests that the incident illumination of this reflection region is not spatially smooth. In addition, the light source regions in (a) go against the assumption that an image is the product of illumination and reflectance. The images in Fig.\ref{fig-rtxcmp}(b) to (d), from left to right, are the reflectance, illumination, and enhanced images produced by three Retinex-based methods DFE, LIME, and PnpRetinex, respectively. It can be seen that the pond regions in the three illumination images are locally smooth. The images in (e), from left to right, are the contrast, residual, and enhanced images produced by our method. The contrast image looks similar to the reflectance images, while the residual image is similar to the illumination images. This phenomenon could be explained by two widely accepted and closely related assumptions about the HVS\cite{soranzo2019layer}: 1) the HVS is mostly sensitive to surface reflectance, and 2) the HVS primarily responds to local scene contrasts. It might be argued that the contrast, to some extent, is a function of perceived reflectance. Although the contrast and residual images are similar to the reflectance and illumination images, their physical implications are fundamentally different. Our model attempts to decompose the light variations from an image in a functionally similar manner to the visual perception of scenes instead of making physical assumptions about the imaging process of the scene.

\begin{figure*}[h]
	\vspace{-2mm}
	\centering	
	\hfil
	\subfloat[Input]{\includegraphics[width=32mm]{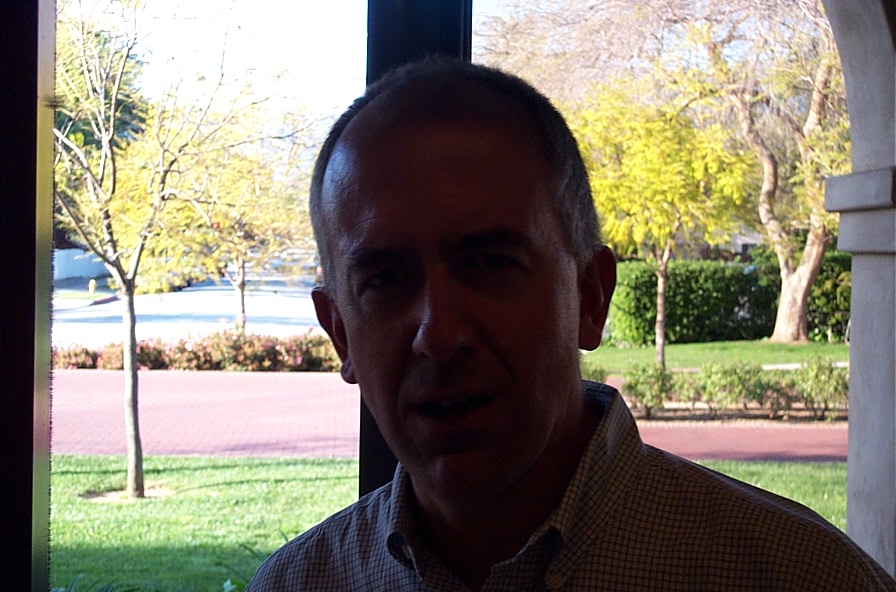}}	
	\hspace{0.2mm}
	\subfloat[Weber]{\includegraphics[width=32mm]{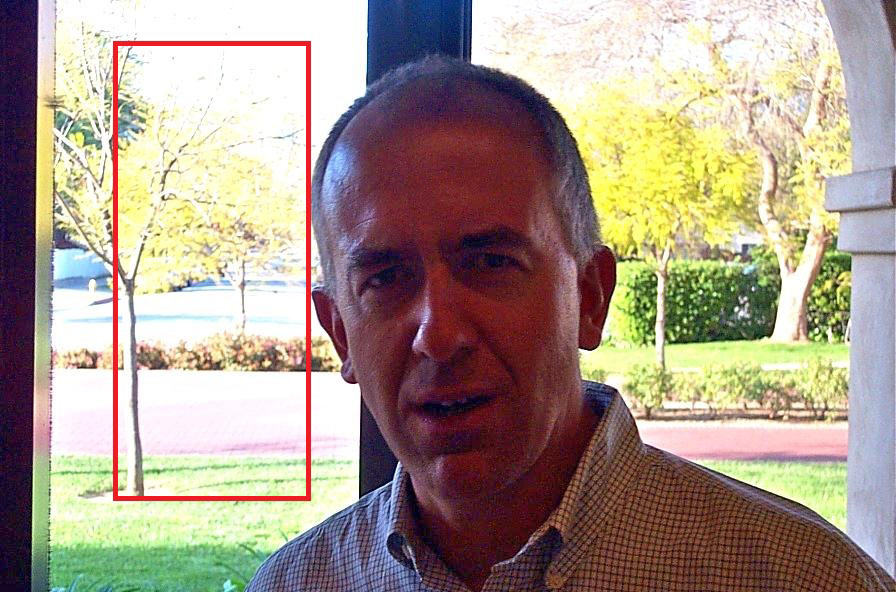}}
	\hspace{0.1mm}
	\subfloat[Michaelson]{\includegraphics[width=32mm]{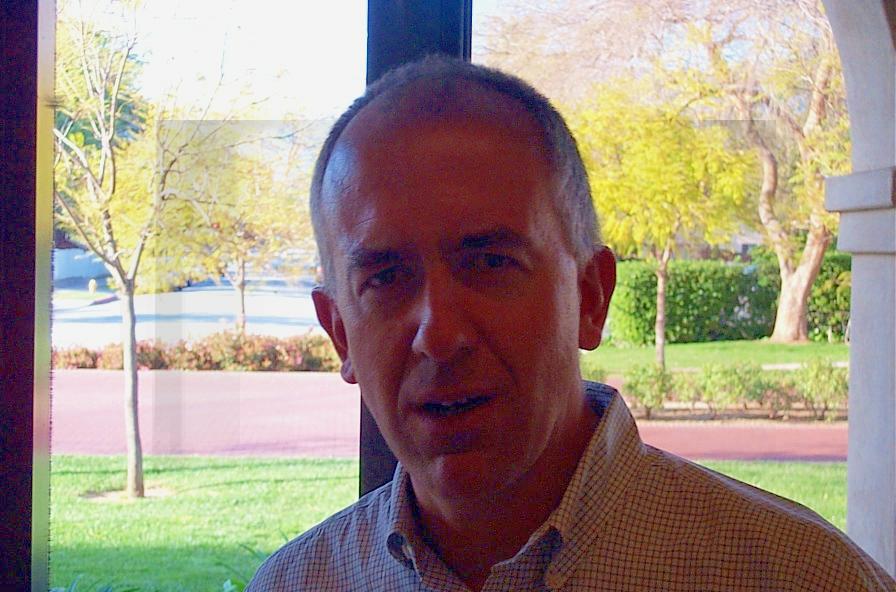}}
	\hspace{0.1mm}
	\subfloat[RMS]{\includegraphics[width=32mm]{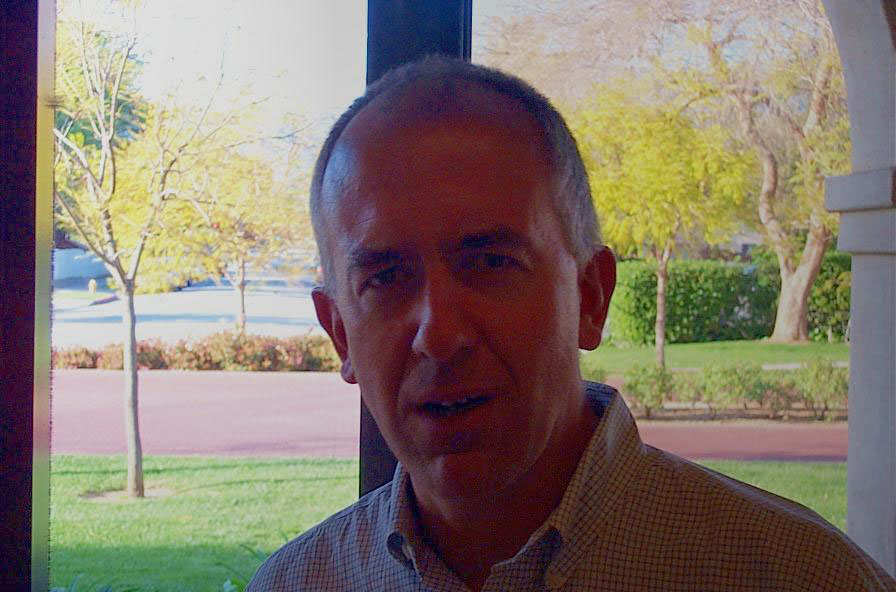}}
	\hspace{0.1mm}
	\subfloat[Proposed]{\includegraphics[width=32mm]{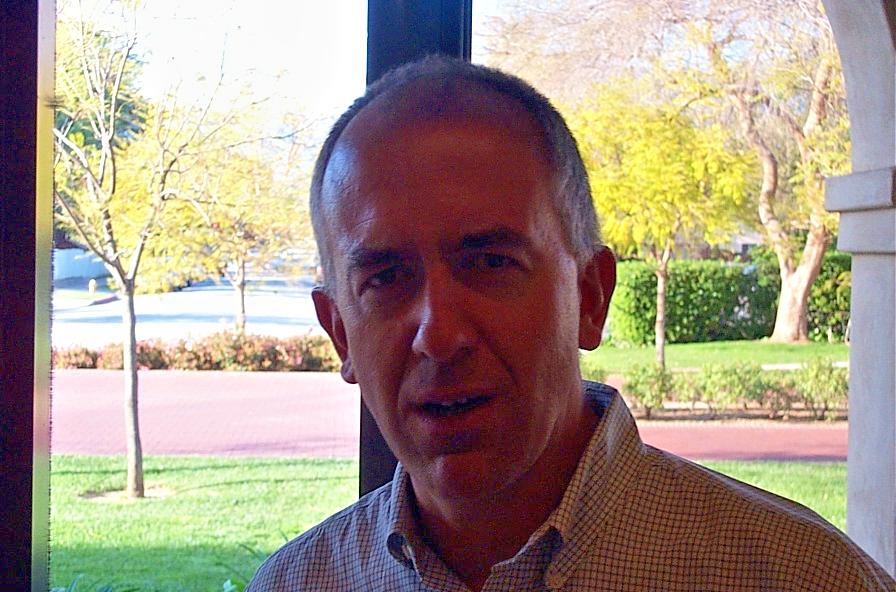}}
	\caption{ Generality of the proposed model. The three ratio-type contrasts achieve different degrees of image enhancement within the proposed model. The Weber contrast obtains comparable performance to the proposed method, but tends to wash out the details, see the marked region in (b).}
	\label{fig_othercontrast}
	\vspace{-4mm}
\end{figure*}

\subsection{Generality of the proposed method}
The core perspective of our method is to decompose an image into the contrast image and the residual image. An interesting question is whether other contrast definitions can be applied to the proposed model. We discuss this question through three well-known contrasts: the Weber contrast, the Michelson contrast, and the root mean square (RMS) contrast \cite{beghdadi2020a}.

The three ratio-type contrasts are defined as follows:
\begin {itemize}
\item 
Weber contrast
\end {itemize}   
\begin{equation}\label{eq_wcimg}
{\bf{C}_{Weber}} = \frac{{{\bf{I}} - {{\bf{I}}_S }}}{{{\bf{I}}_S}}
\end{equation}

\begin {itemize}
\item 
Michelson contrast
\end {itemize}
\begin{equation}\label{eq_mc}
{{\bf{C}}_{Michelson}} = \frac{{{{\bf{I}}_{\max }} - {{\bf{I}}_{\min }}}}{{{{\bf{I}}_{\max }} + {{\bf{I}}_{\min }}}}
\end{equation}

\begin {itemize}
\item 
RMS contrast
\end {itemize}
\begin{equation}\label{eq_rms}
{{\bf{C}}_{RMS}} = \sqrt {\frac{1}{{{M_\Omega }}}\sum\limits_\Omega ^{} {{{\left( {\frac{{{\bf{I}} - {\bf{\bar I}}}}{{{\bf{\bar I}}}}} \right)}^2}} }
\end{equation}

In Eqs. (\ref{eq_wcimg}) to (\ref{eq_rms}), $\bf{I}$ denotes the image intensity, ${{\bf{I}}_S }$ is the local background intensity, ${{\bf{I}}_{\max} }$ and ${{\bf{I}}_{\min} }$  are the highest and lowest intensities in a local image patch, respectively, ${\bf{\bar I}}$ is the mean intensity of the local area $\Omega$, and ${M_\Omega }$ is the number of pixels in $\Omega$. As far as we know, directly applying the three contrasts to enhance unevenly-lit images is less reported. In this paper, we treat these three contrast functions as three different retinal models to generate the contrast images. By testing a number of images, we find that the Weber contrast produces better-enhanced results in terms of fidelity than the other two contrasts and approaches comparable performance to Eq. (\ref{eq_cimg}). However, the Weber contrast sometimes tends to lose the details in bright regions. A representative example is shown in Fig. \ref{fig_othercontrast}. It might be argued that: 1) Eq. (\ref{eq_wcimg}) can be regarded as a simplified version of Eq. (\ref{eq_cimg}). 2) Eq. (\ref{eq_cimg}) and Eq. (\ref{eq_wcimg}) are capable of estimating the contrast in natural images more accurately than the other two contrasts.

\vspace{-1mm}
\subsection{Post-processing for noise suppression}\label{noise_sup}
\begin{figure}[t]
\vspace{0mm}
	\centering
	\captionsetup[subfloat]{labelsep=none,format=plain,labelformat=empty}
	
	\subfloat{\includegraphics[width=0.85in]{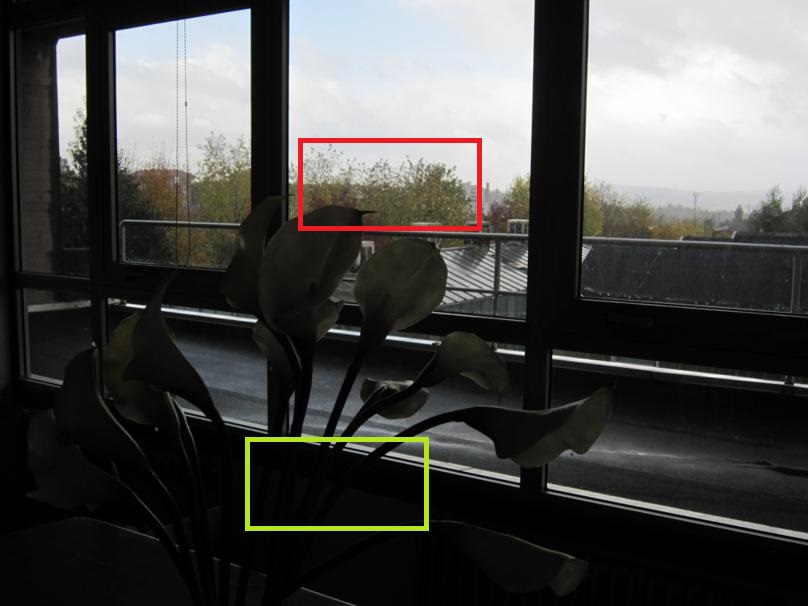}}
	\hfil 
	\subfloat{\includegraphics[width=0.85in]{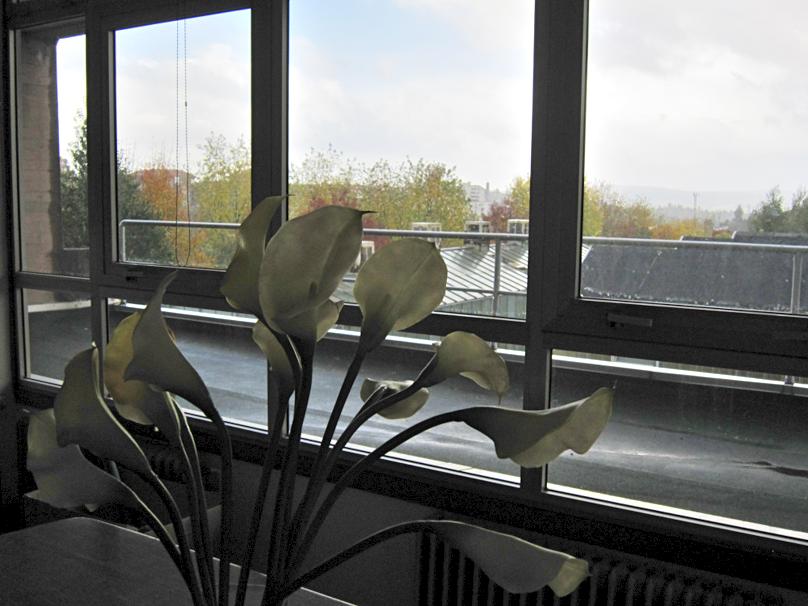}}
	\hfil
	\subfloat{\includegraphics[width=0.85in]{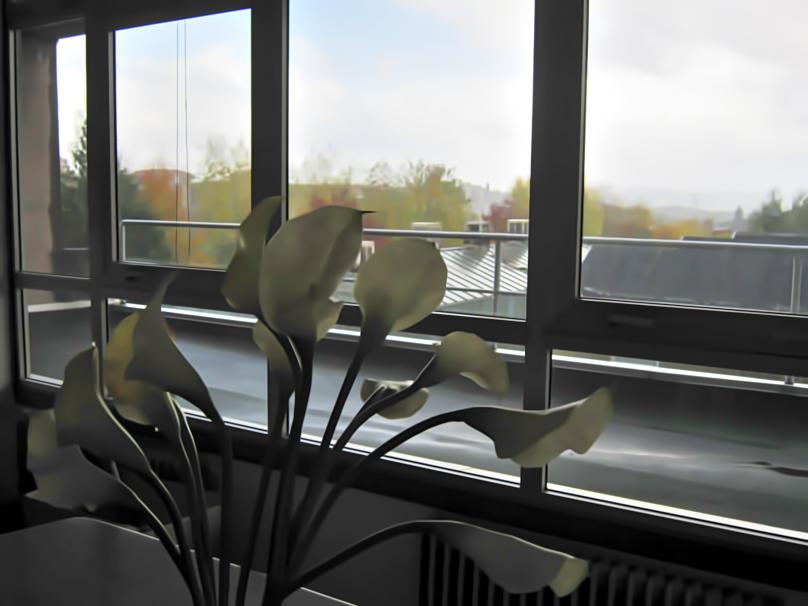}}
	\hfil
	\subfloat{\includegraphics[width=0.85in]{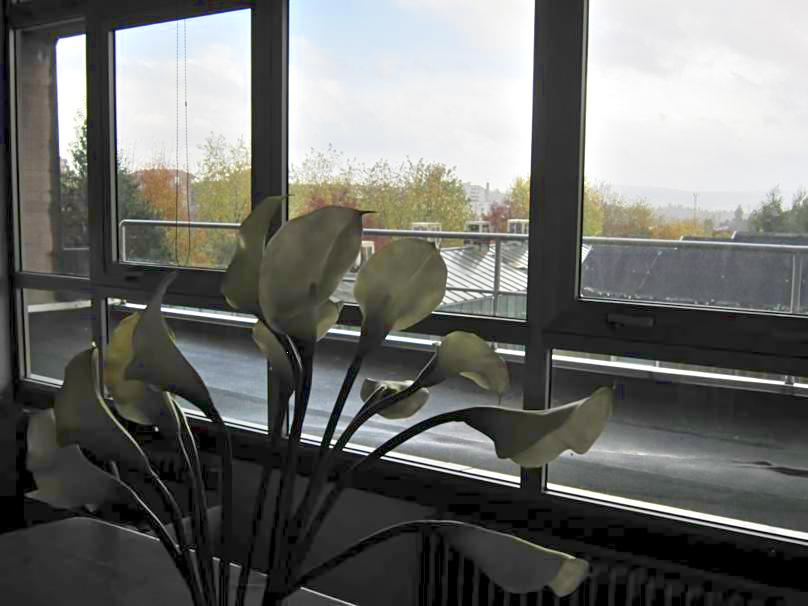}}

	\vspace{-2mm}
	\subfloat{\includegraphics[width=0.85in]{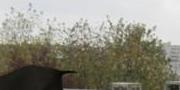}}
	\hfil
	\subfloat{\includegraphics[width=0.85in]{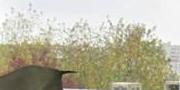}}
	\hfil
	\subfloat{\includegraphics[width=0.85in]{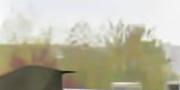}}
	\hfil
	\subfloat{\includegraphics[width=0.85in]{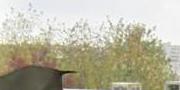}}
	
	\vspace{-3mm}
	\subfloat[(a)]{\includegraphics[width=0.85in]{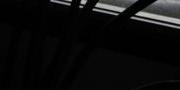}}
	\hfil
	\subfloat[(b)]{\includegraphics[width=0.85in]{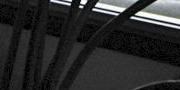}}
	\hfil
	\subfloat[(c)]{\includegraphics[width=0.85in]{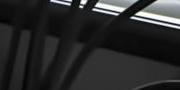}}
	\hfil
	\subfloat[(d)]{\includegraphics[width=0.85in]{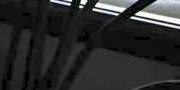}}
	\caption{Example of noise suppression. (a) Input. (b) Enhanced result without denoising. (c) Denoised version of (b). (d) Result of Eq. (\ref{eq_ns}).
	}
	\label{ref_fig_noise}
	\vspace{-4mm}
\end{figure}

Noise in dark regions is easily amplified after enhancement. Therefore, denoising is sometimes required in some applications. Given that noise in different regions of an unevenly-lit image is often amplified to different degrees, we propose a fusion scheme to suppress noise in dark regions while preventing details in bright regions from being over-smoothed, given by
\begin{equation}\label{eq_ns}
{{\bf{T}}_F} = {{\bf{L}}_{NR}}{{\bf{T}}_E} + \left( {1 - {{\bf{L}}_{NR}}} \right){{\bf{T}}_{DE}}
\end{equation}
where ${{\bf{T}}_F}$ is the final recomposed result, ${{\bf{L}}_{NR}}$ is the residual image normalized to [0,1], and ${{\bf{T}}_{DE}}$ is the denoised result of ${{\bf{T}}_{E}}$. This paper adopts the widely used denoising method BM3D \cite{dabov2007image} to achieve ${{\bf{T}}_{DE}}$. Fig. \ref{ref_fig_noise} exhibits an example of noise suppression. Compared to the images in (b) and (c), the result in (d) shows that fine details in the bright regions are nicely preserved, while noise in the dark regions is effectively smoothed out. We would like to mention that the denoising in Eq.(\ref{eq_ns}) can be appended as a post-processing step to any enhancement method for unevenly-lit images.

We must emphasize that noise suppression usually results in a loss of detail in shadows. Suppressing noise while preserving details remains an ongoing research topic in image enhancement. In this paper, we focus on improving contrasts rather than suppressing noise to avoid distraction. Introducing noise estimation methods to determine whether to suppress noise could be one of our following works.

\subsection{Limitations}
It will be of interest to improve the proposed method in our future work in the following aspects: 1) Our method is not good at enhancing over-exposed images because the residual image is adjusted by a fixed gamma curve. The gamma function cannot bring out details in both dark and bright regions. A possible way to alleviate this issue is to train a CNN to fit light enhancement curves under various lighting conditions. 2) The proposed method does not take into account more characteristics of the HVS, such as frequency selectivity and directional selectivity\cite{beghdadi2020a}. Incorporating more visual mechanisms related to natural image analysis into our model may help to improve the fidelity of the enhanced results.

\section{Conclusion}\label{conclusion}
In this paper, we have proposed a biological vision based exploratory data model to enhance images taken under unevenly-lit conditions. The proposed model decomposes the input image into its contrast image and residual image. The perceptually important details in the scene are preserved in the contrast image, while the lighting variations are retained in the residual image. The enhanced result can be achieved by manipulating the two images and recombining them. The major advantage of the proposed method is its simplicity and effectiveness. Unlike existing enhancement methods based on physical lighting models or deep-learning techniques, the proposed method does not require any explicit assumptions and prior knowledge of the natural scenes, nor any learning procedures. Despite its simplicity, experimental results demonstrate that the proposed method is comparable to several state-of-the-art methods.

\bibliographystyle{IEEEtran}
\bibliography{ref}

\end{document}